\documentclass{iopart}

\usepackage[bookmarks=false]{hyperref}
\usepackage{graphicx,epsfig}
\usepackage{longtable}
\usepackage{color}
\usepackage[sort,compress,numbers]{natbib}
\usepackage{doi}

\usepackage{bm}
\usepackage{capt-of}
\usepackage{relsize}
\usepackage{xcolor}
\usepackage{amssymb}
\usepackage[normalem]{ulem}
\usepackage{mathptmx}
\usepackage[export]{adjustbox}
\usepackage{url}

\def\nul{\nu_l}

\def\nubar{\bar\nu}

\def\nulbar{\bar\nu_l}

\newcommand{\eep} { $(e,e^{\,\prime}p)$ }
\newcommand{\ee} { $(e,e^{\,\prime})$ }
\newcommand{\beq}{\begin{eqnarray}}
\newcommand{\eeq}{\end{eqnarray}}
\def\to{\rightarrow}

\def\mix{\leftrightarrow}

\newcommand{\diff}{{\rm\,d}}                    

\newcommand{\np}{{\bf p}}

\newcommand{\nk}{{\bf k}}

\newcommand{\nq}{{\bf q}}

\begin{document}

\topical{Neutral Current Neutrino-Nucleus Scattering. Theory}

\author{Carlotta Giusti$^1$ and Martin V. Ivanov$^2$}

\address{$^1$Dipartimento di Fisica, Universit\`{a} degli Studi di Pavia and \\INFN, Sezione di Pavia, via A. Bassi 6, I-27100 Pavia, Italy}
\address{$^2$Institute for Nuclear Research and Nuclear Energy, Bulgarian Academy of Sciences, Sofia 1784, Bulgaria}

\ead{Carlotta.Giusti@pv.infn.it, martin@inrne.bas.bg}

\begin{abstract}
The treatment of nuclear effects in neutrino-nucleus interactions is one of the main sources of systematic uncertainty for the analysis and interpretation of data of neutrino oscillation experiments. Neutrinos interact with nuclei via charged or neutral currents and both cases  must be studied to obtain a complete information. We give an overview of the theoretical work that has been done to describe nuclear effects in neutral-current neutrino-nucleus scattering in the kinematic region ranging between beam energies of a few hundreds MeV to a few GeV, which is typical of most ongoing and future accelerator-based neutrino experiments, and where quasielastic scattering is the main interaction mechanism. We review the current status and challenges of the theoretical models, the role and relevance of the contributions of different nuclear effects, and the present status of the comparison between the numerical predictions of the models as well as the available experimental data. We discuss also the sensitivity to the strange form factors of the nucleon and the methods and observables that can allow one to obtain evidence for a possible strange quark contribution from measurements of neutrino and antineutrino-nucleus scattering.

\end{abstract}
\noindent{\it Keywords\/}: Neutrino-induced reactions, neutrino interactions with nuclei, weak neutral currents. \\
\submitto{\JPG}

\tableofcontents

\title[Neutral Current Neutrino-Nucleus Scattering. Theory]{}

\newpage
\section{Introduction \label{sec:intro}}

The discovery of neutrino oscillations~\cite{PhysRevLett.81.1562, PhysRevLett.87.071301, PhysRevLett.90.021802} provided unambiguous evidence for nonvanishing neutrino masses and neutrino mixing\footnote[1]{The Nobel Prize in Physics 2015 was awarded jointly to Takaaki Kajita and Arthur B. McDonald ``for the discovery of neutrino oscillations, which shows that neutrinos have mass.''~\cite{nobelprize2015}}. After the discovery, a number of reactor and accelerator-based neutrino oscillation experiments have been carried out over the past few years, with the aim to investigate neutrino properties. The next generation of future accelerator-based long-baseline experiments, such as the Deep Underground Neutrino Experiment (DUNE)~\cite{Acciarri:2016crz, Abi:2018dnh} in the US and Tokai-to-Hyper-Kamiokande (T2HK)~\cite{Abe:2015zbg} in Japan, will make precise measurements of neutrino oscillation parameters, in particular, with the aim to search for leptonic CP violation in appearance mode, thus addressing one of the outstanding fundamental problems of particle physics.

The accurate determination of neutrino properties requires high-statistics, high-precision experiments, which in turn require percent-level control of systematic uncertainties. The treatment of nuclear effects in the interaction of neutrinos with the atomic nuclei in the neutrino detectors is one of the main sources of systematic uncertainty. A deep and detailed knowledge of neutrino-nucleus interactions is therefore mandatory for a proper analysis and interpretation of the experimental data from neutrino oscillation experiments. Reliable models for cross section calculations are required, where all reaction mechanisms and nuclear effects are well under control and theoretical uncertainties are reduced as much as possible.

Neutrino-nucleus scattering has gained in recent years a wide interest that goes beyond the study of the intrinsic properties of neutrinos and extends to different fields. Besides being an appropriate tool to detect neutrinos, it plays an important role in understanding various astrophysical processes, as well as it can be used to answer some cosmological questions and to test the limits of the Standard Model. In hadronic and nuclear physics neutrino-nucleus scattering is a suitable tool to investigate nuclear structure,  to obtain information on the structure of the hadronic weak current, and on the strange quark contribution to the spin structure of the nucleon.

Neutrinos interact with nuclei via charged or neutral currents, transferring energy and momentum to the nucleus. In the charged current (CC) interaction, which occurs through the exchange of a $W^{\pm}$ boson, the neutrino is absorbed by the nucleus and the associated charged lepton is emitted. The charged lepton in the final state is generally the only particle to be detected. In the neutral current (NC) interaction, which occurs through the exchange of a $Z^0$ boson, the neutrino remains a neutrino and is therefore present in the final state. In this case what is detected can be either the recoil target or the reaction products. CC interactions are easier to observe, because the produced charged lepton, usually an electron or a muon, has characteristic signatures in particle detectors and is thus easy to identify. In addition, the identification allows one to deduce the flavor of the incoming neutrino. In contrast, NC interactions do not allow to identify the initial neutrino flavor.

Both CC and NC scatterings must be studied to obtain a complete information on neutrino-nucleus interactions. Weak neutral currents played an important role in the establishment of the Standard Model of electroweak interactions. They were predicted in 1973 by Abdus Salam, Sheldon Glashow, and Steven Weinberg\footnote[2]{The Nobel Prize in Physics 1979 was awarded jointly to Sheldon Lee Glashow, Abdus Salam and Steven Weinberg ``for their contributions to the theory of the unified weak and electromagnetic interaction between elementary particles, including, inter alia, the prediction of the weak neutral current.''~\cite{nobelprize1979}} and confirmed shortly thereafter in 1973, in a neutrino experiment in the Gargamelle bubble chamber at CERN~\cite{HASERT1973138}.

The experimental study of NC neutrino interactions is a demanding task, owing to the difficulties of collecting data on reactions with cross sections even smaller than those of CC processes, in which the outgoing neutrino leaves no signal and the event identification has to rely on the detection of one or more hadrons. NC reactions are relevant for oscillations experiments and contribute as an important event type for neutrino experiments. It is well known that NC $\pi^0$ production events are  a relevant source of background in $\nu_e$ appearance searches with $\nu_\mu$ beams, because they might be misidentified as CC $(\nu_e,e^-)$ interactions.

NC neutrino-nucleus interactions are important to determine the hadronic weak current and the strange quark contribution to the spin of the nucleon. We note that purely isovector CC processes do not depend on the strange form factors. In contrast, NC processes on nucleons and nuclei are sensitive to the nucleon isoscalar weak current. Therefore NC and CC processes give complementary information on nuclear effects in neutrino-nucleus scattering.

Many efforts have been devoted over the last few years and are being devoted to the development of theoretical models able to provide a fully quantitative description of neutrino-nucleus cross sections in the kinematic regime relevant to most ongoing and future accelerator-based neutrino experiments.

Several review papers about the experimental and theoretical status and prospects of neutrino-nucleus scattering have been published in recent years~\cite{Gallagher:2011zza, Morfin:2012kn, Alvarez-Ruso:2014bla, Garvey:2014exa, Mosel:2016cwa, Katori:2016yel, Benhar:2015wva, ALVAREZRUSO20181}.

In this paper we review the theoretical work that has been done to date to describe NC $\nu(\nubar)$-nucleus scattering. The experimental review can be found elsewhere in the present issue~\cite{Tayloe}. Our review focuses on medium-energy interactions, corresponding to beam energies ranging from a few hundreds MeV to a few GeV, {\it i.e.} in the kinematic region typical of  most ongoing and future accelerator-based neutrino experiments, where quasi-elastic (QE) neutrino scattering is the main interaction mechanism. We do not consider neutrino-induced pion-production, which also gives an important contribution to  neutrino-nucleus scattering cross sections and is reviewed elsewhere in the present issue~\cite{SB,JN}.

Our article is organized as follows: in section~\ref{sec:xs} we present the general formalism for $\nu(\nubar)$-nucleus scattering. Both NC and CC scatterings are considered and compared. The theoretical models that have been used to describe NC quasielastic scattering are reviewed and compared in section~\ref{sec:models}. We outline the main assumptions and approximations that have been adopted to describe various nuclear effects and discuss, also with the help of numerical examples in specific kinematic conditions, the role and relevance of their contributions in  different models. Section~\ref{sec:strange} is devoted to the strange nucleon form factors entering the structure of the nucleonic weak neutral current. We discuss their role in the calculated cross sections, as well as the methods and observables that can allow one to obtain evidence for a possible strange quark contribution from measurements of neutrino and $\nu(\nubar)$-nucleus scattering. In section~\ref{sec:results} we discuss how the numerical predictions of a model can be compared with the experimental flux-averaged differential cross sections and give  an overview of the present status of the comparison between theoretical and experimental results. Finally, in section~\ref{sec:conclusions} we draw our concluding remarks.

\section{General formalism for neutrino-nucleus scattering  \label{sec:xs}}

The process of neutrino and antineutrino scattering off a nucleus with mass number $A$ is usually described assuming the Born approximation, {\it i.e.} the one-boson exchange approximation (OBEA), where the exchanged virtual boson is a neutral $Z^0$ boson for the NC process and a charged $W^{\pm}$ boson for the CC one. In NC scattering the final-state lepton is a(n) (anti)neutrino of the same flavor as the incoming one, while in CC scattering the final-state lepton is the charged partner of the incoming flavor:
\beq
\nul (\nulbar) + A  \to  \nul (\nulbar) + X \hspace{0.5cm}  & \mathrm{NC} & \\
\nul (\nulbar) + A  \to  l^- (l^+) + X      \hspace{0.5cm}  & \mathrm{CC}, &
\eeq
where $l=e,\mu,\tau$, the target nucleus $A$ is in its ground state, and $X$ is the hadronic final state.

As in the case of electron scattering, in the lowest order OBEA  the cross section for $\nu$($\nubar$)-nucleus scattering is obtained from the contraction between a leptonic tensor $L^{\mu\nu}$ and a hadronic tensor $W_{\mu\nu}$:
\beq
\mathrm{d} \sigma = K  L^{\mu\nu}\ W_{\mu\nu},
\label{eq.cs1}
\eeq
where $K$ is a kinematic factor. The two situations differ in detail but the general structure is the same~\cite{DONNELLY1985183, Boffi:1993gs, book}.

\subsection{The leptonic tensor\label{sec:letpon}}

The leptonic tensor in the case of NC $\nu(\nubar)$-nucleus scattering is given by
\begin{equation}
\label{eq.lt_new}
 L^{\mu\nu}=
  \Big[\overline{\nu}({\bf k}')
       (\gamma^{\mu}\!\pm\!\gamma^{\mu}\gamma^{5})
       \nu({\bf k})\Big]
  \Big[\overline{\nu}({\bf k}')
       (\gamma^{\nu}\!\pm\!\gamma^{\nu}\gamma^{5})
       \nu({\bf k})\Big]^{*}.
\end{equation}
Here $k^\mu = (E,\nk)$ and $k'^\mu = (E^{\prime},\nk^{\prime})$ are the four-momenta of the incoming and outgoing (anti)neutrino, respectively, and the upper (lower) sign corresponds to $\nu$ ($\nubar$) scattering.

The lepton tensor has a similar structure in electron and neutrino-nucleus scattering~\cite{DONNELLY1985183, WALECKA1975113, Boffi:1993gs, book, Meucci:2003cv, Meucci:2004ip, vanderVentel:2003km}. After projecting into the initial $\nu$ ($\nubar$) and the final-lepton state, one has
\beq
L^{\mu\nu} = \frac {1} {2E E^{\prime}}
\mathrm{Tr} \left[ \gamma \cdot k^{\prime} \ \gamma^\mu \ (1\mp \gamma^5) \
\gamma \cdot k \ \gamma^\nu \right],
\label{eq.lt1}
\eeq
which can be written, by separating the symmetrical and antisymmetrical components, as
\beq
L^{\mu\nu} = \frac {2} {E E^{\prime}}
\left[ l_S^{\mu\nu} \mp l_A^{\mu\nu} \right],
\label{eq.lt2}
\eeq
with
\beq
l_S^{\mu\nu} = k^\mu \ k^{\prime\nu} + k^\nu \ k^{\prime\mu} - g^{\mu\nu} \ k \cdot k^{\prime},
\hspace{0.5cm} l_A^{\mu\nu} = i \ \epsilon ^{\mu\nu\alpha\beta} k_{\alpha} k^{\prime}_\beta,
\label{eq.lt3}
\eeq
where $\epsilon ^{\mu\nu\alpha\beta}$ is the antisymmetric tensor with $\epsilon_{0123} = - \epsilon^{0123} = 1$.

\begin{figure}[tb]
\begin{center}
\includegraphics[width=10cm,valign=m]{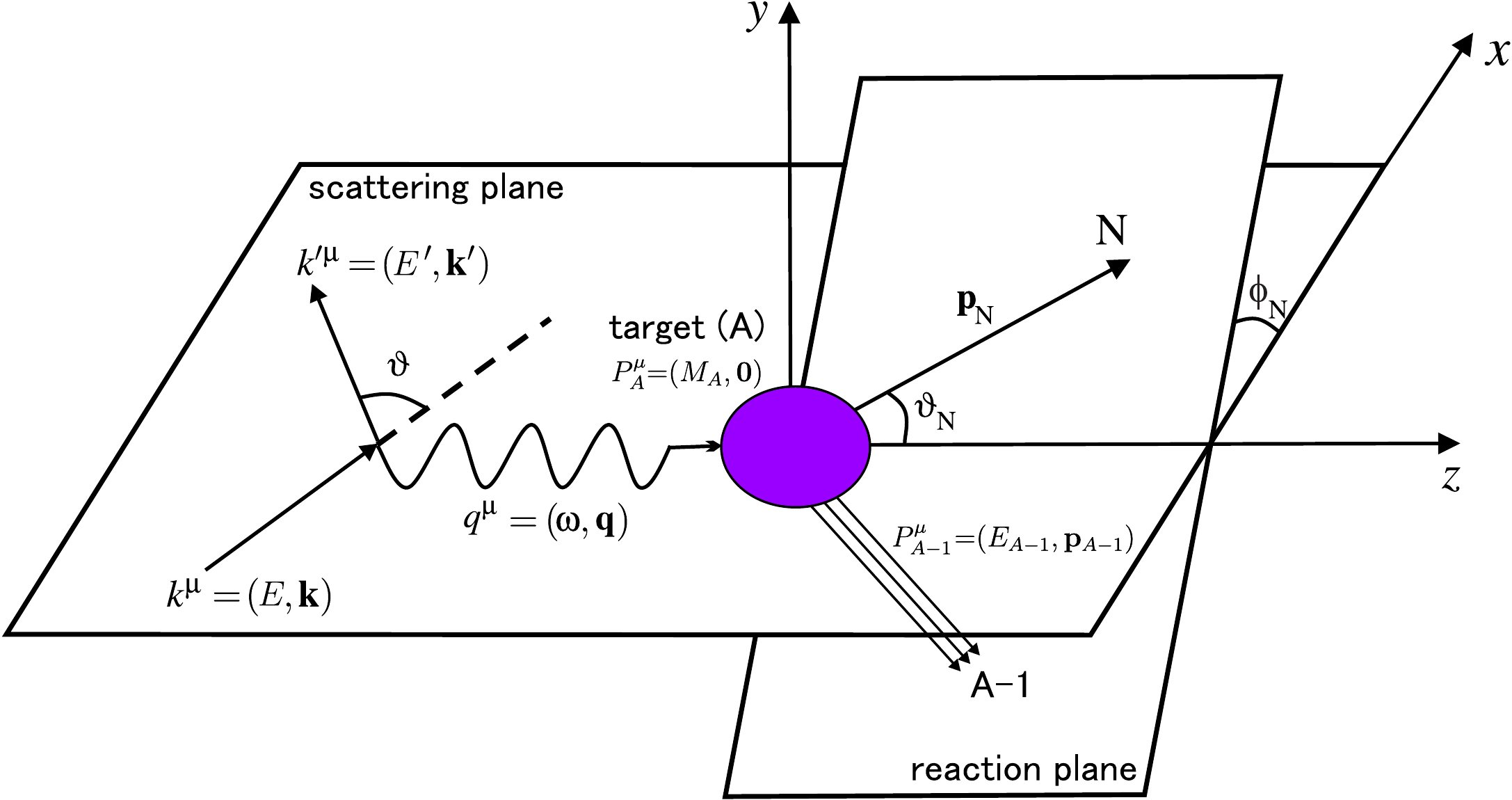}
\end{center}
\vspace{-2mm}
\caption{Kinematics for the quasielastic $\nu(\nubar)$-nucleus scattering process.}
\label{fig:kinematics}
\end{figure}

Assuming the reference frame where the $z$-axis is parallel to the momentum transfer $\nq = \nk - \nk^{\prime}$ and the $y$-axis is parallel to $\nk \times \nk^{\prime}$, the symmetrical components $l_S^{0y}, l_S^{xy}, l_S^{zy}$, and the antisymmetrical ones $l_A^{0x}, l_A^{xz}, l_A^{0z}$, as well as those obtained from them by exchanging their indices, vanish. Our convention for the kinematic variables is defined in figure~\ref{fig:kinematics}.

\subsection{The hadronic tensor\label{sec:hadron}}

The hadronic tensor is given by bilinear products of the matrix elements of the nuclear current operator $J^{\mu}$ between the initial state of the target nucleus $\mid\Psi_0\rangle$, with energy $E_0$, and the final state $\mid \Psi_{\textrm {x}} \rangle$, with energy $E_{\textrm {x}}$, as
\beq
W^{\mu\nu}=\sum_{\textrm {x}} \langle \Psi_{\textrm {x}}\mid J^{\mu}(\nq) \mid \Psi_0\rangle \langle \Psi_0\mid J^{\nu\dagger}(\nq) \mid
\Psi_{\textrm{x}}\rangle \delta (E_0 + \omega - E_{\textrm {x}}),
\label{eq.ha1}
\eeq
where the sum includes all hadronic final states and $ q \equiv q^\mu = (\omega,\nq)$ is the four-momentum transfer, with $Q^2 = -q^2 = |\nq|^2 - \omega^2$.

The hadronic tensor contains all information on nuclear structure and nuclear interactions, {\it i.e.} the entire response of the target, and, as such, it is a very complicated object.

As for the leptonic tensor, the hadronic tensor can be decomposed into two pieces, which are symmetric and antisymmetric under the exchange of the indices $\mu \mix \nu$. The explicit expression of the hadronic tensor depends on the specific process under consideration. For a particular process, its most general covariant form can be constructed from basic symmetry requirements and in terms of the independent four-vectors of the problem. For example, polarization degrees of freedom may be of interest and specific different tensors will have to be considered. Furthermore, the characterization of the process of interest according to which and how many particles are detected leads to different general decompositions of the nuclear response function for inclusive, semi-inclusive, and  exclusive processes.

Different models have been developed to calculate the hadronic tensor, and therefore the cross section, in different neutrino-scattering processes and for different kinematic regimes.

\subsection{Quasielastic scattering \label{sec:quasi}}

In the quasielastic (QE) region (at momentum transfers above approximately 500~MeV) the nuclear response is expected to be dominated by the process where the (anti)neutrino directly interacts with a quasifree nucleon which is then ejected from the nucleus by a direct knockout mechanism. Therefore, in the QE region one nucleon is emitted:
\beq
\nul (\nulbar) + A  \to  \nul (\nulbar) + N + (A-1)
\hspace{0.5cm}  &\mathrm{NC}&,  \\
\nul (\nulbar) + A  \to  l^- (l^+) + p(n) + (A-1) \hspace{0.5cm} &\mathrm{CC}&.
\eeq

The direct knockout mechanism is related to the impulse approximation (IA), which is based on the assumption that the incident particle interacts with the ejectile  nucleon only through a one-body current and the recoiling $(A-1)$-nucleon system acts as a spectator. The final states $\mid \Psi_{\textrm {x}} \rangle$ in equation~(\ref{eq.ha1}) factorize into the product of a discrete (or continuum) state $|n\rangle$ of the residual $(A-1)$-system and the scattering state $\chi^{(-)}_{\np_{\mathrm N}}$ of the emitted nucleon, with momentum $\np_{\mathrm N}$ and energy $E_{\mathrm N}$. Then, the transition amplitudes are obtained as
\beq
\langle n;\chi^{(-)}_{\np_{\mathrm N}}\mid J^{\mu}(\nq) \mid \Psi_0\rangle =
\langle\chi^{(-)}_{\np_{\mathrm N}}\mid   j^{\mu}
(\nq)\mid \varphi_n \rangle  \ , \label{eq.amp}
\eeq
where $j^\mu$ is the single-nucleon current operator and $\varphi_n = \langle n | \Psi_0\rangle$ describes the overlap between the initial nuclear state and the final state of the residual $(A-1)$-nucleon system, corresponding to one hole in the ground state of the target.

The IA has been extensively and successfully adopted for the analysis of QE electron-nucleus scattering data~\cite{Boffi:1993gs, book,RevModPhys.80.189}, both for exclusive and inclusive reactions. In electron-scattering experiments the emitted nucleon can be detected in coincidence with the scattered electron, the residual $(A-1)$-nucleon system is not detected, but it is possible to envisage kinematic situations where we can assume that it is left in a discrete eigenstate. The final nuclear state is, therefore, completely determined: this is an exclusive reaction, for example, the exclusive \eep knockout reaction. If only the scattered electron is detected, the final nuclear state is not determined, the cross section includes all available final nuclear states: this is the inclusive QE \ee scattering. The assumptions underlying the IA reduce the scattering process to collisions involving individual nucleons. In the exclusive scattering only one nucleon and one particular final state $|n\rangle$ are involved; in the inclusive scattering all the nucleons of the target and all the final states $|n\rangle$ are involved, and the cross section is obtained from the incoherent sum of the cross sections describing the scattering off all individual nucleons.

In NC neutrino-scattering experiments only the emitted nucleon can be detected and the cross section is integrated over the energy and angle of the final lepton. Also the state of the residual $(A-1)$-nucleon system is not determined and the cross section is summed over all the available final nuclear states. The same situation occurs for the CC reaction if only the outgoing nucleon is detected. The cross sections are therefore inclusive in the leptonic sector and semi-inclusive in the hadronic sector. In CC scattering the final charged lepton can be detected and the inclusive process where only the charged final lepton is detected (similar to the inclusive \ee scattering) can be considered. The exclusive CC process where the emitted nucleon is detected in coincidence with the charged final lepton can, in principle, be considered as well. However, a coincidence measurement represents in this case a very hard experimental task.

\subsection{Neutral-current and charged-current neutrino-nucleus cross sections\label{sec:NCCC}}

The most general form of the hadronic tensor and its contraction with the leptonic tensor in different lepton-nucleus scattering processes are derived and discussed, for instance,
in Refs.~\cite{WALECKA1975113, DONNELLY1985183, Boffi:1993gs, book, Donnelly:1996ke, Meucci:2003cv, Meucci:2004ip, vanderVentel:2003km, PhysRevC.73.024607, PhysRevC.76.045502}. Here we give the differential cross sections for the inclusive scattering  where only the final lepton is detected~\cite{Meucci:2003cv}
\beq
\frac{\diff \sigma} {\diff E^{\prime} \diff \Omega^{\prime}} =
 k^{\prime} E^{\prime} \ \frac{G^2} {4 \pi^2}
  \Big[ v_0 R_{00} + v_{zz} R_{zz} -  \ v_{0z} R_{0z} + v_T R_T \pm v_{xy} R_{xy} \Big]   \ ,
\label{eq.csin}
\eeq
and for the semi-inclusive process where one nucleon is detected and an integration over its solid angle is performed~\cite{Meucci:2004ip}
\beq
\frac{\diff \sigma} {\diff E^{\prime} \diff \Omega^{\prime}
\diff {T_{\mathrm N}}}& = &  \frac{G^2} {2 \pi^2} \, k^{\prime} E^{\prime}   \frac {|\np_{\mathrm N}| E _{\mathrm N}}  {(2 \pi)^3}\ \nonumber \\
 & \times & \int \diff \Omega_{\mathrm N} \Big [ v_0 R_{00} + v_{zz} R_{zz} - v_{0z} R_{0z} + v_T R_T   \pm v_{xy} R_{xy} \Big] \ .
\label{eq.cssi}
\eeq
In equations~(\ref{eq.csin}) and (\ref{eq.cssi}) $G= G_{\textrm F}$ for NC scattering and $G=G_{\textrm F} \cos \vartheta_{\textrm c}$ for CC scattering, where $G_{\textrm F} \simeq 1.16639 \times 10^{-11}$ MeV$^{-2}$ is the Fermi constant and $\vartheta_{\textrm c}$ is the Cabibbo angle ($\cos \vartheta_{\textrm c}\simeq 0.9749$).

In the case of CC scattering, the coefficients $v$, obtained from the leptonic tensor, are:
\beq
v_0 &=& 1 +  \tilde k^{\prime} \cos\vartheta \ , \nonumber \\
v_{zz} &=& 1 + \tilde k^{\prime} \cos\vartheta - 2 \frac{E |\nk^{\prime}|
\tilde k^{\prime}} {|\nq|^2} \sin^2\vartheta \ , \nonumber \\
v_{0z}&=& \frac{\omega}{|\nq|} \left(1 + \tilde k^{\prime} \cos\vartheta\right) + \frac
                    {m^2}{|\nq|E^{\prime}} \ , \nonumber \\
v_T&=& 1 - \tilde k^{\prime} \cos\vartheta + \frac{E |\nk^{\prime}|\tilde k^{\prime}} {|\nq|^2}
                   \sin^2\vartheta \ , \nonumber \\
v_{xy}&=& \frac {E + E^{\prime}}{|\nq|} \big( 1 - \tilde k^{\prime}
\cos\vartheta \big) -
\frac {m^2}{|\nq|E^{\prime}} \ ,
\label{eq.leptoncc}
\eeq
where $\tilde k^{\prime} = |\nk^{\prime}|/E^{\prime}$ and  $\vartheta$ is the lepton scattering angle. In the case of NC scattering the coefficients $v$ are obtained from those in equation~(\ref{eq.leptoncc}) by assuming $m=0$ and $\tilde k^{\prime} =1$,
{\it i.e.}
\begin{eqnarray}
v_0 &=& 2 \cos^2 \frac{\vartheta}{2} = \beta \ \ , \ \
v_{zz} = \frac {\omega^2} {|\nq|^2}\ \beta \ , \ \
v_{0z} = \frac{\omega} {|\nq|} \ \beta \quad ,  \nonumber \\
v_T&=&\tan^2\frac {\vartheta}{2} + \frac{Q^2} {2|\nq|^2} \ \beta \ , \ \
v_{xy}= \tan \frac {\vartheta}{2} \left[ \tan^2\frac {\vartheta}{2} +
 \frac{Q^2} {|\nq|^2} \right]^{\frac {1} {2}} \ \beta \ .
\label{eq.lepton}
\end{eqnarray}

The structure functions are given in terms of the components of the hadronic tensor as
\beq
R_{00} &=& W^{00}\ , \nonumber \\
R_{zz} &=& W^{zz}\ , \nonumber \\
R_{0z} &=& W^{0z} + W^{z0}\ , \nonumber \\
R_T  &=& W^{xx} + W^{yy}\ , \nonumber \\
R_{xy} &=& i\left( W^{yx} - W^{xy}\right)\ .
\label{eq.rf}
\eeq

\subsection{The single-nucleon weak current\label{sec:one-body}}

The components of the hadronic tensor are given in equation~(\ref{eq.ha1}) by products of the matrix elements of the nuclear current operator $J^\mu$, which in general is the sum of one-nucleon and two-nucleon contributions. In the IA the matrix elements are obtained in equation~(\ref{eq.amp}) in the single-particle representation and the nuclear current is assumed to be a one-body single-nucleon current.

Within the relativistic IA (RIA) the single-nucleon operator related to the neutral and charged weak current is
\beq
  j^{\mu} &=&  F_1^{\textrm V}(Q^2) \gamma ^{\mu} +
             i\frac {\kappa}{2M} F_2^{\textrm V}(Q^2)\sigma^{\mu\nu}q_{\nu}
	     -G_{\textrm A}(Q^2)\gamma ^{\mu}\gamma ^{5}
 \ \hspace{0.3cm} \ ({\mathrm {NC}})
\ , \nonumber
	     \\  	
  j^{\mu} &=&  \Big[F_1^{\textrm V}(Q^2) \gamma ^{\mu} +
             i\frac {\kappa}{2M} F_2^{\textrm
	     V}(Q^2)\sigma^{\mu\nu}q_{\nu} -G_{\textrm A}(Q^2)\gamma ^{\mu}\gamma ^{ 5} \nonumber \\
	    & + &
 F_{\textrm P}(Q^2)q^{\mu}\gamma ^{5}\Big]\tau^{\pm} \ \hspace{5cm}  \ ({\mathrm {CC}}) \ ,
	     \label{eq.nc1b}
\eeq
where $\tau^{\pm}$ are the isospin operators, $\kappa$ is the anomalous part of the magnetic moment, $M$ is the nucleon mass, and  $\sigma^{\mu\nu}=\left(i/2\right)\left[\gamma^{\mu},\gamma^{\nu}\right]$.

The weak isovector Dirac and Pauli form factors, $F_1^{\textrm V}$ and $F_2^{\textrm V}$, are related to the corresponding electromagnetic form factors by the conservation of the vector current (CVC) hypothesis~\cite{WALECKA1975113} plus, for NC reactions, a possible isoscalar strange-quark contribution $F_i^{\mathrm s}$, {\it i.e.},
\beq
F_i^{\mathrm {V,p(n)}} (Q^2) &=& \left(\frac{1}{2} -
2\sin^2{\theta_{\mathrm W}}\right)
 F_i^{\mathrm {p(n)}} (Q^2) -\frac{1}{2} F_i^{\mathrm {n(p)}} (Q^2)  -
 \frac{1}{2} F_i^{\mathrm s} (Q^2) \ ({\mathrm {NC}}), \nonumber \\
F_i^{\mathrm V} (Q^2) &=&
 F_i^{\mathrm p}(Q^2) - F_i^{\mathrm n}(Q^2)
 \ \hspace{3.5cm} ({\mathrm {CC}}) \ , \label{eq.nc1c}
\eeq
where $\theta_{\mathrm W}$ is the Weinberg angle $(\sin^2{\theta_{\mathrm W}} \simeq 0.23143)$.

The present knowledge of the nucleon electromagnetic form factors is rather precise, particularly in the proton case, through electron-scattering measurements. Different models are able to provide a proper description of data, but some ambiguities emerge from the analysis of different experiments, in particular, concerning the behavior of the form factors as a function of $Q^2$~\cite{GonzalezJimenez:2011fq}. The present knowledge about the vector strange form factors is rather limited, although some general considerations can be made~\cite{GonzalezJimenez:2011fq}. With regard to the functional dependence with $Q^2$, a standard procedure is to assume the usual dipole form as \cite{Alberico:2001sd}
\beq
F_1^{\mathrm s}(Q^2) =  \frac {(\rho^{\mathrm s} +
\mu^{\mathrm s}) \tau}{(1+\tau) (1+Q^2/M_{\mathrm V}^2)^2}\ , \
F_2^{\mathrm s}(Q^2) =  \frac {\left(\mu^{\mathrm s}-\tau \rho^{\mathrm s}
\right)}{(1+\tau) (1+Q^2/M_{\mathrm V}^2)^2}\ ,
\label{eq.sform}
\eeq
where $\tau = Q^2/(4M^2)$ and $M_{\mathrm V} = 0.843$~GeV. The constants $\rho^{\mathrm s}$ and $\mu^{\mathrm s}$ describe the strange quark contribution to the electric and magnetic form factors, respectively, A functional dependence of the strange vector form factors with $Q^2$,  based on a monopole form $(1+Q^2/M_{\mathrm V}^2)^{-1}$, has been explored  as well in Ref.~\cite{GonzalezJimenez:2011fq}.

The axial form factor $G_{\mathrm A}$ in equation~(\ref{eq.nc1b}) is usually expressed as~\cite{Musolf:1992xm}
\beq
G_{\mathrm A} (Q^2) &=& \frac{1}{2} \left( \tau_3 g_{\mathrm A} -
g^{\mathrm s}_{\mathrm A}\right) G(Q^2) \  \ \ ({\mathrm {NC}}) \ , \nonumber \\
G_{\mathrm A} (Q^2) &=&   g_{\mathrm A} G\ (Q^2) \ \ ({\mathrm {CC}}) \ , \label{eq.ga}
\eeq
where $g_{\mathrm A} \simeq 1.2723(23)$~\cite{Patrignani}, $g^{\mathrm s}_{\mathrm A}$ describes possible strange-quark contributions, and $\tau_3 = +1 (-1)$ for proton (neutron) knockout.

Concerning the dependence of the axial form factor on $Q^2$, the usual procedure is to assume the standard dipole form, {\it i.e.},
\beq
G(Q^2) = (1+Q^2/M_{\mathrm A}^2)^{-2} \ .
\label{eq.ma}
\eeq
Recent studies~\cite{PhysRevC.76.025202, Beckannurev.nucl.51.101701.132312, Musolf:1993tb} have addressed the importance of including radiative corrections in the axial-vector term. For the axial mass $M_{\mathrm A}$ the standard value extracted from CCQE experiments on deuterium bubble chambers is $M_{\mathrm A} = 1.026\pm0.021$~GeV~\cite{Bernard:2001rs}. The value of the axial mass raised a strong debate in connection with the description of MiniBooNE CCQE data~\cite{PhysRevD.81.092005}, which seemed to require a larger value of $M_{\mathrm A}$. A new extraction of the axial form factor, performed in Ref.~\cite{PhysRevD.93.113015} by analyzing neutrino-deuteron scattering world data using a model-independent representation of the axial form factor,  confirms the value in Ref.~\cite{Bernard:2001rs} but with a much larger uncertainty~\cite{Katori:2016yel}. It was pointed~\cite{PhysRevD.81.092005} that the very small uncertainty often quoted for $G_{\mathrm A}(Q^2)$  may be related to the artificial constraints of the dipole ansatz. A recent theoretical work~\cite{PhysRevD.93.053002} models the axial form factor by considering the axial-vector dominance using the masses and widths of the lightest axial mesons as parameters, as obtained from the averaged Particle Data Group values~\cite{Agashe:2014kda}. Precise determinations of $G_{\mathrm A}(Q^2)$  are expected from lattice QCD calculations~\cite{Katori:2016yel, Meyer:2016kwb, Alexandrou:2018zdf, ALVAREZRUSO20181}.

The induced pseudoscalar form factor $F_{\mathrm P}(Q^2)$ can be connected to the axial form factor $G_{\mathrm A}(Q^2)$  making use of the partially conserved axial current (PCAC) hypothesis~\cite{Thomas:2001kw} as
\beq
F_{\mathrm P} (Q^2) = \frac{2M G_{\mathrm A} (Q^2)}{m_{\pi}^2 + Q^2}  \ , \label{eq.fp}
\eeq
where $m_{\pi}$ is the pion mass. In the cross section the contribution of  $F_{\mathrm P}(Q^2)$ is suppressed by a factor $m^2$ (for free nucleons) and therefore it is much less important than $G_{\mathrm A}(Q^2)$.

\section{Theoretical models \label{sec:models}}

Quasielastic scattering is the main interaction mechanism for neutrinos with energy up to about 1~GeV, in the energy region at the core of the energy distribution for many neutrino experiments. A wide variety of theoretical models has been used to describe QE $\nu(\nubar)$-nucleus scattering.

In this section models for NC quasielastic scattering are reviewed and compared. Usually the models adopted for NC and CC scattering are the same. The two processes have many similar aspects, but the kinematic situation is different. In NC scattering experiments the final (anti)neutrino is not observed and only the emitted nucleon can be detected, while in CC scattering either the emitted nucleon or the final charged lepton, or, in principle, both can be detected. Thus, the two processees could show a different sensitivity to different nuclear effects.

Most of the models applied to QE $\nu(\nubar)$-nucleus scattering were originally developed for QE electron-nucleus scattering, where a large amount of accurate data provided, in the comparison with the results of sophisticated models, a wealth of detailed information on nuclear properties. The extension to neutrino scattering of the formalism and of the models developed for electron scattering is straightforward. In spite of many similar aspects, however, the two situations are different. There are differences in the nuclear current and, in general, in the kinematic situation, and it is not guaranteed that a model able to describe electron-scattering data will be able to describe neutrino-scattering data with the same accuracy. Since, for instance, the vector part of the weak response is related to the electromagnetic response through CVC, the comparison with electron-scattering data is a first necessary but not sufficient test of the validity of a nuclear model. Although not sufficient, such a test represents, in any case, a necessary prerequisite to assess the validity and the predictive power of a model: if the model is able to describe electron-scattering data it can be extended to $\nu(\nubar)$-nucleus scattering; if the model is unable to describe  electron-scattering data its extension to $\nu(\nubar)$-nucleus scattering is meaningless.

\subsection{The relativistic impulse approximation \label{sec:ria}}

In the QE region it is usually assumed that the nuclear response is dominated by the process where the incoming particle directly interacts, through a one-body current, only with the ejectile nucleon and the recoiling $(A-1)$-nucleon system acts as a spectator. This is the basis of the IA and corresponds to one-particle-one hole (1p-1h) excitations. In spite of the simplicity of the elementary reaction mechanism, the IA represents a complicated many-body problem, which involves a proper treatment of nuclear structure, including nuclear correlations, and of the final-state interactions (FSI) between the outgoing nucleon and the residual nucleus.

Nonrelativistic and relativistic models can be considered, but in the kinematic regime of ongoing and future neutrino experiments, with typical energies of the order or larger than the nucleon mass, relativistic effects are important and a relativistic model is required, where not only a relativistic kinematics, but also relativistic nuclear dynamics and current operators are taken into account. Some of the available models are based on the relativistic IA (RIA), other theoretical approaches include relativistic effects on models based on a nonrelativistic reduction of nuclear dynamics.

The simplest nuclear model used for QE lepton-nucleus scattering is the relativistic Fermi gas (RFG) model~\cite{Smith:1972xh, PhysRevC.38.1801, PhysRevC.48.3078, Alberico:1997vh}, where the nucleus is viewed as a gas of non-interacting free nucleons described by Dirac spinors. The RFG is a very schematic model that just takes into account the average kinetic energy of the nucleons in the nuclear medium. Its main advantage is its simplicity, while maintaining important aspects of the problem, such as, Lorentz covariance and gauge invariance, in a fully relativistic way. Due to its simplicity, the RFG model is used in most Monte Carlo generators and as a basis of more sophisticated  theoretical models where different types of nuclear correlations are included. Its extreme simplicity, however, makes the RFG inadequate to describe data for QE electron- and neutrino-nucleus scattering. The model ignores important effects, due to nuclear shell structure and correlations (only correlations due to the Pauli principle are included), FSI, and two-body meson-exchange currents (MEC).

Nuclear shell structure and detailed single-particle (s.p.) properties are accounted for in a relativistic shell model (RSM), where relativistic shell-model wave functions are assumed for the initial and for the residual nucleus and a scattering state for the emitted nucleon. In the RSM the overlap  function, $\varphi_n$ in equation~(\ref{eq.amp}), is self-consistent solution of the Dirac-Hartree equation derived within a mean-field approach from a Lagrangian including $\sigma, \omega$, and $\rho$ mesons~\cite{Serot:1984ey, Rein:1989, Ring:1996, Lalazissis:1996rd, Serot:1997xg}. The scattering wave function of the emitted nucleon, $\chi^{(-)}$ in equation~(\ref{eq.amp}), includes the FSI between the outgoing nucleon and the residual $(A-1)$-nucleon nucleus. In the simplest approach FSI are neglected and a plane-wave (PW) approximation is adopted for the scattering wave function. This is the so-called RPWIA. In many calculations FSI are accounted for by a complex energy-dependent (ED) optical potential (OP) and the outgoing-nucleon wave function is a distorted wave (DW) function obtained by solving the  Dirac equation with a strong scalar and vector relativistic OP (ROP). This is the so-called RDWIA.

We notice that RPWIA and RDWIA, as well as the corresponding nonrelativistic PWIA and DWIA, do not necessarily entail that a RSM or SM approximation is assumed for the overlap  function $\varphi_n$, which can and, in principle, should be obtained from the calculation of the hole spectral function (SF)~\cite{Boffi:1993gs, book}. The assumption of SM or RSM wave functions for $\varphi_n$ represents an additional approximation to the IA, which gives a simple description of the SF. However, in this paper, as well as in many other papers on neutrino-nucleus scattering, RPWIA and RDWIA denote RIA within the RSM.

In standard RDWIA calculations phenomenological ROPs are usually adopted, whose parameters are obtained through  a fit to elastic proton-nucleus scattering data. A series of global ({\it i.e.} spanning a large range of kinetic energies of the nucleon, up to $1040$~MeV) ED ROPs are available in $A$-dependent (AD, fitted to data on a wide range of nuclei) and $A$-independent (AI, fitted to data on a specific nucleus) versions~\cite{Cooper:2009,Cooper:1993nx}.

The RDWIA has been successfully tested against a large amount of high-quality data for the coincidence single-nucleon knockout $(e,e{^\prime}p)$ reaction~\cite{Udias:1996iy, Udias:1993xy, PhysRevLett.84.3265, Meucci:2001qc} and it has then been extended to QE $\nu(\nubar)$-nucleus scattering~\cite{Alberico:1997vh, Alberico:1997rm, Maieron:2003df, Meucci:2004ip, Meucci:2006ir, PhysRevC.73.024607, PhysRevC.76.045502, Kim:2007ez, PhysRevC.77.054604, Butkevich:2011fu}.

In electron scattering experiments it is possible to consider the exclusive reaction, where the scattered electron and the knocked-out nucleon are detected in coincidence and the residual nucleus is left in a discrete eigenstate. In this case the final nuclear state is completely determined and in the RSM the cross section is obtained from the knockout of a nucleon from a particular shell-model state. In $\nu(\nubar)$-nucleus scattering RPWIA and RDWIA calculations within a RSM framework can be performed for the corresponding  exclusive reaction, but experimentally it is extremely hard to achieve a situation where the final nuclear state is completely determined. The beam energy is in general not known and in the final state only the emitted nucleon can be detected in NC scattering, while in CC scattering either the final charged lepton or the emitted nucleon can be detected. Thus, even if we assume a definite value for the incident (anti)neutrino energy, the final nuclear state is not determined, the cross section contains the contributions of all the available final nuclear states, and in the IA it is obtained from the incoherent sum of one-nucleon knockout processes over all the individual nucleons of the target, over all the occupied SM states.

Electron scattering studies have shown the validity but, at the same time, also  the limitation of a pure SM description for nuclear structure. Accurate data for the exclusive $(e,e{^\prime}p)$ reaction provided evidence for the SM states, but the fact that the spectroscopic factors, {\it i.e.}, the norm of the overlap functions,  extracted for these states from the comparison between experimental data and DWIA and RDWIA results, are lower than predicted by the SM is a clear indication of the need to include correlations beyond the mean-field approximation (MFA).

\subsection{The spectral function \label{sec:sf}}

Models based on realistic spectral functions are well suited to identify mean-field  and correlation effects. The hole SF is a crucial ingredient of the cross section of the $(e,e{^\prime}p)$ reaction in the IA and contains the information on nuclear structure and correlations available from the  one-nucleon knockout process.

In the SM, where  the nucleons in the nuclear ground state occupy the $A$ lowest-energy eigenstates of the nuclear Hamiltonian, the SF is obtained as
\beq
S(\np,E) = \sum_n \mid \varphi_n(\np) \mid^2 \delta (E - E_n)  \ , \label{eq.sfsm}
\eeq
where the sum is over all the occupied SM states, labeled by the index $n$, with eigenvalues $E_n$.

In a  more sophisticated and realistic nuclear model, where the effects of nuclear correlations are taken into account, more complex states, where one or more of the spectator nucleons are excited,  are included in the SF.

Calculations of  $\nu(\nubar)$-nucleus cross sections have been carried out using SFs obtained within different correlation approaches: the correlated basis function (CBF) theory~\cite{Benhar:2010nx, Benhar:2011wy, PhysRevC.74.054316, Ankowski:2012ei, Ankowski:2013gha, Ankowski:2015lma, Benhar:2015wva, Benhar:2015xga, Rocco:2015cil, Benhar:2016jkq, Vagnoni:2017hll}, the self-consistent Green's function (SCGF) approach~\cite{Rocco:2018mwt}, and  the realistic spectral function model~\cite{PhysRevC.83.045504, Ivanov:2013saa, Ivanov:2015wpa, Ivanov:2018nlm} (for more details, see Section~\ref{sec:susa}). These correlation models are inherently nonrelativistic and their extension to the kinematic region relevant to accelerator-based neutrino experiments, where a relativistic model is required, presents serious conceptual difficulties.

The SF model for QE lepton-nucleus scattering is based on the factorization ansatz, which amounts to writing  the nuclear final state as a product of a plane wave, describing the motion of the knocked-out nucleon, and the $(A-1)$-nucleon recoiling spectator system. Within this scheme, and neglecting FSI, the cross section reduces to the convolution of the elementary lepton-nucleon cross section with the target SF, yielding the energy and momentum distribution of the struck nucleon in the initial state~\cite{Ankowski:2015lma}. Therefore in this model the SF, describing the intrinsic properties of the target, can be obtained from a nonrelativistic many-body theory and the matrix elements of the nucleon current can be computed using its fully relativistic form.

In the SF model of Refs.~\cite{Benhar:2005dj, Benhar:2010nx, Benhar:2011wy, Ankowski:2013gha, Ankowski:2015lma, Benhar:2015wva, Benhar:2015xga, Rocco:2015cil, Benhar:2016jkq, Vagnoni:2017hll} the hole SF is given by  the sum of two terms: a mean-field contribution, accounting for the shell structure of the nucleus, like in equation~(\ref{eq.sfsm}) but with spectroscopic factors extracted from $(e,e^{\prime}p)$ data,  and the contribution of nucleon-nucleon ($NN$) short-range correlations, obtained from nuclear matter calculations performed in the CBF theory~\cite{PhysRevC.41.R24,BENHAR1994493}. Those two components are consistently combined within the framework of the local-density approximation (LDA), which is based on the tenet that $NN$ correlations are largely unaffected by surface and shell effects, and allows one to obtain the correlation contribution for a finite nucleus from the corresponding results computed for uniform nuclear matter at different densities.

With this approach the SF can in principle be obtained for any nucleus, provided accurate $(e,e^{\prime}p)$ data are available for that nucleus. The SF formalism has been used to calculate cross sections for NC and CC $\nu(\nubar)$ scattering off $^{12}$C and $^{16}$O nuclei~\cite{Benhar:2010nx, Benhar:2011wy, Ankowski:2013gha, Ankowski:2015lma, Benhar:2015wva, Benhar:2015xga, Rocco:2015cil, Benhar:2016jkq, Vagnoni:2017hll, Rocco:2018mwt}, for which accurate $(e,e^{\prime}p)$ data are available. Since future neutrino experiments, such as DUNE~\cite{Acciarri:2016crz, Abi:2018dnh} and SBN~\cite{Antonello:2015lea} programs, will use large liquid argon detectors, it is very important to obtain the SF of $^{40}$Ar, a nucleus for which no $(e,e^{\prime}p)$ data is available till now. A new measurement of the coincidence $^{40}$Ar$(e,e^{\prime}p)$ cross section at Jefferson Lab~\cite{Benhar:2014nca} will provide the experimental input indispensable to construct the argon SF.

\begin{figure}[tb]
\centering
\includegraphics[width=\textwidth]{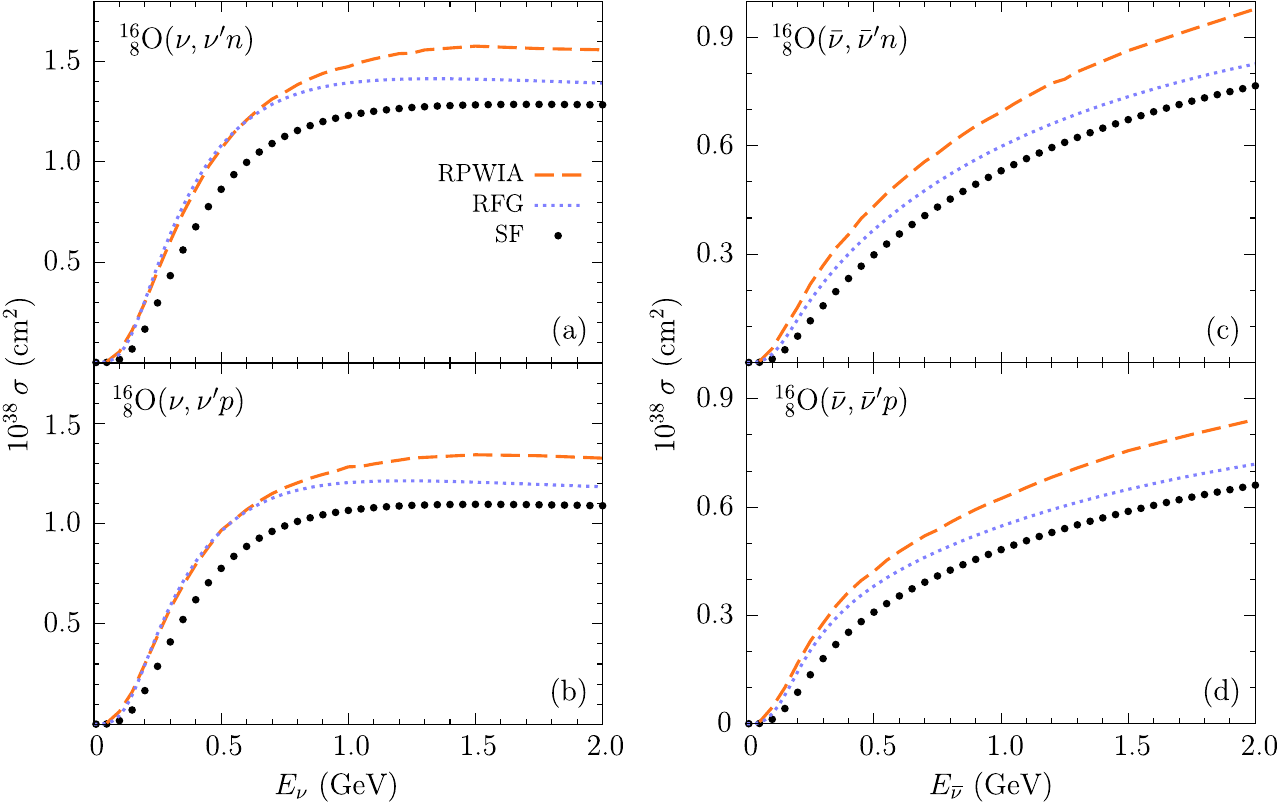}
\vspace{-2mm}
\caption{Cross sections calculated in the RFG, RPWIA, and SF models for neutron [(a) and (c)] and proton [(b) and (d)] knockout from $^{16}$O induced by NC QE interaction of neutrino [(a) and (b)] and antineutrino [(c) and (d)]. Taken from Ref.~\cite{Ankowski:2015lma}.}
\label{fig:RPWIA-RFG-SF}
\end{figure}

The NCQE $\nu(\nubar)$-$^{16}$O cross sections calculated for neutron and proton knockout in the RPWIA, RFG, and SF~\cite{PhysRevC.41.R24,Benhar:2005dj} models are compared in figure~\ref{fig:RPWIA-RFG-SF} as a function of the incident neutrino and antineutrino energy. Since FSI are neglected, the comparison provides information on the uncertainties associated with the description of the initial state. For neutrino scattering such uncertainties are within $\sim$10 and $\sim$25\%, depending on the energy, and turn out to be similar for neutron and proton knockout. Larger uncertainties, between $\sim$20 and $\sim$40\%, are obtained for antineutrino scattering.

The cross sections of NC and CC scattering off $^{12}$C and $^{16}$O calculated with two SFs, derived within the framework of the nuclear many-body theory using the CBF formalism~\cite{BENHAR1994493} and the SCGF theory~\cite{Dickhoff:2004xx, Barbieri:2016uib} have been compared in Ref.~\cite{Rocco:2018mwt}. The two approaches start from different nuclear Hamiltonians to describe the interactions between nucleons and make use of different approximations to calculate the hole SF.

As a consequence of the presence of the contribution of correlations in the SF, the cross section contains, in addition to single-nucleon knockout processes, where the residual nucleus is left in a bound 1p-1h state, also contributions of interactions leading to the excitation of 2p-2h final states. In the SF model, however, these contributions originate from initial-state dynamics and would be vanishing without ground-state correlations. Additional 2p-2h contributions arise from FSI.

\subsection{Final-state interactions \label{sec:fsi}}

The outgoing nucleon, in its way out of the nuclear target, can give energy to other nucleons, can change direction, scatter and rescatter, and excite, and even emit, other nucleons. These FSI processes can mask the initial process. Therefore, a reliable model for cross section calculations requires a reliable description of FSI. The role of FSI depends on the specific reaction under investigation and on the kinematic conditions, but a consistent theory should be able to describe FSI in exclusive and inclusive reactions.

In the simplest RFG model FSI are neglected. In Ref.~\cite{Co:2002jiq} a phenomenological convolution model was applied to the RFG to account for nucleon rescattering, showing that it can produce a reduction of the QE cross section as large as $15\%$ at incoming neutrino energies of about 1 GeV.

In calculations based on the SF model~\cite{Benhar:2010nx, Benhar:2011wy, Ankowski:2013gha, Ankowski:2015lma, Benhar:2015wva, Benhar:2015xga, Rocco:2015cil, Benhar:2016jkq, Vagnoni:2017hll, Rocco:2018mwt} FSI can be taken into account through a convolution scheme~\cite{PhysRevC.44.2328,PhysRevD.91.033005}, which amounts to integrating the IA prediction with a folding function. In addition, the propagation of the knocked-out particle in the mean-field generated by the spectator system is described by modifying  the energy spectrum of the knocked-out nucleon with the real part of a ROP derived from the Dirac phenomenological fit of Ref.~\cite{Cooper:1993nx}.

In the RDWIA FSI are taken into account in the outgoing-nucleon scattering wave function, which is eigenfunction of a complex energy-dependent ROP. We note that in the RDWIA the factorization between the SF and the matrix elements of the nuclear current is destroyed. The basic ingredients of the calculation are the matrix elements of equation~(\ref{eq.amp}), where the three main ingredients of the model, {\it i.e.}, the  overlap function, the one-body nuclear current, and the scattering wave function, are intertwined.

\begin{figure}[tb]
\centering
\includegraphics[width=0.8\textwidth]{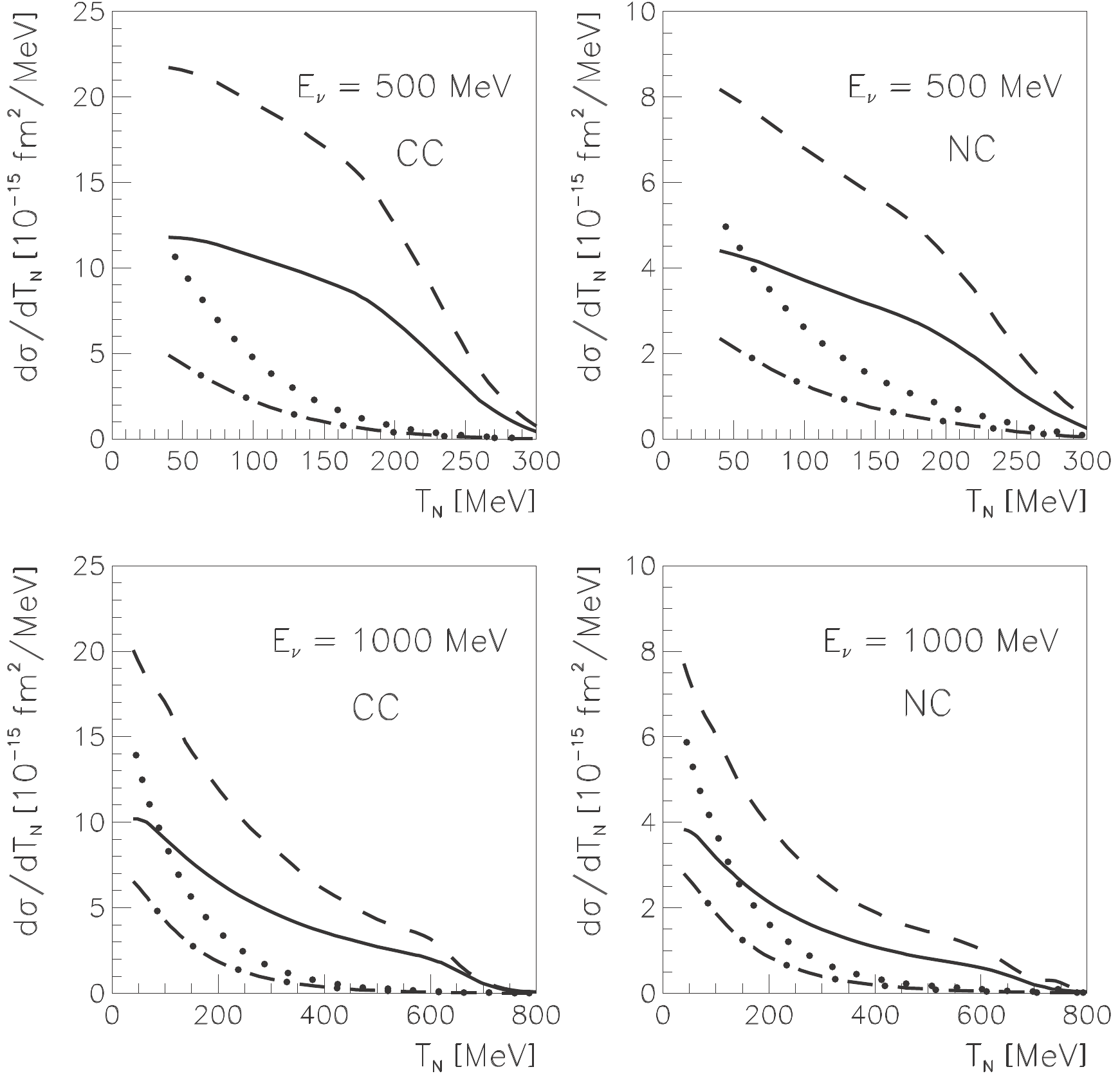}
\vspace{-2mm}
\caption{Differential cross sections of the CC and NC QE $\nu$ ($\nubar$) scattering on $^{12}$C as a function of outgoing nucleon kinetic energy $T_{\mathrm N}$. Solid and dashed lines are the results in RDWIA and RPWIA, respectively, for an incident neutrino. Dot-dashed and dotted lines are the results in RDWIA and RPWIA, respectively, for an incident antineutrino.
Taken from Ref.~\cite{Giusti:2009ym}.}
\label{fig:RPWIA-RDWIA}
\end{figure}

An example of the comparison between RPWIA and RDWIA results is shown in figure~\ref{fig:RPWIA-RDWIA}, where the cross sections of the $^{12}$C$\left(\nu,\mu^-p\right)$ and $^{12}$C$\left(\bar\nu,\mu^+n\right)$ CC reactions and of the $^{12}$C$\left(\nu,\nu p\right)$ and $^{12}$C$\left(\bar\nu,\bar\nu p\right)$ NC reactions are compared at $500$ and $1000$ MeV. The cross sections for CC scattering are much larger than the ones for NC scattering, but the effects of FSI, which reduce the cross sections by $\sim 50$\%, are similar in the two cases. The reduction is due to the imaginary part of the ROP and it is in agreement with the reduction found in the RDWIA calculations for the  exclusive $(e,e^{\prime}p)$ reaction. We note that the cross sections for
an incident neutrino are larger than for an incident antineutrino.

\begin{figure}[tb]
\centering
    \includegraphics[width=\textwidth]{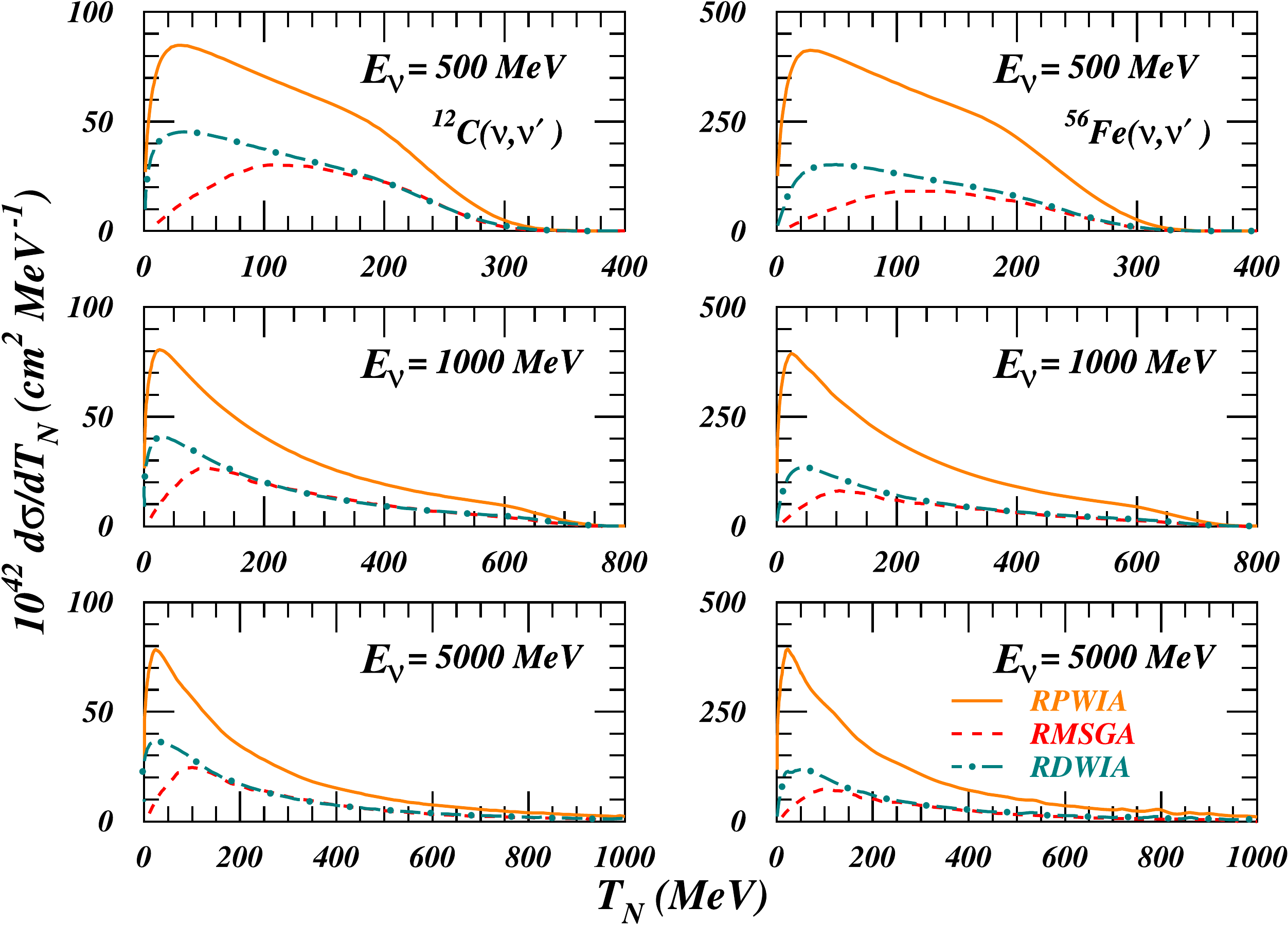}
\vspace{-2mm}
\caption{NC $^{12}$C$({\nu},{\nu}')$ (left panels) and $^{56}$Fe$({\nu},{\nu}')$ (right panels) cross sections as a function of the kinetic energy of the outgoing nucleon $T_{\mathrm N}$ at different incoming neutrino energies ($E_{\nu}$ in the figure) calculated in RPWIA (solid lines), RDWIA (dot-dashed lines), and RMSGA (dashed lines). Taken from Ref.~\cite{PhysRevC.73.024607}.}
\label{fig:RDWIA-RMSGA}
\end{figure}

The results obtained for the single-nucleon knockout contribution to CC and NC $\nu(\nubar)$-nucleus scattering in RDWIA  and with an alternative description of FSI based on a relativistic multiple-scattering Glauber approximation (RMSGA)  have been compared in Ref.~\cite{PhysRevC.73.024607}. The Glauber approach relies on the eikonal and the frozen-spectators approximation. It allows formulation of a full-fledged multiple-scattering theory for the emission of a fast nucleon from a system consisting of $(A-1)$ temporarily ``frozen'' nucleons~\cite{RYCKEBUSCH2003226}. At lower energies, the RDWIA approach is considered to be the optimum choice, whereas at high energies a Glauber approach, which is based in the eikonal approximation, appears more natural and should provide reliable results. An example in shown in figure~\ref{fig:RDWIA-RMSGA} for NC neutrino scattering on $^{12}$C and $^{56}$Fe at different neutrino energies. The effects of FSI increase with growing atomic number. The two ways of dealing with FSI are consistent down to remarkably low outgoing nucleon kinetic energies of about 200~MeV~\cite{PhysRevC.73.024607}. Below this energy, the RMSGA predictions are not realistic.

Both RDWIA and RMSGA were originally designed to describe FSI in the exclusive scattering and have been successfully tested against $(e,e^{\prime}p)$ data. In the case of the semi-inclusive NC and CC processes, where only the outgoing nucleon is detected, the RDWIA cross section is obtained from the incoherent sum of the exclusive cross sections (integrated over the phase space of the final lepton and of the outgoing nucleon) for one-nucleon knockout  from all the SM states $n$.

In the RDWIA FSI produce a large reduction of the calculated cross sections that is essential to reproduce $(e,e^{\prime}p)$ data. The reduction is given by the absorptive imaginary part of the ROP. The optical potential describes elastic nucleon-nucleus scattering and its imaginary part accounts for the fact that, if other channels are open besides the elastic one, part of the incident flux is lost in the elastically scattered beam and goes to the inelastic channels that are open. In the RDWIA the imaginary part removes the contribution of inelastic channels. This approach can be correct for the exclusive reaction, where the residual nucleus is left in the final state $n$, and it is reasonable to assume that the experimental cross section receives contributions mainly from the one-nucleon knockout process where  the outgoing nucleon scatters elastically with the residual nucleus in the selected final state $n$. In the inclusive scattering, where only the final lepton is detected, a model based on the RDWIA (where the cross section is obtained from the sum of integrated one-nucleon knockout processes and FSI are described by a complex optical potential with an absorptive imaginary part) is conceptually wrong. Since all the final-state channels are included, the flux lost in a channel must be recovered in the other channels, and in the sum over all the channels the flux can be redistributed but must be conserved.

Different models have been proposed to describe FSI in the inclusive scattering within the RIA. There are models which make use of real potentials, either retaining only the real part of the ROP (rROP), or by using for the scattering state  $\chi^{(-)}$ the same strong real relativistic mean field potential used for the description of the initial bound state $\varphi_n$ (RMF). With a real potential the flux is conserved, but the rROP is conceptually wrong, because the optical potential is energy dependent, its energy dependence reflects the different contributions of the inelastic channels which are open for each energy, and, under such conditions, dispersion relations dictate that the potential should have a nonzero imaginary term~\cite{hori}. In contrast, the RMF model, which  is based on the use of the same energy-independent potential for bound and scattering state, fulfills the dispersion relations and the continuity equation~\cite{Maieron:2003df, PhysRevC.74.015502, Caballero:2005sn}.

In the relativistic Green's function (RGF) model FSI are described, consistently with exclusive process, by the same complex energy-dependent ROP, but the complex ROP redistributes the flux in all the final-state channels and in the sum over all the channels the flux is conserved. The model was originally developed within a nonrelativistic~\cite{Capuzzi:1991qd, Capuzzi:2004au} and then within a relativistic framework ~\cite{Meucci:2003uy, Meucci:2005pk} for the inclusive QE electron scattering, it was successfully applied to electron scattering data~\cite{Capuzzi:1991qd, Capuzzi:2004au, Meucci:2003uy, Boffi:1993gs, Meucci:2009nm, esotici2}, and it was later extended to neutrino-nucleus scattering~\cite{Meucci:2011vd, Meucci:2003cv, Meucci:2011pi, Meucci:ant, Meucci:2013gja, PhysRevD.89.117301, Meucci:2011nc, PhysRevC.88.025502, PhysRevD.89.057302,Ivanov:2016pon}.

In the RGF model the components of the hadron tensor are written in terms of the s.p. optical-model Green's function. This is the
result of suitable approximations, such as, the assumption of a one-body current, as well as subtler approximations related to the RIA. The explicit calculation of the s.p. Green's function can be avoided by its spectral representation, which is based on a biorthogonal expansion in terms of a non Hermitian optical potential and of its Hermitian conjugate. The components of the hadron-tensor are obtained in a form  which contains matrix elements of the same type as the RDWIA ones of equation~(\ref{eq.amp}), but they involve the eigenfunctions of the ROP and of its Hermitian conjugate, where the opposite sign of the imaginary part gives in one case an absorption and in the other case a gain of strength. The RGF formalism makes it possible to reconstruct the flux lost into nonelastic channels, in the case of the inclusive response, starting from the complex ROP which describes elastic  nucleon-nucleus scattering data. Moreover, it provides a consistent treatment of FSI in the exclusive and in the inclusive scattering, and, because of the analyticity properties of the ROP, it fulfills the Coulomb sum rule.

With the use of a complex ROP the model can recover contributions of nonelastic channels, such as, for instance, some multi-nucleon processes, rescattering, non nucleonic contributions, that are not included in usual descriptions of FSI based on the RIA: all the available final-state channels are included, not only direct one-nucleon emission processes. The energy dependence of the OP reflects the different contribution of the different inelastic channels that are open at different energies and makes the results very sensitive to the kinematic conditions of the calculation.

The RGF is appropriate for the inclusive scattering, where only the final lepton is detected, and it had indeed some success in the description of data for inclusive QE $(e,e^{\prime})$ and CCQE scattering. In NC $\nu(\nubar)$-nucleus scattering, where only the final nucleon is detected and the cross sections are semi-inclusive in the hadronic sector, the RGF can recover important contributions that are not present in the RDWIA, but it may include also channels that are present in the inclusive but may be not present in the semi-inclusive measurements.

\begin{figure}[tb]
\centering
\includegraphics[width=8cm]{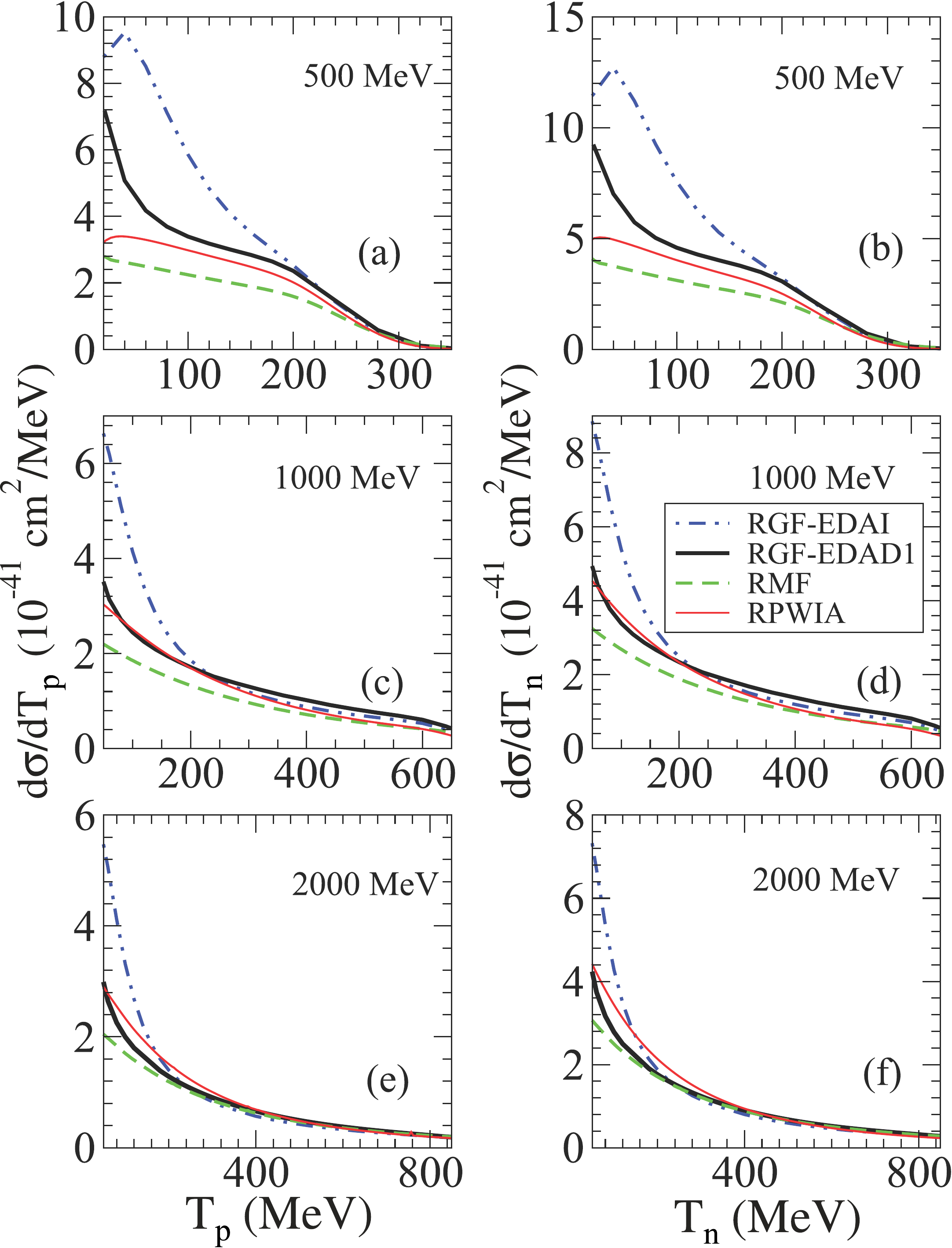}
\vspace{-2mm}
\caption{Differential cross sections of the NC elastic neutrino scattering on $^{12}$C as a  function of the kinetic energy of the emitted proton [panels (a), (c), and (e)]  or neutron [panels (b), (d), and (f)] at different neutrino energies  calculated in RPWIA (thin solid lines), RMF (dashed lines), RGF-EDAD1 (thick solid lines), and RGF-EDAI (dash-dotted lines).
Taken from Ref.~\cite{PhysRevC.88.025502}.}
\label{fig:RMF-RGF}
\end{figure}

An example of the comparison of the RPWIA, RMF, and RGF results is presented in figure~\ref{fig:RMF-RGF} for NC elastic (NCE) $\nu-^{12}$C scattering at different neutrino energies as a function of $T_{\mathrm N}$. The RMF gives cross sections that are generally $30\%$ lower than the RPWIA ones at small nucleon kinetic energies, but with a longer tail extending towards larger values of $T_{\mathrm N}$, {\it i.e.}, higher values of the transferred energy, that is attributable to the strong energy-independent scalar and vector potentials adopted in the RMF approach. The RGF results are generally larger than the RPWIA and the RMF ones. The enhancement is due to the inelastic contributions included in the imaginary part of the ROP. The RGF results obtained with two different parametrizations of the ROP are compared in the figure to give an idea of how  the predictions of the model are affected by uncertainties in the determination of the phenomenological ROP. The differences depend on kinematics and are essentially due to the different imaginary part of the two potentials, which accounts for the overall effects of inelastic channels and is not univocally determined  only from elastic proton-nucleus scattering  phenomenology.

Nucleon rescattering, leading to energy losses, charge exchange, and multiple nucleon emission, have been taken into account in CC and NC QE scattering using a Monte Carlo (MC) simulation method~\cite{Nieves:2004wx, Nieves:2005rq, PhysRevD.85.113008, PhysRevD.88.113007}. Starting from a local Fermi gas (LFG) picture of the nucleus, which accounts for Pauli blocking, several nuclear effects are taken into account: a correct energy balance, medium polarization, random phase approximation (RPA), including $\Delta$-hole degrees of freedom and explicit pion and $\rho$ exchanges in the vector-isovector channel of the effective nucleon-nucleon force, short-range correlation (SRC) effects; FSI are accounted for by using nucleon  propagators dressed in the nuclear medium, which amounts to work with a LFG of interacting nucleons. The model also accounts for reaction mechanisms where the exchanged boson is absorbed by two nucleons.

FSI are one of the main ingredients of the coupled-channels semiclassical Giessen Boltzmann-Uehling-Uhlenbeck (GiBUU) transport  model~\cite{Leitner:2006ww, Leitner:2006sp, PhysRevC.79.034601, BUSS20121, Gallmeister:2016dnq}, where nucleon knockout and pion production are investigated. The model, originally developed to describe heavy-ion collisions, has been extended to describe the interactions of pions, photons, electrons, and neutrinos with nuclei. In the GiBUU model the space-time evolution of a many-particle system under the influence of a mean-field potential and a collision term is described by a BUU equation for each particle species. A collision term accounts for changes (gain and loss) in the phase-space density, due to elastic and inelastic collisions between the particles, and also due to particle decays into other hadrons whenever it is allowed by Pauli blocking. In between the collisions, all particles (also resonances) are propagated in their mean-field potential according to their BUU equation. The GiBUU model differs from standard Glauber approaches because the collision term allows not only for absorption but also for side-feeding and rescattering: FSI lead to absorption, charge exchange, a redistribution of energy and momentum, as well as to the production of new particles.

\begin{figure}[tb]
\begin{minipage}{0.495\textwidth}
\centering\includegraphics[width=\textwidth]{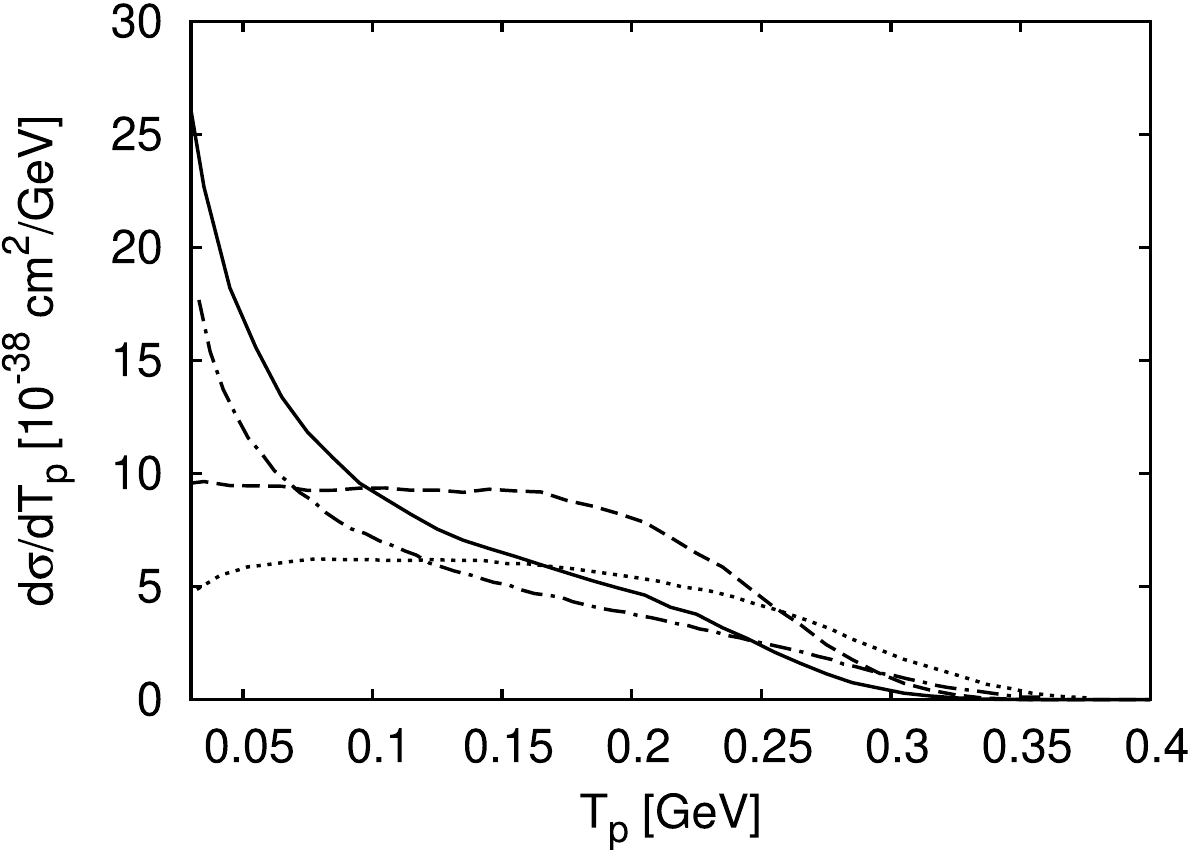}
\end{minipage}\hfill
\begin{minipage}{0.495\textwidth}
\centering\includegraphics[width=\textwidth]{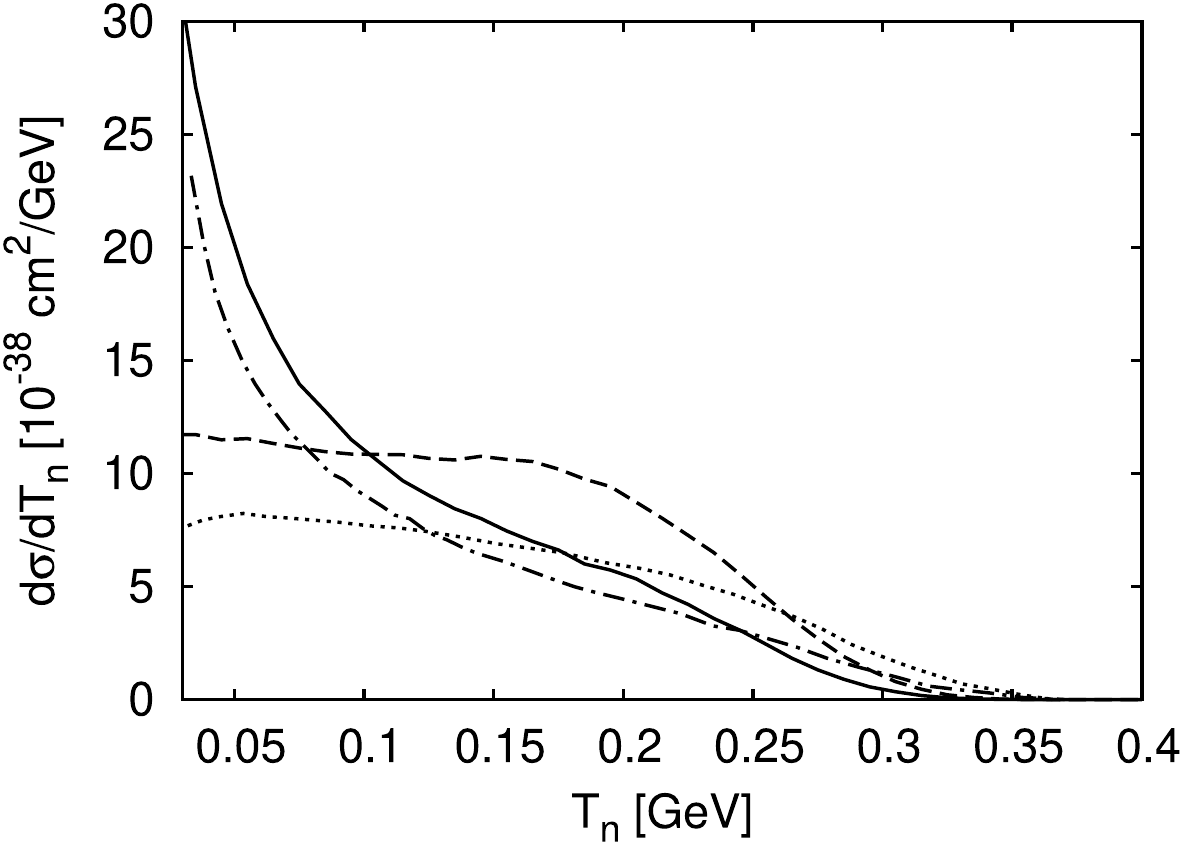}
\end{minipage}
\vspace{-2mm}
\caption{Differential cross section for NC neutrino scattering on $^{40}$Ar at $E_{\nu}=500$ MeV for proton (left panel) and neutron (right panel) knockout. The solid (dashed) lines denote the result of the GiBUU model~\cite{Leitner:2006sp} with (without) FSI. The dash-dotted (dotted) lines denote the results of Ref.~\cite{Nieves:2005rq} with (without) $NN$ rescattering. Taken from Ref.~\cite{Leitner:2006sp}.}
\label{fig:GiBUU-Nieves}
\end{figure}

An example of the comparison between the results of the GiBUU~\cite{Leitner:2006sp} and the LFG-RPA model of Nieves \emph{et al.}~\cite{Nieves:2005rq} is shown in figure~\ref{fig:GiBUU-Nieves}, where the calculated differential cross sections for NC scattering on $^{40}$Ar at an incident neutrino energy of 500~MeV are plotted as a function of the kinetic energy of the knocked-out nucleon for proton and neutron knockout. Since the LFG-RPA model~\cite{Nieves:2005rq} does not include any resonances, the initial resonance excitation has been switched off in the GiBUU calculation, so only nucleon knockout induced by initial QE events is considered. The differences between the GiBUU results without FSI and the results of Nieves \emph{et al.} without $NN$ rescattering could be attributed to the RPA correlations, included in Ref.~\cite{Nieves:2005rq}, which lead to a reduction of the cross section and a spreading of the spectrum. When FSI and rescattering are included, in both models  there is a reduction of the calculated cross section for higher energetic nucleons and a large number of secondary low energy nucleons, and the final results of the two models are very similar.

\subsection{Superscaling \label{sec:susa}}

The SuperScaling Approach (SuSA), based on the superscaling properties of inclusive electron scattering~\cite{PhysRevLett.82.3212, PhysRevC.60.065502, PhysRevC.65.025502}, has been extensively used to predict neutrino and antineutrino NC~\cite{Ankowski:2015lma, GONZALEZJIMENEZ20131471, PhysRevLett.100.052502, PhysRevC.77.064604, PhysRevC.73.035503} and CC~\cite{Caballero:2005sn, PhysRevC.74.015502, PhysRevLett.98.242501, PhysRevC.75.034613, PhysRevD.89.093002, PhysRevC.90.035501} cross sections for complex nuclei. The detailed description of the model can be found, {\it e.g.}, in Refs.~\cite{PhysRevC.73.035503, Amaro:2004bs, PhysRevC.90.035501}. Here we recall the main features of the SuSA model.

In the QE peak region the basic ingredient of the model is a phenomenological superscaling function
\begin{equation}\label{scf_L}
f_L^{QE} = k_F \frac{R_L^{QE}}{G_L^{QE}}
\end{equation}
extracted from the world electromagnetic $(e,e')$ data by dividing the {\it longitudinal} response $R_L^{QE}$ times the Fermi momentum $k_F$ by the single-nucleon elementary function $G_L^{QE}$. The data show that $f_L^{QE}$ is to a large extent a function of only one variable, the scaling variable $\psi^\prime_{QE}$, and it is independent of the momentum transfer $q$ (scaling of first kind) and of the nucleus, represented by the Fermi momentum $k_F$ (scaling of second kind).

The function $f_L^{QE}$ embeds most of the nuclear effects, both in the initial and in the final state, and can therefore be used to predict the weak charged current quasielastic $(\nu_l,l)$ cross section. In its original version the SuSA model assumes that the superscaling function $f^{QE}$ is the same in the longitudinal ($L$) and transverse ($T$) channels, a property referred to as scaling of zeroth-kind.

Scaling of first and second kind are clearly observed in the longitudinal channel, whereas violations associated with reaction mechanisms not taken into account within the IA scheme, such as, inelastic scattering and processes involving MEC, show up in the transverse channel, mainly at energy transfer larger than corresponding to single-nucleon knockout.

The main merit of the SuSA model is the reasonable agreement, required by construction, with electron scattering data over a very wide range of kinematics and for a large variety of nuclei. Such an agreement is a crucial test for any nuclear model to be applied to neutrino reactions. The model accounts for both kinematic and dynamic relativistic effects and can therefore safely be used in a broad energy range. Although phenomenological, the model has firm microscopic foundations in the RMF model, which is able to reproduce both the height and the asymmetric shape of the experimental superscaling function~\cite{CABALLERO2007366}.

Within SuSA, $\nu(\nubar)$-nucleus cross sections are simply obtained by multiplying the phenomenological superscaling function by the appropriate elementary weak cross sections.

\begin{figure}[tb]
\centering \includegraphics[width=\textwidth]{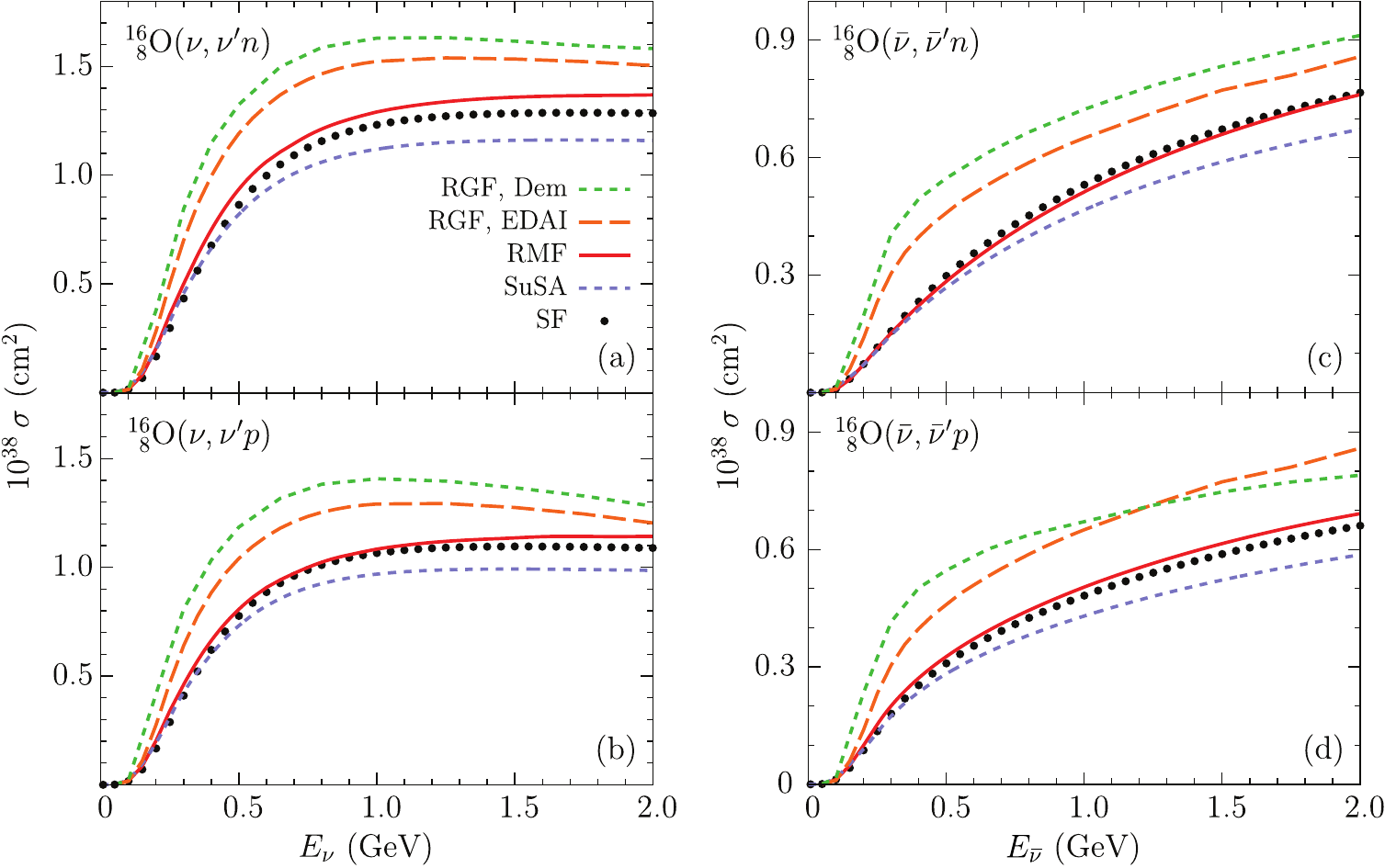}
\vspace{-2mm}
\caption{Cross sections calculated with the SuSA, RMF, SF, and RGF models for neutron [(a) and (c)] and proton [(b) and (d)] knockout from $^{16}$O induced by NC QE interaction of neutrino [(a) and (b)] and antineutrino [(c) and (d)]. The RGF results obtained with two different phenomenological ROPs (EDAI) and ``democratic" (Dem) are compared.  Taken from Ref.~\cite{Ankowski:2015lma}.}
\label{fig:SuSA-RMF-SF-RGF}
\end{figure}

The NCQE $\nu(\nubar)$-$^{16}$O cross sections calculated for neutron and proton knockout in the SuSA, RFG, and SF~\cite{PhysRevC.41.R24, Benhar:2005dj} models are compared in figure~\ref{fig:SuSA-RMF-SF-RGF} as a function of the incident neutrino and antineutrino energy. FSI are explicitly taken into account in the RMF and RGF models, with real scalar and vector relativistic mean field potentials (RMF) and with complex ROPs (RGF). The RGF results obtained with EDAI~\cite{Cooper:1993nx} and the so-called ``democratic" (Dem)~\cite{Cooper:2009} ROP are compared in the figure. The SF results are identical to those displayed in figure~\ref{fig:RPWIA-RFG-SF}. The RMF and SF approaches, while being based on very different models for nuclear dynamics, yield remarkably similar results. The RGF cross sections obtained with the two ROPs lay significantly above the ones obtained with the RMS, SuSA, and SF models. In addition, they show a sizable sensitivity to the choice of the ROP, the differences between the two results ranging between $5-10\%$ at $E_\nu \sim 2$ GeV and $10-25\%$ for $E_\nu \sim 0.3-0.5$ GeV. The larger cross sections in the RGF may be associated with the redistribution of the strength arising from reaction mechanisms other than
single-nucleon knockout, such as rescattering of the outgoing nucleon, possibly leading to the excitation of non-nucleonic degrees of freedom, or scattering off a nucleon belonging to a correlated pair. These channels, although not explicitly included in the RGF model, may be phenomenologically included by the imaginary part of the optical potential.

In figure~\ref{fig:SuSA-RMF-SF-RGF} the SuSA cross section are sizeably lower than the RMF ones. This is likely to be ascribed to the fact that the RMF model predicts a transverse superscaling function
\begin{equation}\label{scf_T}
f_T^{QE} = k_F \frac{R_T^{QE}}{G_T^{QE}}
\end{equation}
which is higher than the longitudinal one, a result supported by the separated $L/T$ data analysis~\cite{PhysRevC.29.2230, PhysRevLett.82.3212} and strictly linked to the relativistic nature of the  model~\cite{IVANOV2013265}. The transverse enhancement, clearly observed in electron scattering data, is not reproduced by the SuSA model adopted to obtain the results shown in the figure, in which the same scaling function is used in the longitudinal and transverse channels.

The result of equation~(\ref{scf_T}) has been used to improve the ingredients of the SuSA model by constructing a new version (SuSAv2; see~\cite{PhysRevC.90.035501}) where RMF effects~\cite{Caballero:2005sn, PhysRevC.74.015502, CABALLERO2007366} in the longitudinal and transverse nuclear responses, as well as in the isovector and isoscalar channels are incorporated. This is of great interest in order to describe CC neutrino reactions that are purely isovector. We note that in this approach the enhancement of the transverse nuclear response emerges naturally from the RMF theory as a genuine relativistic effect. Moreover in SuSAv2 the effects of Pauli blocking, initially neglected, have been implemented. We point out that the effects of Pauli blocking are included in the SuSA results displayed in figure~\ref{fig:SuSA-RMF-SF-RGF}.

The SuSAv2 has been validated against all existing $(e,e')$ data sets on $^{12}$C, yielding excellent agreement over the full range of kinematics spanned by experiments, except for the very low energy and momentum transfers, where all approaches based on the IA are bound to fail. The scaling approach can then be inverted and predictions can be made for CCQE $\nu$($\bar{\nu}$)-nucleus reactions by replacing the elementary electromagnetic vertex, $\gamma^*NN$, with the weak one, $WNN$.

Ingredients beyond the IA, namely 2p-2h MEC effects, are essential in order to explain the neutrino-nucleus cross sections of interest for neutrino oscillation experiments. In particular, 2p-2h MEC effects produce an important contribution in the ``dip'' region between the QE and $\Delta$ peaks, giving rise to a significant enhancement of the IA responses in the case of inclusive electron-nucleus~\cite{PhysRevD.94.013012, Barbaro:2019vsr, PhysRevD.91.073004, Megias_2018} and neutrino-nucleus~\cite{PhysRevD.94.093004, PhysRevD.91.073004, Megias_2018} scattering processes. The 2p-2h MEC model is developed in Ref.~\cite{Simo:2016ikv} and it is an extension to the weak sector of the seminal papers~\cite{VanOrden:1980tg, DePace:2003spn, Amaro:2010iu} for the electromagnetic case. The calculation is entirely based on the RFG model and it incorporates the explicit evaluation of the five response functions involved in inclusive neutrino scattering.

As in the case of CC processes, predictions of NC processes based on scaling ideas, when possible, are clearly demanded. The identification of CC events is relatively simple via the outgoing charged lepton, similar to what happens in inclusive ($e, e'$) scattering. This means that the energy and momentum transferred at the leptonic vertex are known, and thus the scaling analysis of CC neutrino-nucleus cross sections proceeds in a way identical to the electron case. However, in the case of NC events, the scattered neutrino is not detected, and identification of the NC event is usually made when (i) no final charged lepton is found and (ii) a nucleon ejected from the nucleus is detected. Even in the case that the nucleon energy and momentum can be measured, the transferred energy and momentum at the leptonic vertex will remain unknown. The kinematics of the NC process is thus different from both electron scattering and its CC neutrino counterpart, rendering the derivation of scaling less obvious. The translation of the scaling analysis to NC processes was outlined in Ref.~\cite{PhysRevC.73.035503}. There, it was shown that NCQE neutrino cross sections exhibit superscaling properties even in presence of strong FSI and how the scaling analysis can be extended to NC processes, which was realized in Ref.~\cite{GONZALEZJIMENEZ20131471}.

It will be interesting, as a future work, to extend the SuSAv2 model to NC processes and to evaluate the role of 2p-2h MEC contribution to the NC cross sections.

\begin{figure}[tb]
\centering\includegraphics[width=.8\textwidth]{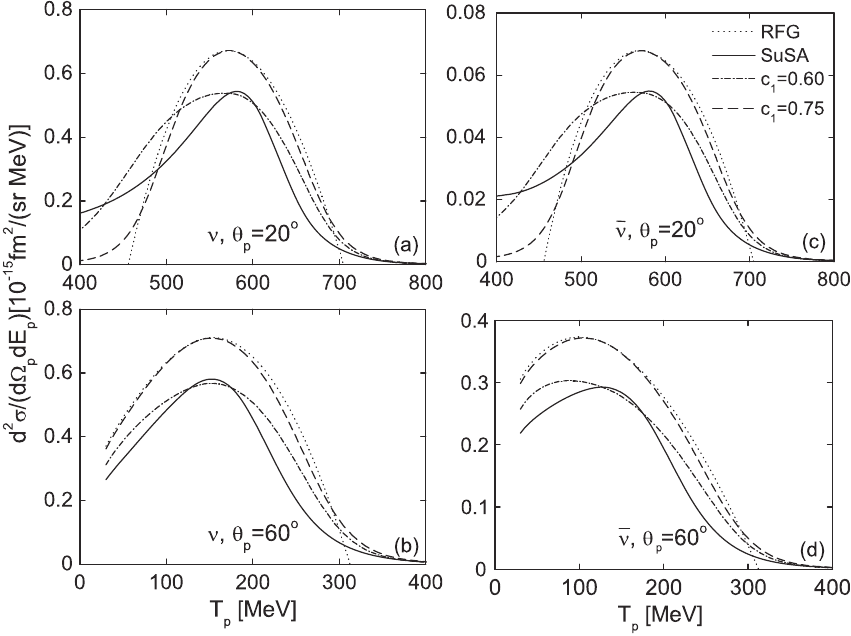}
\caption{NCQE differential cross section for $\nu$ and $\nubar$ scattering from $^{12}$C at $1$~GeV, for proton knockout at $\theta_p = 20^\circ$ (a,c) and $60^\circ$ (b,d), obtained using the CDFM scaling function with $c_1 = 0.60$ (dash-dotted lines) and $c_1 = 0.75$ (dashed lines). The RFG results are given by the dotted lines and the results given by the SuSA scaling function (see figure~\ref{fpsi}) are given  by the solid lines. Taken from Ref.~\cite{PhysRevC.75.064617}\label{fig:nccdfm}}
\end{figure}

In Ref.~\cite{PhysRevC.73.035503} it was shown that the superscaling analysis of NC reactions is feasible. The RFG ($e,e'$) response exhibits perfect superscaling by definition~\cite{PhysRevC.38.1801}, but it is not in accord with the magnitude or with the shape of the experimental scaling function. Thus, the necessity to consider superscaling in theoretical methods which go beyond the RFG model has arisen. One of such approaches is the coherent density fluctuation model (CDFM) (see, \emph{e.g.}, Refs.~\cite{anton1, anton2}), which is a natural extension of the Fermi gas model based on the generator coordinate method and which includes long-range correlations of collective type. The QE scaling function $f(\psi')$ is deduced in the CDFM on the basis of realistic density and momentum distributions in nuclei, and it agrees with the empirical data for the scaling function for negative values of $\psi'\leq -1$~\cite{PhysRevC.69.044321, PhysRevC.71.014317, PhysRevC.73.047302, PhysRevC.74.054603, PhysRevC.75.034319, PhysRevC.77.034612}. This agreement is related to the realistic high-momentum tail of the nucleon momentum distribution $n(k)$ in the CDFM, which is similar for a wide range of nuclei, in contrast with the sharp behavior of $n(k)$ as a function of $k$ in the RFG model (see, \emph{e.g.}, Refs.~\cite{PhysRevC.69.044321, PhysRevC.71.014317, PhysRevC.75.034319}).
It was pointed out in Ref.~\cite{PhysRevC.75.034319} that the behaviour of the scaling function $f(\psi^\prime)<-1$ obtained in the inclusive QE electron scattering from nuclei is related to the power-law asymptotics of the nucleon momentum distribution $n(k) \longrightarrow [V_{NN}(k)/k^2]^2$, where $V_{NN}(k)$ is the \emph{NN} interaction in the momentum space. It has been shown that $n(k) \sim 1 / k^{4+m}$, with $m \simeq 4.5$ at large $k$, and in this way some information of the \emph{NN} forces in the nuclear medium can be obtained from the superscaling phenomenon. The superscaling approach with  the scaling function obtained within the CDFM was applied to  electron~\cite{PhysRevC.74.054603, PhysRevC.79.044602}, CC~\cite{PhysRevC.74.054603} and NC~\cite{PhysRevC.75.064617} $\nu$ and $\nubar$ scattering from $^{12}$C, at an energy of $1$~GeV, in the QE and $\Delta $ regions~\cite{PhysRevC.77.034612}. In the superscaling approach the constructed realistic CDFM scaling function is an essential ingredient for the description of lepton-scattering processes from nuclei.

The differential NC $\nu$ and $\nubar$ cross sections at $1$~GeV, shown in figure~\ref{fig:nccdfm} for proton knockout from $^{12}$C at $\theta_p = 20^\circ$ and $60^\circ$, are obtained using the CDFM scaling function with a parabolic form at positive $\psi$ values and for two values of $c_1=0.6$ and $c_1=0.75$ (more details are given in Ref.~\cite{PhysRevC.75.064617}). The results at $60^\circ$ show that $\nu$ and $\nubar$ cross sections are roughly in a $2 : 1$ ratio, for larger values of the scattering angle they are closer, diminishing the above ratio. At forward scattering angles the $\bar{\nu}$ cross sections are strongly suppressed (by an order of magnitude or more). This is observed for both proton and neutron knockout~\cite{PhysRevC.75.064617}. Moreover, the neutron knockout cross sections are somewhat larger than the proton knockout cross sections, due to the behavior of the NC single-nucleon form factors~\cite{PhysRevC.75.064617}. From the comparison with the RFG and SuSA results, also shown in figure~\ref{fig:nccdfm}, it can be seen that the use of the asymmetric CDFM scaling function ($c_1=0.6$) gives results which are close to those of the original SuSA approach, while the symmetric scaling function ($c_1=0.75$) leads to a similarity with the RFG model results. We point out that the constructed realistic CDFM scaling function is an essential ingredient in this approach for the description of processes of lepton scattering from nuclei.

In a series of papers~\cite{PhysRevC.83.045504, Ivanov:2013saa, Ivanov:2015wpa, Ivanov:2018nlm} the realistic spectral function model was developed and applied to the electron, CC, and NC $\nu(\nubar)$-nucleus scattering in the QE region. The area of analyses of the scaling function, the spectral function, and their connection (see, \emph{e.g.}, Refs.~\cite{PhysRevC.81.055502, PhysRevC.83.045504}) provides insight into the validity of the MFA and the role of $NN$ correlations, as well as into the effects of FSI. Though in the MFA it is possible, in principle, to obtain the contributions of different shells to $S(\np,E)$ [see equation~(\ref{eq.sfsm})] and $n_i(\np)=\mid \varphi_i(\np) \mid^2$  for each s.p. state, owing to the residual interaction, the hole states are not eigenstates of the residual nucleus but are mixtures of several s.p. states. This leads to the spreading of the shell structure, and a successful description of the relevant experimental results requires a spectral function obtained from theoretical methods going beyond the MFA. To this aim, a realistic spectral function $S(\np,E)$, which incorporates effects beyond MFA and is in agreement with the scaling function $f(\psi')$ obtained from the ($e,e'$) data, has been constructed~\cite{PhysRevC.83.045504}. The procedure includes: (i) the account for effects of a finite energy spread and (ii) the account for \emph{NN} correlation effects considering s.p. momentum distributions $n_i(\np)$ [that are components of $S(\np,E)$] beyond the MFA, such as those related to the use of natural orbitals (NOs)~\cite{PhysRev.97.1474} for the s.p. wave functions and occupation numbers, within methods in which short-range \emph{NN} correlations are included. For the latter the Jastrow correlation method~\cite{PhysRevC.48.74} has been considered. FSI are accounted for in Ref.~\cite{PhysRevC.83.045504} by a complex optical potential, which gives an asymmetric scaling function, in accordance with the experimental scaling function, thus showing the essential role of FSI in the description of electron scattering data.

\begin{figure}[tb]
\centering\includegraphics[width=8cm]{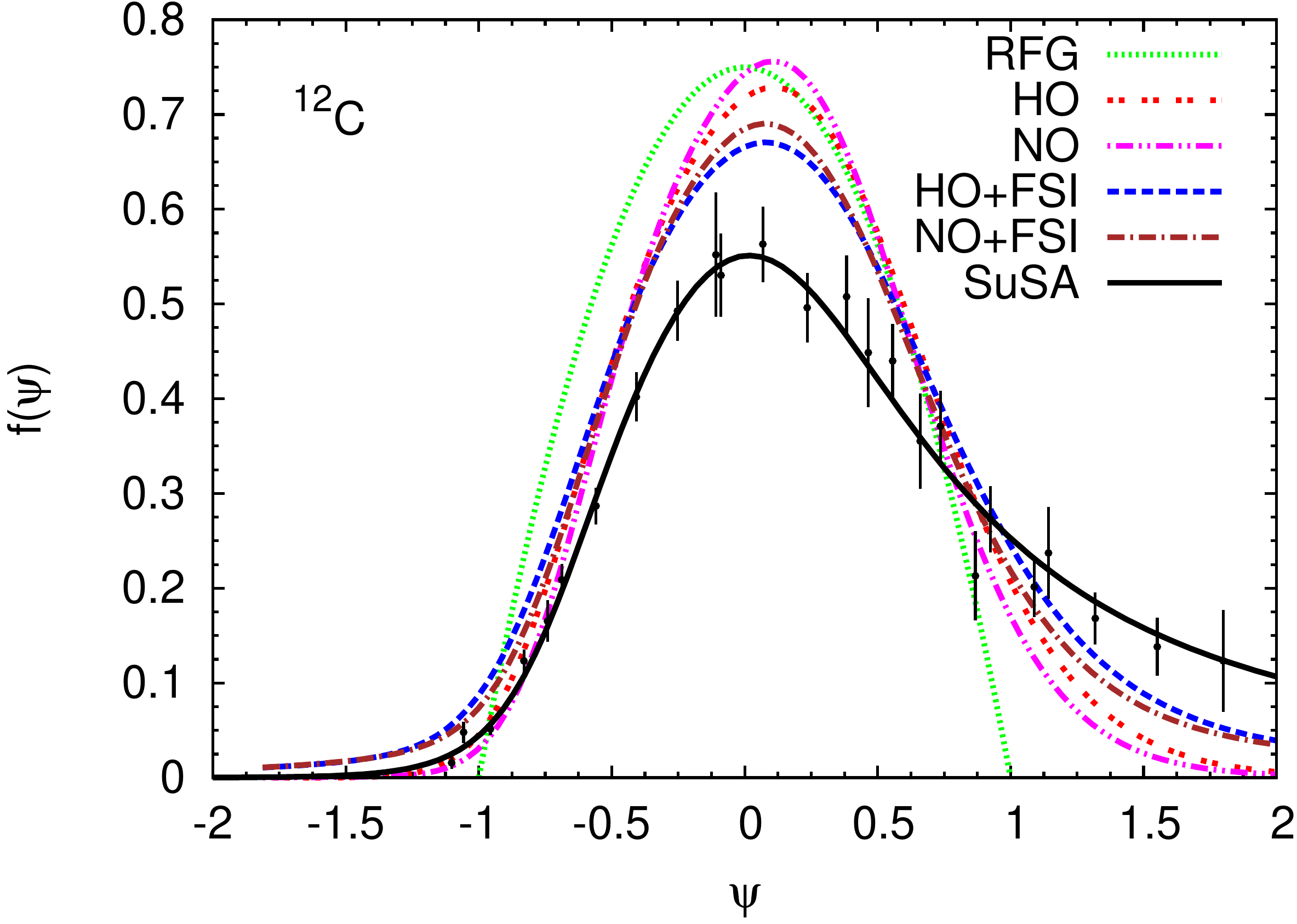}
\vspace{-2mm}
\caption{The scaling function $f(\psi)$ of $^{12}$C obtained using the HO and NO, with (HO+FSI and NO+FSI) and without (HO and NO) FSI, RFG, and SuSA approaches, compared with the longitudinal experimental superscaling function~\cite{Amaro:2004bs}. Taken from Ref.~\cite{Ivanov:2013saa}. \label{fpsi}}
\end{figure}

In Refs.~\cite{Ivanov:2013saa, Ivanov:2015wpa, Ivanov:2018nlm} the spectral function $S(\np,E)$ is constructed from the momentum distribution provided by two different  nuclear models, denoted as HO (harmonic oscillator) and NO (natural orbital). In figure~\ref{fpsi} the scaling function of $^{12}$C obtained with the two models, with (HO+FSI and NO+FSI) and without (HO and NO) FSI, are compared with the RFG result and with  the superscaling function within the SuSA model, which is a phenomenological fit to the longitudinal experimental superscaling function~\cite{Amaro:2004bs}, that is also shown in the figure. The effects of FSI lead to a redistribution of the strength, with lower values of the scaling function at the maximum and an asymmetric shape around the peak position, {\it viz.,} where $\psi'=0$. The two spectral function models, including FSI, clearly give a much more realistic representation of the data than the RFG model.

Examples of the numerical predictions of NC cross sections calculated from these scaling functions in comparison with available experimental data are given  in Section~\ref{sec:results}.

\subsection{RPA and np-nh contributions \label{sec:npnh}}

In the IA the effects of the spectator $(A-1)$-nucleon system is only present in the hole SF and in FSI. For small values of the momentum transfer, when $1/q$ becomes comparable to the internucleon distance, the IA  does not hold and collective effects can come into play. Collective effects can be handled within the random phase approximation (RPA).

RPA 1p-1h excitations are included in some models starting from a LFG picture of the nucleus. In the model of Nieves \emph{et al.}~\cite{Nieves:2004wx, Nieves:2005rq, PhysRevD.85.113008, PhysRevD.88.113007} several nuclear effects and different reaction mechanisms are taken into account. The model was originally developed for real~\cite{Carrasco:1989vq} and virtual~\cite{Gil:1997jg, Gil:1997bm} photons, where it is able to describe inclusive electron-nucleus scattering, total nuclear  photoabsorption data, and also measurements of photo- and electronuclear production of pions, nucleons, pairs of nucleons,  pion-nucleon pairs, and it has been extended to CC and NC neutrino-nucleus interactions.

The building blocks of the model are: a gauge invariant model for the interaction of the exchanged virtual boson with nucleons, mesons, and nucleon resonances, with parameters determined from  the  vacuum  data; a  microscopic treatment of nuclear effects, including long and short range nuclear correlations; FSI, explicit meson and $\Delta$ degrees of freedom, two and even three-nucleon absorption channels The model is a nonrelativistic one and assumes that the nucleons propagate semiclassically in the nucleus, it includes QE processes, pion production, and two-body processes from the QE region to the region beyond the $\Delta$ peak, and double pion production and higher nucleon resonance degrees of freedom induced processes.

A  similar model, where RPA excitations are included starting from a LFG basis, has been proposed by Martini \emph{et al.}~\cite{Martini:2009uj,Martini:2010ex,Martini:2011wp}. Beyond QE processes and coherent and incoherent pion production, the model incorporates multi-nucleon emission processes through 2p-2h and 3p-3h excitations. For the description of 2p-2h processes the parametrization deduced from Ref.~\cite{Alberico:1983zg} is used, while the 3p-3h contribution is taken from Ref.~\cite{Oset:1987re}. Relativistic corrections have been added in Ref.~\cite{Martini:2011wp}.

Other models~\cite{Botrugno:2005kn,PhysRevC.59.3246,Jachowicz:2002rr,PhysRevC.89.024601,Pandey:2014tza,PhysRevC.98.054603} adopt the continuum RPA (CRPA), which is the theory most often used to describe the excitation of the nuclear giant resonances.

In Refs.~\cite{Botrugno:2005kn,CO1985392} the secular equations of the CRPA theory are solved using the Fourier-Bessel expansion technique~\cite{DEHARO1982265}. The formalism requires s.p. states both in the discrete and in the continuum part of the spectrum, for both of which the same Woods-Saxon potential is used. The s.p. energies are a crucial ingredient for the solution of the CRPA equations, for which, in the spirit of the Landau-Migdal theory of the finite Fermi systems, the experimental s.p. energies have been used, when available. The other ingredient of the CRPA calculations is the residual particle-hole interaction. The sensitivity to the choice of the residual interaction has been studied. The CRPA theory is implemented by a phenomenological description of the effects of FSI based on a folding integral and using the parameters of a phenomenological optical potential.

In the CRPA model of Refs.~\cite{PhysRevC.59.3246, Jachowicz:2002rr, PhysRevC.89.024601, Pandey:2014tza, PhysRevC.98.054603} the nucleus is described in a mean field generated by a self-consistent Hartree-Fock method using the extended Skyrme force (SkE2) for the nucleon-nucleon interaction. The CRPA extends the HF approach taking into account long-range correlations consistently by using the same nucleon-nucleon interaction. The knocked-out nucleon is modeled as a continuum state of the residual nucleus. By using the same HF potential for the calculation of the outgoing-nucleon wave function, the initial and final states are orthogonal. The HF-CRPA model was originally developed for exclusive electron~\cite{RYCKEBUSCH1989694} and photo-induced~\cite{RYCKEBUSCH1988237} nucleon knockout reactions and it was later extended to CC and NC neutrino scattering.

A comparison of the results of the HF-CRPA model, which is able to describe giant resonances and QE excitations, with those of the RPA-based model of Refs.~\cite{Martini:2009uj,Martini:2010ex,Martini:2011wp}, which is able to describe multi-nucleon emission and coherent and incoherent pion production, as well as QE excitations, is presented in Ref.~\cite{PhysRevC.94.015501} for CC electron neutrino cross sections on $^{12}$C. The two approaches are compared in the genuine QE channel, where they give a satisfactory agreement at large energies, while at low energies the collective giant resonances show up only in the CRPA approach. The double differential cross section is quenched in the HF model, featuring more strength for larger energy transfers.  The RPA model leads to a quenching of the cross section, for a large part due to the admixture of nucleon states with the $\Delta$ resonance, which makes the double differential cross section in the RPA comparable to the results found in the HF model in the QE region.  The effect of the CRPA is twofold: firstly there is a slight suppression of the cross section in the QE region up to the QE peak. On the other hand, for low energy transfers, the CRPA approach leads to an enhancement of the cross section from the contributions of giant resonances, while these collective states are not included in the RPA approach.

In the case of neutrino experiments a delicate point concerns the definition of a QE process. The definition is clear in the case of electron scattering experiments, where the values of the energy and momentum transfer are determined. In the case of neutrino scattering, the experiment does not distinguish genuine QE processes and np-nh contributions. The role of multi-nucleon processes and two-body MEC can be important for the comparison with the experimental data and should be carefully evaluated.

\subsection{Meson-exchange currents and np-nh excitations \label{sec:mec}}

The treatment of the multi-nucleon emission channels (related to $NN$ correlations and MEC contributions) is extremely difficult and computationally demanding. Different approximations are necessarily employed by the different theoretical approaches. The main aspects of the microscopic models and a discussion of the  analogies and differences among them can be found in Ref.~\cite{Katori:2016yel}.

In the SuSAv2-MEC model~\cite{PhysRevD.94.013012, PhysRevD.94.093004, Simo:2016ikv} 2p-2h MEC effects are added to the SuSAv2 model. Contrary to the SuSAv2 approach, that is based on the RMF predictions in the QE domain and also on the phenomenology of electron scattering data, the 2p-2h MEC calculations are entirely performed within the RFG model. This is due to the technical difficulties inherent to the calculation of relativistic two-body contributions even in the simple RFG model. The model neglects nuclear correlations in the 2p-2h sector as well as the interference between $NN$-correlations and MEC. It is, however, noteworthy to point out that the 2p-2h MEC contributions correspond to a fully relativistic calculation, needed for the extended kinematics involved in neutrino reactions. The SuSAv2-MEC model has been applied to the inclusive electron scattering and to CC $\nu(\nubar)$-nucleus scattering. The extension to NC processes has not been done yet.

In Ref.~\cite{Butkevich:2017mnc} the same MEC contribution as in the SuSAv2-MEC approach has been added to the genuine QE interaction treated within the RDWIA. Also the RDWIA-MEC model neglects $NN$ correlations and interference between correlations and MEC, and it has been applied only to the inclusive electron scattering and to CCQE scattering.

Relativistic two-body MEC corrections to the IA have been presented in Refs.~\cite{PhysRevLett.74.4993, PhysRevC.52.3399} for QE electron $(e,e^\prime)$, CC, and NC neutrino scattering reactions, using a single unified formalism obtained generalizing the method developed in Ref.~\cite{CHEMTOB19711}. Calculations were performed assuming a simple RFG model and considering only 1p-1h final states.

The contribution of MEC to one and two-nucleon knockout has been investigated in Ref.~\cite{PhysRevC.95.054611}. The formalism provides a framework for the treatment of MEC and short-range correlations in the calculation of exclusive, semi-exclusive, and inclusive single-nucleon  and two-nucleon knockout cross sections. Bound and emitted nucleons are described as Hartree-Fock wave functions, For the vector seagull and pion-in-flight two-body currents the standard expressions are used, while  for the axial current three different prescriptions are adopted and compared. Calculations have been done only for CC neutrino-nucleus scattering.

The factorization scheme underlying the IA and the SF formalism has been generalized to include relativistic MEC arising from pairs of interacting nucleons~\cite{PhysRevC.92.024602}. The contribution of MEC has been investigated in electron scattering~\cite{Rocco:2015cil} and, more recently, in CC and NC scattering~\cite{Rocco:2018mwt}, adopting two-body currents derived~\cite{Simo:2016ikv} from the weak pion-production model of Ref.~\cite{Hernandez:2007qq}, which is able to provide results consistent with those of Refs.~\cite{DePace:2003spn, Rocco:2015cil}. The aim of the work is to devise the formalism for including relativistic MEC with two different realistic models of the target SF, {\it i.e.} the CBF formalism and the SCGF theory, which both rely upon a nonrelativistic nuclear Hamiltonian to describe the interaction among nucleons. The semi-phenomenological Hamiltonian AV18+UIX~\cite{PhysRevC.51.38, PhysRevC.64.014001} employed in the CBF calculation has been derived from a fit to the properties of the exactly solvable two- and three-nucleon systems and describes the $NN$ scattering phase shifts up to 300~MeV, but  fails to provide an accurate description of the spectra and radii of nuclei wih $A > 4$. The chiral Hamiltonian NNLO$_\textrm{sat}$~\cite{PhysRevC.91.051301} employed in the SCGF calculation reproduces the properties of the light and medium-mass nuclei, but fails to describe $NN$ scattering above 35~MeV. The approximations adopted in the calculations neglect the interference between one-body and two-body current matrix elements. In addition, also FSI have been neglected.

\begin{figure}[tb]
\centering\includegraphics[width=0.9\textwidth]{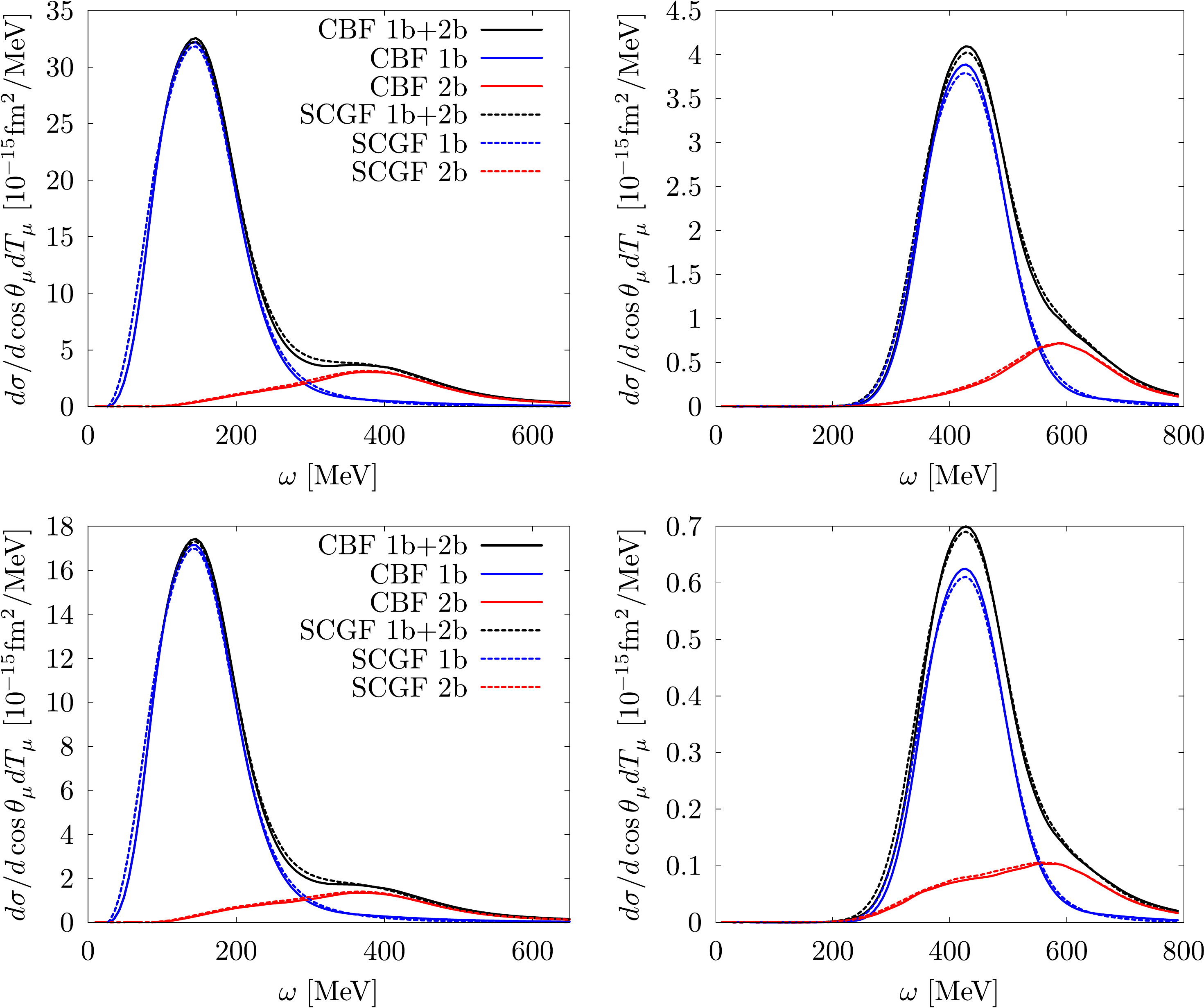}
\vspace{-2mm}
\caption{NC differential cross section of $\nu_\mu$ (upper panels) and $\nubar_\mu$ (lower panels) scattering on $^{12}$C, for an incident  $\nu_\mu$ and $\nubar_\mu$ energy of 1~GeV and a scattering angle of  $30^\circ$ (left panels) and $70^\circ$ (right panels), as a function of the energy transfer $\omega$. The blue (red) lines give the results including only one-body  (two-body) current contributions, the black lines display the total result. Dotted and solid lines show the results obtained from the SF computed with the SCGF and CBF methods, respectively. Taken from Ref.~\cite{Rocco:2018mwt}.}
\label{sf-mec}
\end{figure}

An example is shown in figure~\ref{sf-mec} for the NC differential $\nu$-$^{12}$C and $\nubar$-$^{12}$C cross sections at an incident energy of $1$~GeV and two different values of the scattering angle, $30^\circ$ and $70^\circ$.  The solid and dashed lines have been obtained using the CBF and SCGF hole SFs, respectively. The black lines give the full calculations, which include both one-body and two-body currents, the red and blue curves show the separate contribution of the one-body and of the two-body currents. The CBF and SCGF results, although obtained from different models and different Hamiltonians, are in remarkably good agreement. MEC primarily enhance the cross sections in the dip region, between the QE peak and the resonance-production
region, although for antineutrino scattering at $70^\circ$ they also provide excess strength in the QE-peak region, and give a larger contribution to the cross section for larger values of the scattering angle.

A recent {\it ab initio} quantum Monte Carlo (QMC) calculation of the NC differential $\nu$-$^{12}$C cross section, based on the Green's function Monte Carlo (GFMC) methods, has been recently reported~\cite{PhysRevC.97.022502}. While limited to the QE region, it has the advantage of relying on a first-principles description of nuclear dynamics, where the nucleons interact via two-body, AV18~\cite{PhysRevC.51.38}, and three-body, IL7~\cite{annurev.nucl.51.101701.132506}, potentials and with electroweak fields via effective currents, including one- and two-body terms. The GFMC methods then allow one to fully account, without approximations, for the complex many-body spin and isospin dependent correlations induced by these nuclear potentials and currents, and for the interaction effects in the final nuclear states.

A GFMC calculation of the $^{12}$C longitudinal and transverse electromagnetic response functions within a similar theoretical framework~\cite{PhysRevLett.117.082501} is in good agreement with the experimental data obtained from the Rosenbluth separation of inclusive $(e,e^\prime)$ cross sections. This result validates the dynamical framework and the model for the vector currents. Two-body currents significantly increase the transverse response in the QE peak. Other GFMC calculations for the electromagnetic and neutral-weak response functions of $^4$He and $^{12}$C show that two-body currents generate excess transverse strength, from threshold to the QE to the dip region and beyond~\cite{PhysRevC.91.062501,PhysRevLett.112.182502}.

\begin{figure}[tb]
\centering\includegraphics[width=\textwidth]{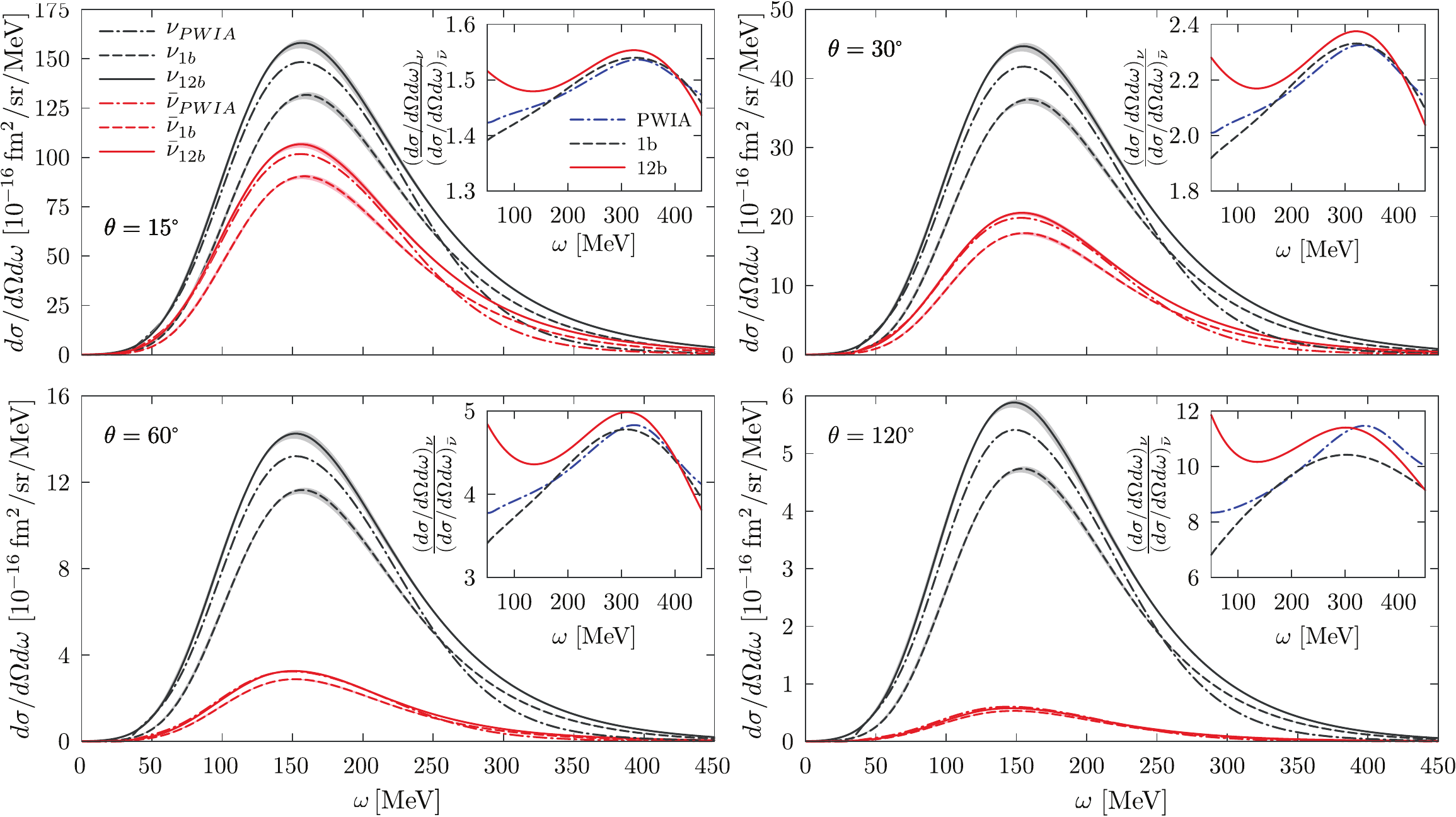}
\vspace{-2mm}
\caption{NC differential cross section for $\nu$ (black curves) and $\nubar$ (red curves) scattering on $^{12}$C at $q=570$~MeV
obtained with GFMC methods with one-body current (1b) and with the sum one-body and two-body currents (12b) as functions of $\omega$ and for different values of the scattering angle. The insets show ratios of the $\nu$ to $\nubar$ cross sections. Also shown are the PWIA results. Taken from Ref.~\cite{PhysRevC.97.022502}
}
\label{gfmc}
\end{figure}

In figure~\ref{gfmc} the NC differential cross sections obtained with the GFMC methods for $\nu$ and $\nubar$ scattering on $^{12}$C are displayed, as a function of the energy transfer $\omega$, for a fixed value of the momentum transfer, $q = 570$~MeV, and for different values of the scattering angle. The initial and final $\nu$ and $\nubar$ energies are obtained in terms of $\omega, q$, and $\theta$. The insets in the figure show the ratio of the $\nu$ to $\nubar$ cross sections. The results indicate substantial two-body current contributions, over the entire QE region, which significantly impacts the magnitude of the cross sections, their energy dependence, and the ratio of $\nu$ to $\nubar$ cross sections. The contribution of two-body currents produces a substantial enhancement of the cross section calculated with only the one-body current. The enhancement is mostly due to the constructive interference between the one- and two-body current matrix elements. The $\nubar$ cross section decreases rapidly relative to the $\nu$ cross section, as the scattering angle changes from the forward to the backward direction. Two-body current contributions are smaller for the $\nubar$ than for the $\nu$ cross section. For comparison, in the figure also the PWIA results are presented, in which only one-body currents are retained. The comparison shows how correlations and FSI quench the QE peak, redistributing strength to the threshold and high-energy transfer regions.

In the LFG-RPA model the np-nh contributions are obtained by Martini~\emph{et al.}~\cite{Martini:2009uj, Martini:2010ex, Martini:2011wp} starting from the microscopic calculations of the transverse response in electron scattering of Ref.~\cite{Alberico:1983zg}, from the results of pion and photon absorption of Ref.~\cite{Oset:1987re}, and from the results of pion absorption at threshold of Ref.~\cite{Shimizu:1980kb}. The approach of Nieves \emph{et al.}~\cite{Nieves:2004wx, Nieves:2005rq, PhysRevD.85.113008, PhysRevD.88.113007} can be considered as a generalization of the work of Ref.~\cite{Gil:1997jg}, developed for electron scattering, to neutrino scattering. The contributions related to the non-pionic $\Delta$ decay are taken, as in the case of Martini~\emph{et al.}, from Ref.~\cite{Oset:1987re}. Both models are based on the Fermi gas and 2p-2h calculations are performed in a basis of uncorrelated nucleons. If a basis of uncorrelated nucleons is used in the 1p-1h sector, also $NN$-correlation contributions must be considered. In the framework of an independent particle model, like a FG model or a MFA, these correlations are included by considering an additional two-body correlation current. Even in a simple FG model an exact relativistic 2p-2h calculation is computationally extremely demanding, since it requires to perform multi-dimensional  integrals for a huge number of 2p-2h diagrams. In addition, divergences in the $NN$-correlation sector and in the angular distribution of the ejected nucleons may appear and need to be regularized~\cite{PhysRevD.90.033012}. Different approximations are employed by the different groups to reduce the dimension of the integrals and to regularize the divergences. MEC contributions, $NN$ correlations, and $NN$ correlations-MEC interference are present  both in Martini~\emph{et al.} and Nieves~\emph{et al.} models, although with some differences~\cite{Katori:2016yel}. Therefore, although in general able to give a reasonable description of the experimental data, the models can give different results.

\section{Strange nucleon form factors \label{sec:strange}}

The strange quark contribution to the elastic form factors of the nucleon is an open and interesting question which has been the subject of intense experimental activity for several decades. Neutral-current QE neutrino-nucleus scattering is a useful tool to probe the strange quark content of the nucleon~\cite{Musolf:1993tb, Alberico:2001sd}.

It is well known that the net strangeness of the nucleon is zero. It is also known, however, that according to the quantum field theory in the cloud of a physical nucleon there must be pairs of strange particles. From the viewpoint of QCD the nucleon consists of $u$ and $d$ quarks and of a sea of $q{\bar q}$ pairs produced by virtual gluons. Then, the question is: how do the sea quarks, in particular strange quarks, contribute to the observed properties of the nucleon? The first evidence that the constant $g^{\mathrm s}_{\mathrm A}$ of equation~(\ref{eq.ga}), which is the value of the strange axial form factor at $Q^2=0$ and, equivalently, the net strange spin contribution ${\Delta}s$, is different from zero and large was found  by the EMC experiment at CERN~\cite{Ashman:1989ig}, in a measurement of deep inelastic scattering of polarized muons on polarized protons. This result triggered new experiments and a lot of theoretical work.

It is very important that different and alternative methods are used to determine the matrix elements of the strange current. NC elastic (NCE) $\nu(\nubar)$-nucleon scattering is one of these methods and a suitable tool to investigate $g^{\mathrm s}_{\mathrm A}$. Actually, in neutrino experiments nuclear targets are generally used. Therefore, neutrinos interact with nucleons embedded in a nucleus, where nuclear effects can affect the cross section and the observed final state. In the QE region the neutrino-nucleus scattering process roughly corresponds to the knockout of individual nucleons out of the nucleus. This is the region where the strange quark contribution to the elastic form factors of the nucleon can be investigated. A precise study, however, requires that nuclear effects in $\nu(\nubar)$-nucleus interactions, as well as nuclear model uncertainties, are well under control.

The sensitivity of NCQE $\nu(\nubar)$-nucleus cross sections calculated with different theoretical models to the strange nucleon form factors has been investigated in a series of papers~\cite{Meucci:2004ip, Meucci:2006ir, PhysRevC.88.025502, vanderVentel:2003km, vanderVentel:2005ke, Leitner:2006sp, PhysRevC.77.054604, Cheoun_2008, PhysRevC.54.1954, PhysRevC.76.055501, PhysRevLett.74.4993, PhysRevC.52.3399}. The determination of strange form factors from an absolute cross section measurement is, however, a very hard experimental task due to difficulties in the determination of the neutrino flux. In addition, theoretical uncertainties on the approximations and on the ingredients of the models are usually larger than the uncertainty related to the strangeness content of the nucleon.

Different nucleon form factors contribute to the s.p. weak current operator of the NC scattering of equation~(\ref{eq.nc1b}). A combination of different measurements is required for a complete information. A determination of the strange form factors through a combined analysis of $\nu p, \nubar p$, and ${\vec e} p$ elastic scattering has been performed in Ref.~\cite{PhysRevLett.92.082002}. The electromagnetic form factors, $F_1$ and $F_2$ in equation~(\ref{eq.nc1c}), can be investigated in electron scattering. The value of the Weinberg angle $\theta_{\mathrm W}$ can be obtained from measurements of NC processes. CCQE scattering can give information on the axial form factor $G_{\mathrm A}$, whose determination is very important in general and, in particular, if we want to determine $g^{\mathrm s}_{\mathrm A}$, that is highly correlated to $G_{\mathrm A}$ and thus to the axial mass $M_{\mathrm A}$. The strange form factors, $F_1^{\mathrm s}$, $F_1^{\mathrm s}$, and $G^{\mathrm  s}_{\mathrm A}$, can be investigated in NC $\nu$ scattering and in parity-violating electron scattering (PVES). PVES is  essentially sensitive to $F_1^{\mathrm s}$ and $F_1^{\mathrm s}$ or, equivalently, to the strange electric and magnetic from factors $G^{\mathrm s}_{\mathrm E}$ and $G^{\mathrm s}_{\mathrm M}$.  A determination of $G^{\mathrm s}_{\mathrm A}$ in PVES is hindered by radiative corrections. In contrast, NC $\nu$ scattering is primarily sensitive to $G^{\mathrm s}_{\mathrm A}$. The interference with the strange vector form factors can be resolved by complementary experiments of PVES.

The role of strangeness contribution to the electric and magnetic nucleon form factors has been recently analyzed for PVES~\cite{PhysRevC.88.025502}. Specific values for the electric and magnetic strangeness were provided making use of all available data at different transferred momenta $Q^2$. The analysis of $1\sigma$ and $2\sigma$ confidence ellipses showed that zero electric and magnetic strangeness were excluded by most of the fits. However, the values of the strangeness in the electric and magnetic sectors compatible with previous studies lead to very minor effects in the separate proton and neutron contributions to the cross section for neutrino and antineutrino scattering. Moreover, these small effects tend to cancel, being negligible for the total differential cross sections. Although this cancellation also works for the axial-vector strangeness, its relative contribution to the separate proton and neutron cross sections is much larger than the one associated with the electric and magnetic channels. Therefore, we can consider only the influence of the axial-vector strangeness and how the NCE antineutrino cross sections change when the description of the axial-vector form factor of the nucleon is modified.

It is a common prescription to apply the dipole parametrization to the strange axial form factor and to use the same value of the axial mass  used for the non-strange form factor as a cutoff; the strange axial coupling constant  at $Q^2=0$ is $\Delta s$. A measurement of $\nu (\bar{\nu})$-proton elastic scattering at the Brookhaven National Laboratory at low $Q^2$ suggested a nonzero value for $\Delta s$ \cite{PhysRevD.35.785, PhysRevC.48.761}. The MiniBooNE Collaboration used the ratio of proton-to-nucleon NCE cross sections to extract $\Delta s = 0.08 \pm 0.26$~\cite{PhysRevD.81.092005} based on the RFG with $M_A=1.35$~GeV. The analysis performed in Ref.~\cite{GONZALEZJIMENEZ20131471} with the RMF model led to $\Delta s = 0.04 \pm 0.28$, while the COMPASS Collaboration reported a negative $\Delta s = -0.08 \pm 0.01$ (stat.) $\pm 0.02$ (syst.) as a result of a measurement of the deuteron spin asymmetry~\cite{ALEXAKHIN20078}, in agreement with the HERMES results~\cite{PhysRevD.75.012007}.

\begin{figure}[tb]
\centering
\includegraphics[width=.8\textwidth]{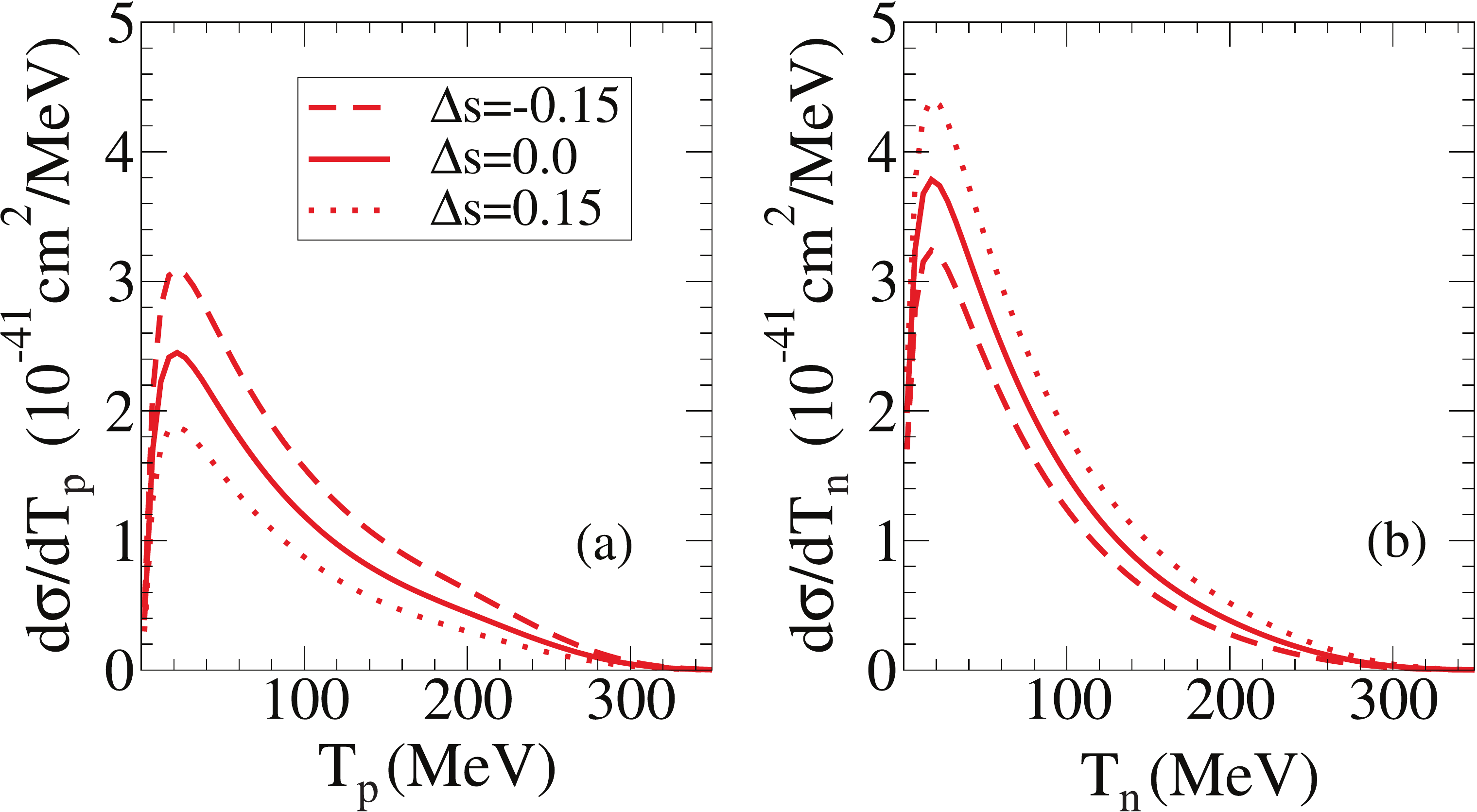}
\vspace{-2mm}
\caption{NCE antineutrino cross section at $E_{\bar \nu} = 500$~MeV as a function of the emitted proton [panel (a)] or neutron [panel (b)] kinetic energy. Calculations are performed in the RPWIA. Solid lines are the results with $\Delta s=0.0$, dashed lines with $\Delta s=-0.15$, and dotted lines with $\Delta s=+0.15$. Taken from Ref.~\cite{PhysRevC.88.025502}.}
\label{fig:L_T_Tp_antine3}
\end{figure}

An example of the NCE differential cross section is displayed in figure~\ref{fig:L_T_Tp_antine3} for antineutrino scattering off $^{12}$C. In order to avoid complications related to the description of FSI and to uncertainties due to the particular model, calculations have been performed in the RPWIA. The cross section for an emitted proton decreases when increasing $\Delta s$, while the cross section for an emitted neutron has the opposite behavior. Thus, the total proton$+$neutron cross section is almost independent of $\Delta s$ in the range $-0.15 \div 0.15$. This result is obtained for both neutrino and antineutrino scattering and it is rather independent of the incident energy.

\subsection{Ratios of cross sections \label{sec:ratio}}

Since the determination of the strangeness contribution to the axial form factor from measurements of NCE cross section is an extremely difficult experimental task, ratios of cross sections have been proposed as an alternative and useful tool to search for strangeness effects. Difficulties due to the determination of the absolute neutrino flux are reduced in the ratios. Moreover, also the sensitivity to FSI and to other  nuclear effects and to theoretical  uncertainties on these effects can be negligible or, in any case, strongly reduced in the ratios, where they can give a similar contribution to the numerator and to the denominator. In contrast, strangeness effects can be emphasized in the ratios, where form factors may contribute in a different way, for instance, with a different sign in the numerator and in the denominator.

The following ratios have been proposed~\cite{Alberico:2001sd,Alberico:1997vh,Alberico:1995bi}:
\begin{itemize}
\item
The ratio of neutrino-to-antineutrino cross section
\begin{equation}
R(\nu/\nubar)=
{{\left (\mathrm{d} \sigma /\mathrm{d} T_{\mathrm N} \right)_\nu}
\over {\left (\mathrm{d} \sigma /\mathrm{d} T_{\mathrm N} \right)_{\nubar}}
} \,\, .
\label{eq.rationunub}
\end{equation}
A measurement of this ratio is difficult to perform due to difficulties in dealing with antineutrinos.
\item
The ratio of proton-to-neutron (p/n) NC cross section
\begin{equation}
R(\mathrm{p}/\mathrm{n})=
{\left(\mathrm{d} \sigma /\mathrm{d} T_{\mathrm p} \right)_{\nu({\nubar})}
\over \left(\mathrm{d} \sigma /\mathrm{d} T_{\mathrm n} \right)_{\nu({\nubar})}}
\label{eq.ratiopn}
\end{equation}
is sensitive to the strange-quark contribution as the interference between $g^{\mathrm s}_{\mathrm A}$ and $g_{\mathrm A}$ contributes with an opposite sign in the numerator and in the denominator [see equation~(\ref{eq.ga})]. A precise measurement of this ratio appears, however, problematic due to the difficulties  associated with neutron detection.
\item
The ratio of NC over CC cross sections
\begin{equation}
R(\mathrm{NC}/\mathrm{CC}) =
{\left(\mathrm{d} \sigma /\mathrm{d} T_{\mathrm N} \right)^{\mathrm{NC}}_{\nu({\nubar})}
\over \left(\mathrm{d} \sigma /\mathrm{d} T_{\mathrm N} \right)^{\mathrm{CC}}_{\nu({\nubar})}
} \,\, .
\label{eq.ratioNCCC}
\end{equation}
Although sensitive to strangeness only in the numerator, a measurement of $R(\mathrm{NC}/\mathrm{CC})$ may appear more feasible since it is easier to measure CC cross sections.
\end{itemize}

In order to show up the effect of the strange-quark contribution and to allow an unambiguous determination of the presence of the magnetic and/or axial strange nucleon form factors, the study of the $\nu$-$\nubar$ asymmetry
\begin{equation}
{\cal{A}}
=
{
\left [ \left( \mathrm{d} \sigma /\mathrm{d} T_{\mathrm N} \right)_{\nu}-
\left( \mathrm{d} \sigma /\mathrm{d} T_{\mathrm N} \right )_{\nubar} \right ]^{\mathrm{NC}}
\over
\left [ \left( \mathrm{d} \sigma / \mathrm{d} T_{\mathrm N}  \right)_{\nu} -
\left( \mathrm{d} \sigma /\mathrm{d} T_{\mathrm N} \right)_{\nubar} \right ] ^{\mathrm{CC}} }
\label{eq.asym}
\end{equation}
was proposed in Ref.~\cite{Alberico:1995bi}.

\subsection{Numerical examples \label{sec:strangeex}}

Numerical examples for the above defined ratios calculated in different theoretical models and of their sensitivity to the strange quark content of the nucleon form factors can be found in many papers. Here we present only a few examples.

\begin{figure}[tb]
\centering
\includegraphics[width=0.8\textwidth]{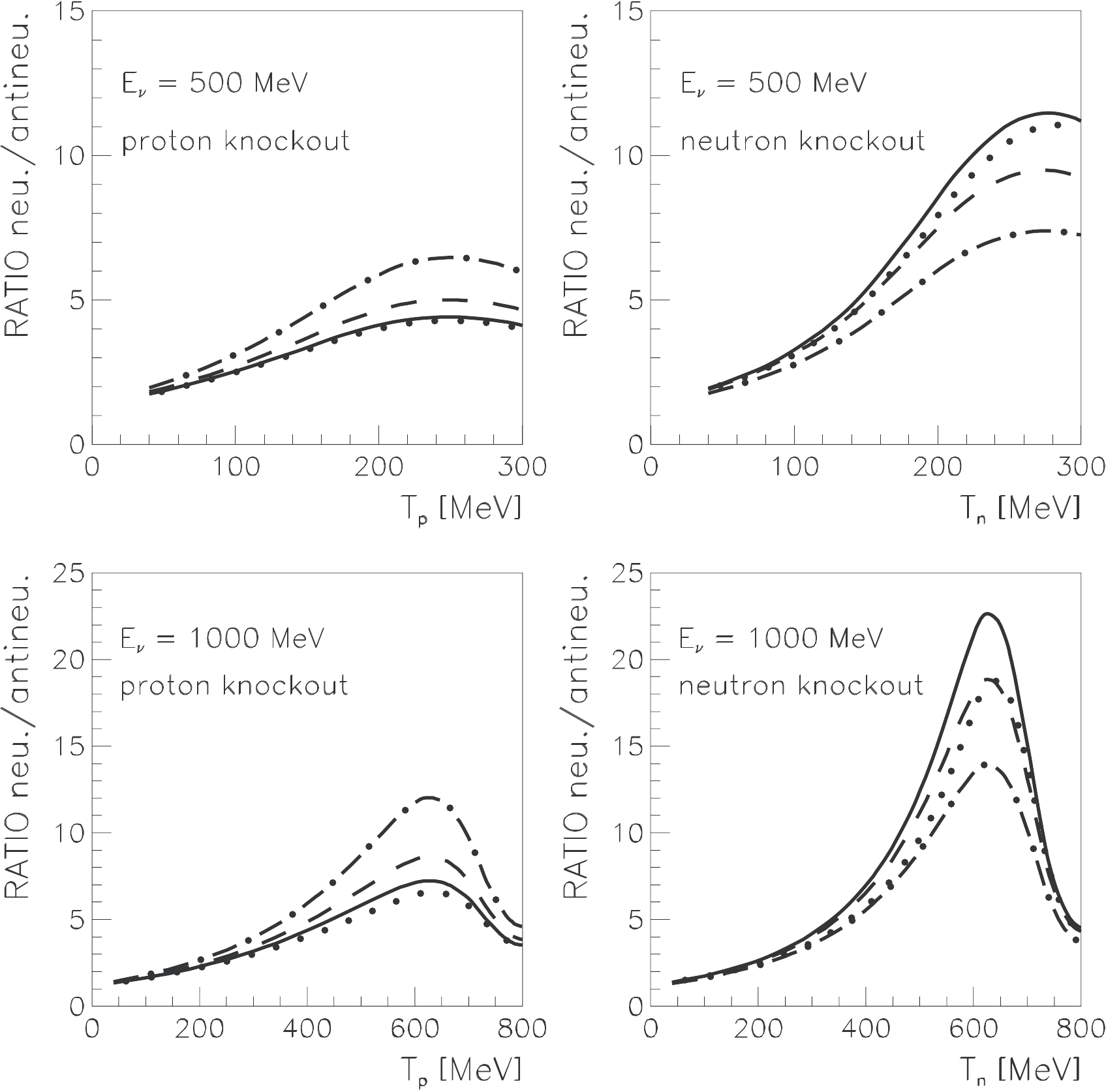}
\vspace{-2mm}
\caption{Ratio of neutrino-to-antineutrino NC cross sections, $R(\nu/\nubar)$ in  equation~(\ref{eq.rationunub}), on $^{12}$C as a function of T$_{\mathrm N}$. Dashed lines are the results with no strangeness contribution, solid lines with $g^{\mathrm s}_{\mathrm A} = -0.10$, dot-dashed lines with $g^{\mathrm s}_{\mathrm A} = -0.10$ and $\mu^{\mathrm s} = -0.50$, dotted lines with $g^{\mathrm s}_{\mathrm A} = -0.10$ and $\rho^{\mathrm s} = +2$. Taken from Ref.~\cite{Meucci:2006ir}.} \label{fig:rationunub}
\end{figure}

\begin{figure}[tb]
\centering
\includegraphics[width=.8\textwidth]{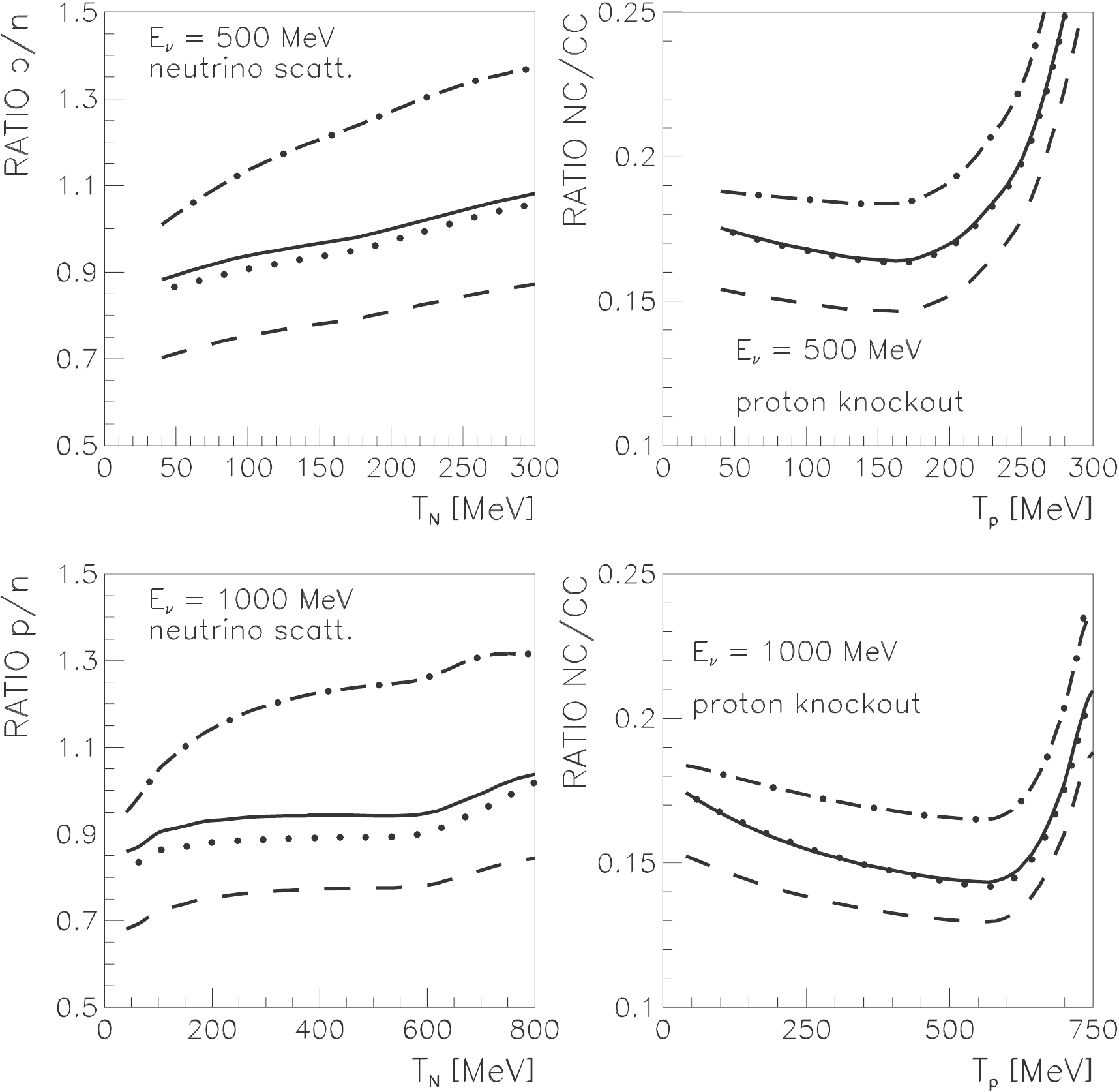}
\vspace{-2mm}
\caption{ Ratio of proton-to-neutron NC cross sections, $R(\mathrm{p}/\mathrm{n})$ in  equation~(\ref{eq.ratiopn}), (left panels) and of NC-to-CC cross sections,  $R(\mathrm{NC}/\mathrm{CC})$ in equation~(\ref{eq.ratioNCCC}), (right panels) for QE $\nu$ scattering off $^{12}$C as a function of the kinetic energy of the outgoing nucleon. Dashed lines are the results with no strangeness contribution, solid lines with $g^{\mathrm s}_{\mathrm A} = -0.10$, dot-dashed lines with $g^{\mathrm s}_{\mathrm A} = -0.10$ and $\mu^{\mathrm s} = -0.50$, dotted lines with $g^{\mathrm s}_{\mathrm A} = -0.10$ and $\rho^{\mathrm s} = +2$. Taken from Ref.~\cite{Meucci:2006ir}.} \label{fig:rationpcc}
\end{figure}

Examples of the ratios $R(\nu/\nubar)$, $R(\mathrm{p}/\mathrm{n})$, and   $R(\mathrm{NC}/\mathrm{CC})$ calculated in the RDWIA approach of Ref.~\cite{Meucci:2006ir} for $\nu$ and $\nubar$  scattering from $^{12}$C at incident energies of $500$ and $1000$~MeV are presented in figures~\ref{fig:rationunub} and \ref{fig:rationpcc}. The effects of a non-zero strange quark contribution to the nucleon form factors are shown by comparing  results without strange form factors and with $g^{\mathrm s}_{\mathrm A} = -0.10$, $g^{\mathrm s}_{\mathrm A} = -0.10$ and $\mu^{\mathrm s} = -0.50$, and $g^{\mathrm s}_{\mathrm A} = -0.10$ and $\rho^{\mathrm s} = +2$.

The ratio $R(\nu/\nubar)$  shown in figure~\ref{fig:rationunub} for proton and neutron knockout is sensitive to $\Delta {\mathrm s}$ and presents a maximum at T$_{\mathrm N} \simeq 0.6$ E$_{\nu}$. In the case of proton knockout the ratio is reduced by $g^{\mathrm s}_{\mathrm A}$ and enhanced by $\mu^{\mathrm s}$. In contrast, for neutron knockout  the ratio is enhanced by $g^{\mathrm s}_{\mathrm A}$ and reduced by $\mu^{\mathrm s}$. The ratio is reduced by $\rho^{\mathrm s}$, this effect is very small for proton knockout and larger for neutron knockout and increases with the incident energy. A measurement of $R(\nu/\nubar)$ is, however, difficult, due to difficulties in dealing with antineutrinos.

The ratio $R(\mathrm{p}/\mathrm{n})$ for neutrino scattering, shown in the left panels of figure~\ref{fig:rationpcc}, is enhanced by a factor $\simeq 20-30$\% when $g^{\mathrm s}_{\mathrm A}$ is included and by $\simeq 50$\% when both  $g^{\mathrm s}_{\mathrm A}$ and $\mu^{\mathrm s}$ are included. A minor effect is produced by $\rho^{\mathrm s}$, which gives only a slight reduction of the ratio. $R(\mathrm{p}/\mathrm{n})$ is sensitive to strangeness, in particular, to the strange axial form factor and thus to  $g^{\mathrm s}_{\mathrm A}$~\cite{Alberico:1997vh, PhysRevC.48.1919, PhysRevC.48.3078, vanderVentel:2003km, Meucci:2006ir, PhysRevC.88.025502}, but a precise measurement of this ratio requires neutron detection.

The strange quark contribution is expected to be less important in the  ratio $R(\mathrm{NC}/\mathrm{CC})$, which is sensitive to strangeness only in the numerator. A measurements of this ratio, however, appears, in principle, more feasible since it is easier to measure CC cross sections.  The results for proton knockout and an incident neutrino, in the right panels of  figure~\ref{fig:rationpcc}, show similar features at both incident energies. The enhancement of the ratio at large values of the kinetic energy of the knocked-out proton is due to the fact that, because of the muon mass, the CC cross section goes to zero more rapidly than the corresponding NC cross section. The strange-quark contribution produces a somewhat constant  enhancement of the ratio with respect to the case in which the strange-quark contribution is omitted. The simultaneous inclusion of $g^{\mathrm s}_{\mathrm A}$ and $\mu^{\mathrm s}$ gives an enhancement that is about a factor of $2$ larger than the one corresponding to the case where only $g^{\mathrm s}_{\mathrm A}$ included. The effect of $\rho^{\mathrm s}$ is very small. The effects of strangeness are somewhat energy dependent. In this case, the larger effect, {\it i.e.}, a reduction of the ratio, is obtained when only $g^{\mathrm s}_{\mathrm A}$ is included. The global effect is reduced when also $\mu^{\mathrm s}$ is considered.

\begin{figure}[tb]
\centering
\includegraphics[width=0.49\textwidth]{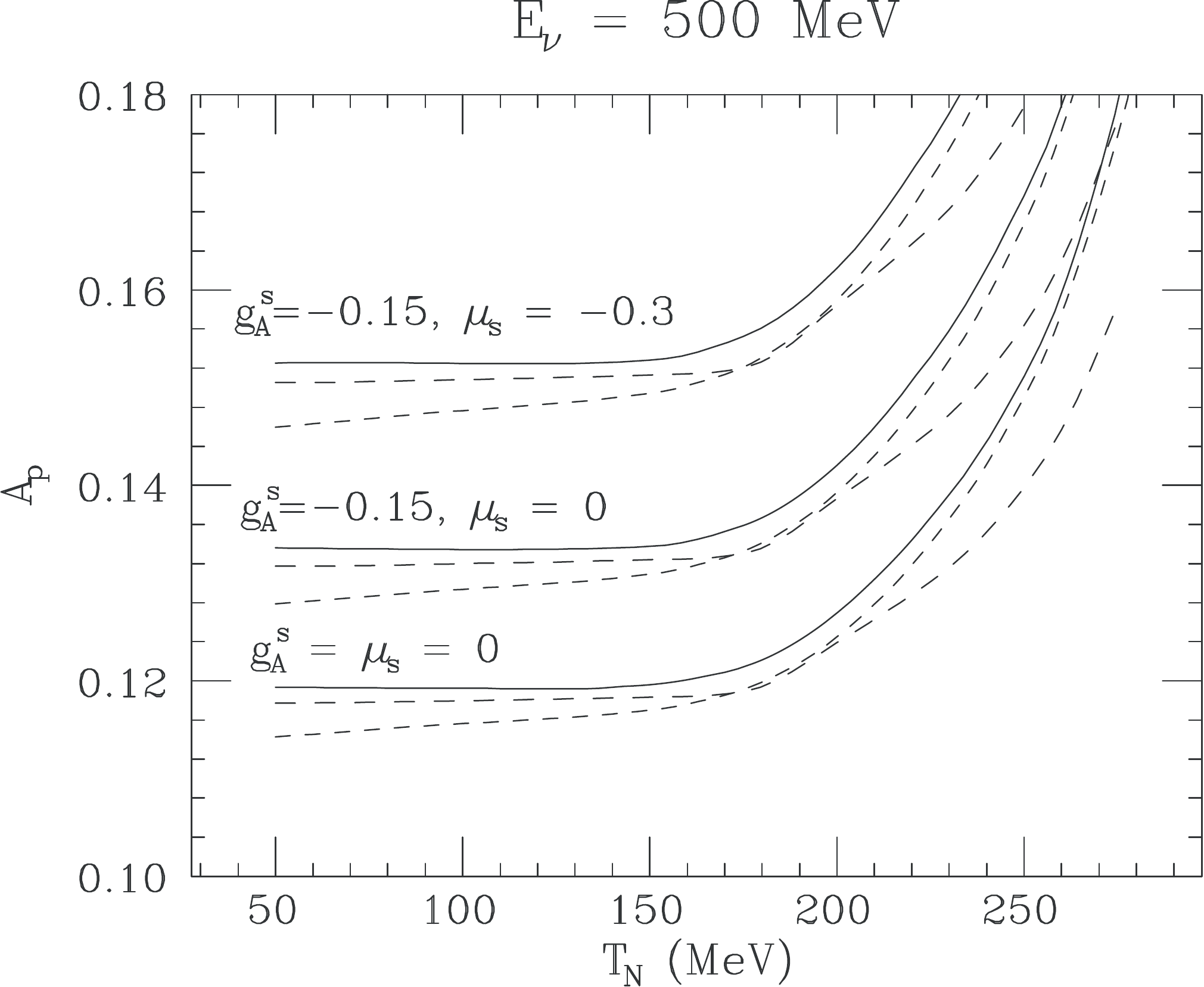}\hfill
\includegraphics[width=0.49\textwidth]{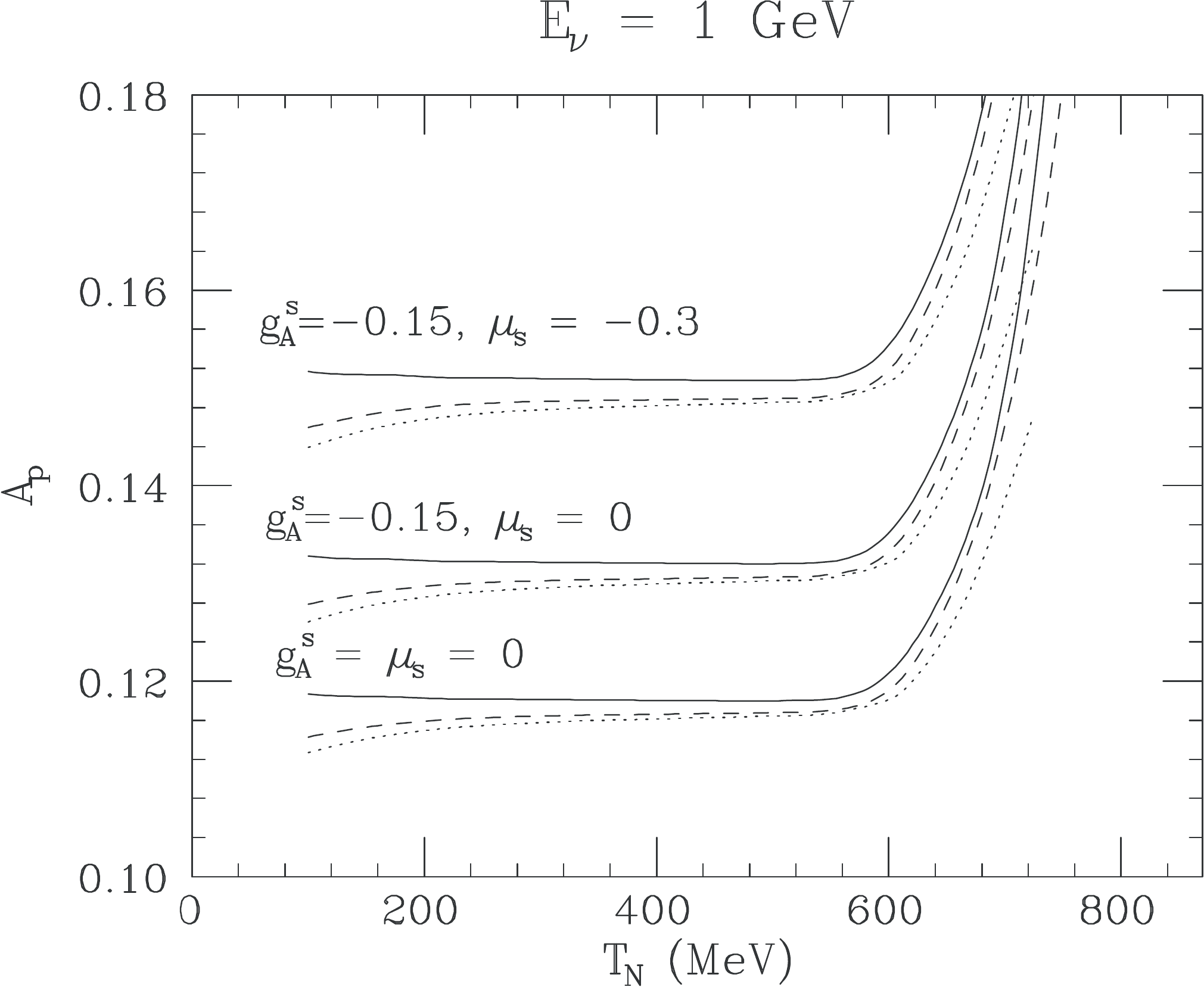}
\vspace{-2mm}
\caption{The  asymmetry,  ${\cal A}_p$ in equation~(\ref{eq.asym}), as a function of T$_{\mathrm N}$ calculated for $\nu$ and $\nubar$ scattering on $^{12}$C at the incident energies of $500$ MeV and $1$ GeV. The results obtained with different models are compared: RSM (solid lines), RDWIA (dot-dashed lines), RFG (dashed lines in the left panel), and RSM corrected by FSI and Coulomb distortion of the muon in  the CC processes (dotted lines in the right panel). The three sets of curves correspond to different choices of strangeness parameters: $g_A^s=\mu_s=0$ (lower lines), $g_A^s=-0.15$, $\mu_s=0$ (intermediate  lines) and $g_A^s=-0.15$, $\mu_s=-0.3$ (upper lines). Taken from Ref.~\cite{Alberico:1997vh}.}
\label{fig:asym}
\end{figure}

Numerical examples for the  asymmetry  ${\cal A}_p$, defined in equation~(\ref{eq.asym}) as the ratio of the differences between neutrino and antineutrino NC and CC cross sections, are presented in figure~\ref{fig:asym}. The results obtained  for $\nu$ and $\nubar$ scattering on $^{12}$C at incident energies of $500$~MeV and $1$~GeV, with and without strange form factors and with different values of the strangeness parameters $g_A^s$ and $\mu_s$, are compared for different relativistic models: RFG, RSM, and RDWIA~\cite{Alberico:1997vh}. The finite muon mass, which, as already noticed in the discussion of figure~\ref{fig:rationpcc}, brings the CC cross sections down to zero at lower $T_{\mathrm N}$ values with respect to the NC ones, produces a rapid increase of the asymmetry for $T_{\mathrm N} \ge 150$~MeV, thus leaving a reasonable energy range, below $150$~MeV, in which the asymmetry has a fairly constant value. The differences obtained between the results of the various models for the cross sections are strongly reduced in the ratio, the largest difference, within 8\%, is associated to FSI. The results show a stronger  sensitivity to the choice of the strangeness parameters and confirm that the asymmetry can be an appropriate quantity to investigate strange nucleon form factors.

\begin{figure}[tb]
\centering
\includegraphics[width=\textwidth]{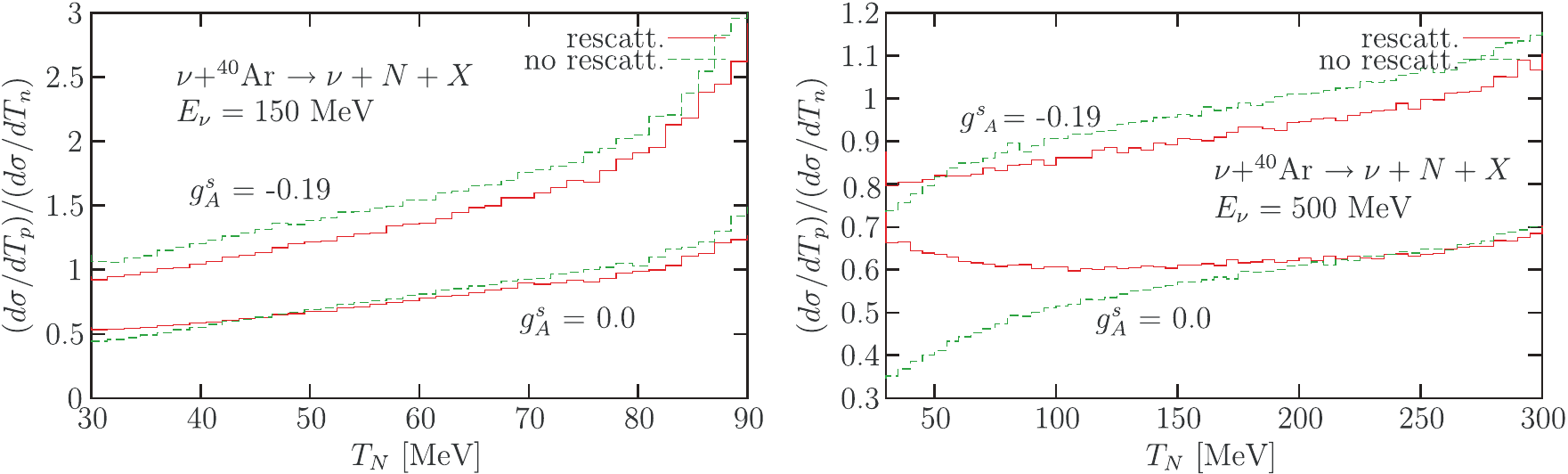}
\vspace{-2mm}
\caption{Ratio of proton-to-neutron NC cross sections, $R(\mathrm{p}/\mathrm{n})$ in  equation~(\ref{eq.ratiopn}), for NC scattering on $^{40}$Ar at $E_\nu=150$ MeV (left panel) and  $E_\nu=500$ MeV (right panel) as a function of T$_{\mathrm N}$. Calculations are performed in the LFG-RPA model of Nieves \emph{et al.} ~\cite{Nieves:2005rq} with (solid histogram) and without (dashed histogram) nucleon rescattering. Taken from Ref.~\cite{Nieves:2005rq}.}
\label{fig:resc}
\end{figure}

The ratio $R(\mathrm{p}/\mathrm{n})$ is sensitive to the parameter $g^s_A$ and it is therefore a useful observable to obtain information on the strange axial form factor of the nucleon. A numerical example for $R(\mathrm{p}/\mathrm{n})$,  calculated for NC scattering on $^{40}$Ar at $E_\nu=150$~MeV and $E_\nu=500$~MeV in the LFG-RPA model of Ref.~\cite{Nieves:2005rq} is presented in figure~\ref{fig:resc}. The results of the full model and those obtained without nucleon rescattering are compared for $g^s_A = 0$ and $-0.19$. The contribution of nucleon rescattering produces only minor changes for light nuclei, because of the smaller average density, and increases for heavier nuclei~\cite{Nieves:2005rq} and for larger neutrino energies, as shown in the figure, where the importance of the secondary nucleons shows up at the low energies side of the spectrum. The results confirm the sensitivity of the  ratio $R(\mathrm{p}/\mathrm{n})$ to $g^s_A$.

\begin{figure}[tb]
\centering
\includegraphics[width=0.8\textwidth]{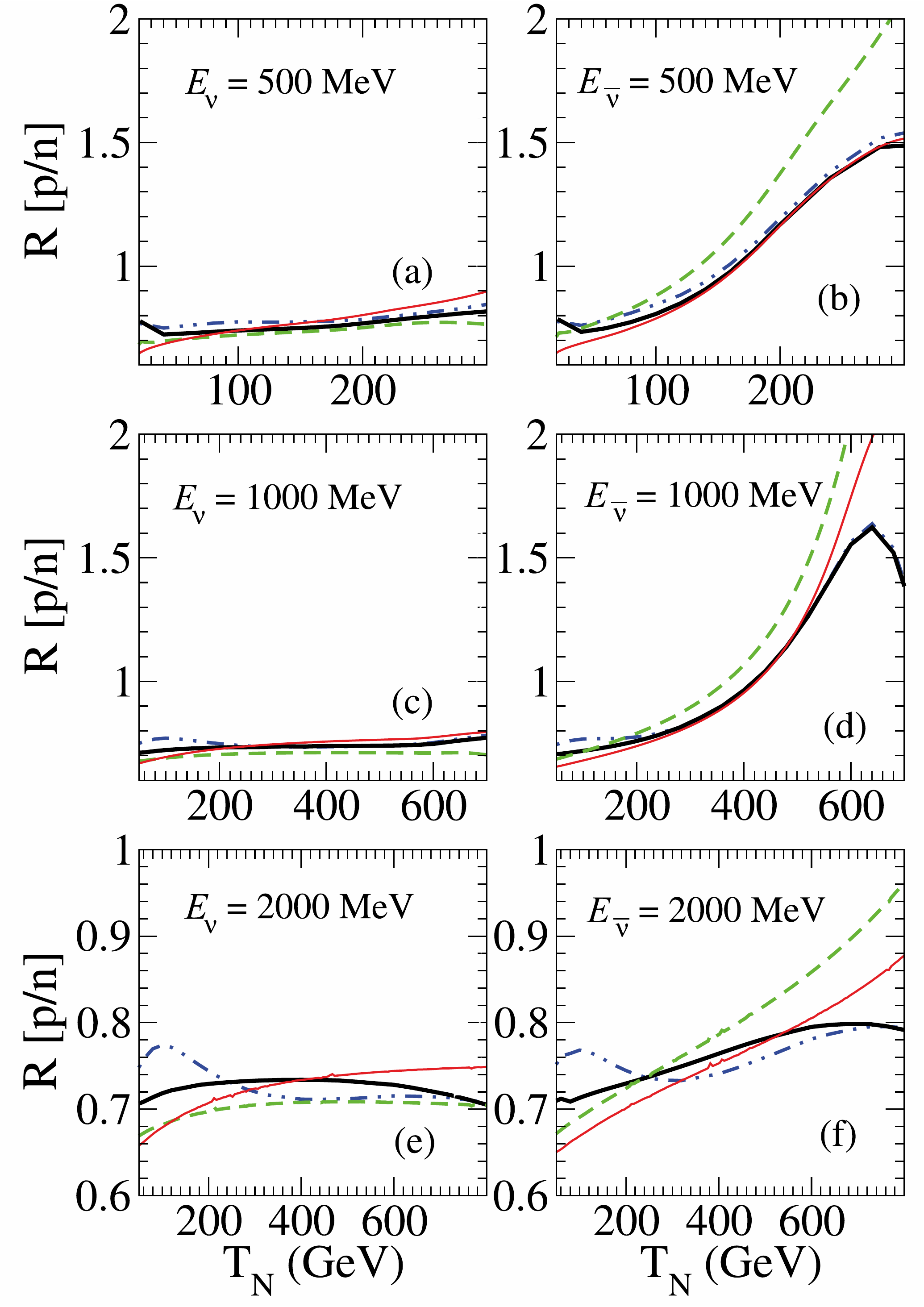}
\vspace{-2mm}
\caption{Ratio of proton-to-neutron cross sections $R(\mathrm{p}/\mathrm{n})$ for NC neutrino [panels (a), (c), and (e)] and antineutrino [panels (b), (d), and (f)] scattering on $^{12}$C and for different incident energies as a function of $T_{\mathrm N}$.  Results of different descriptions of FSI are compared: RPWIA (thin solid lines), RMF (dashed lines), RGF-EDAD1 (thick solid lines), and RGF-EDAI (dash-dotted lines). The vector and axial-vector strange form factors have been fixed to zero. Taken from Ref.~\cite{PhysRevC.88.025502}.}
\label{fig:ratiopn_models}
\end{figure}

In figure~\ref{fig:ratiopn_models} we display $R(\mathrm{p}/\mathrm{n})$ for three different neutrino and antineutrino energies. The results obtained with the same RMF for nuclear structure and with different models to describe FSI are compared: RPWIA, where FSI are neglected, RMF, where the same RMF potential is used for bound and scattering states, and RGF, where FSI are described by a complex ROP. The RGF results obtained with two different parametrizations of the phenomenological ROP are compared in the figure. The strange form factors have been fixed to zero for an easier comparison of the results obtained with different models. The differences between the results of the different models can be significant on the cross sections, where the effects of FSI are generally large~\cite{PhysRevC.88.025502}. These effects are largely reduced in the ratio, where the three models give similar results. In the case of neutrino scattering the ratio is almost constant and the RPWIA, RMF, and RGF results coincide up to a few percent, but for a small bump at low values of $T_{\mathrm N}$ in the RGF-EDAI model, which is due to the different behaviour of the EDAI optical potential at low energies. For larger $T_{\mathrm N}$ the ratio stabilizes, being the discrepancy between the different models at most of $\sim 4-5\%$. Finally, the differences increase at the largest values of $T_{\mathrm N}$. Note that in this region the cross sections are very small and show a significant sensitivity to FSI and/or the thresholds used. The maximum uncertainty in the ratio, linked to the different models, is of the order of $\sim 15\%$ ($E_\nu=500$ MeV) and $\sim 8\%$ ($E_\nu=1$ and $2$~GeV). Larger differences are obtained in the case of antineutrino scattering, in particular for the RMF model, whose results are significantly enhanced with respect to the RGF ones for large values of $T_{\mathrm N}$, where, however, the cross section becomes significantly lower than its maximum and a very precise measurement is required to obtain a clear result.

\begin{figure}[tb]
\centering
\includegraphics[width=0.8\textwidth]{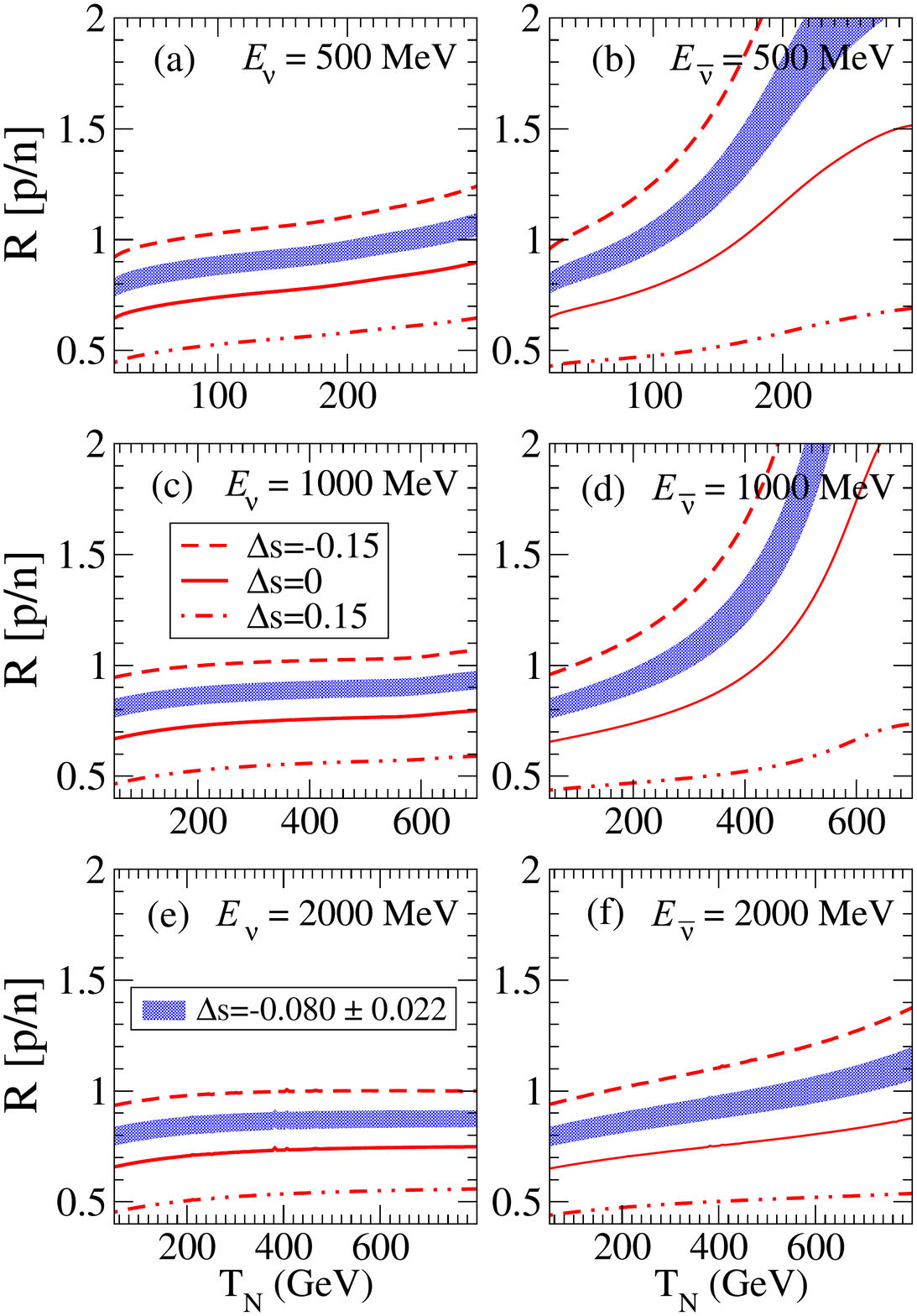}
\vspace{-2mm}
\caption{Ratio of proton-to-neutron cross sections $R(\mathrm{p}/\mathrm{n})$ for NC neutrino [panels (a), (c), and (e)] and antineutrino [panels (b), (d), and (f)] scattering on $^{12}$C and for different incident energies as a function of $T_{\mathrm N}$. Calculations are performed in the RPWIA and with different values of $\Delta s$.  The shadowed band refers to results corresponding to the COMPASS-HERMES measurement for the axial strangeness. Taken from Ref.~\cite{PhysRevC.88.025502}.}
\label{fig:ratiopn_strange}
\end{figure}

The dependence of $R(\mathrm{p}/\mathrm{n})$ on the strange quark-contribution is shown in figure~\ref{fig:ratiopn_strange}. The results obtained in the RPWIA with different values of $\Delta s$ are compared in the figure. The ratio is enhanced when calculations are performed with a negative $\Delta s$  and suppressed when a positive $\Delta s$ is considered. In the case of antineutrino scattering the role of strangeness contribution is particularly significant when a negative $\Delta s$ is assumed, with a large peak at $T_{\mathrm N} \approx 0.7 E_{\bar{\nu}}$. It was found in Ref.~\cite{Meucci:2006ir} that a moderately large and negative strangeness contribution to the magnetic moment of the nucleon can cancel the peak in $R(\mathrm{p}/\mathrm{n})$. Although a large strangeness contribution to the vector form factors is not supported by any available experimental evidence \cite{GonzalezJimenez:2011fq}, it would be intriguing to look for possible strangeness effects with a direct measurement of this quantity. Even if a precise measurement of $R(\mathrm{p}/\mathrm{n})$ represents a very hard experimental task, the first measurement of the MiniBooNE Collaboration~\cite{miniboonenc} has proven the validity of this experimental technique and, hopefully, new data will be available in the future.

In the results of figure~\ref{fig:ratiopn_strange}, the uncertainty in $R(\mathrm{p}/\mathrm{n})$ associated to the axial strangeness is quite large. This large range  of variability of  $\Delta s$ is in accordance with $\nu (\bar{\nu})$ Brookhaven data~\cite{PhysRevD.35.785, PhysRevC.48.761} and also with the MiniBooNE results~\cite{miniboonenc}, but the COMPASS measurements suggest a narrower interval for the axial strangeness~\cite{ALEXAKHIN20078}, which results in a reduced range of variation of the
ratio. This is represented in figure~\ref{fig:ratiopn_strange} by the shadowed band, that is of the same order of magnitude as
the uncertainties related to the FSI effects in figure~\ref{fig:ratiopn_models}.

This sensitivity to $\Delta s$ gets much larger for antineutrinos, where the ratio goes up very fast with increasing values of $T_{\mathrm N}$. However, as in the case of neutrinos, the range of variation in the ratio associated to the COMPASS measurement is similar to the uncertainty introduced by nuclear models and/or FSI effects.

The results in figure~\ref{fig:ratiopn_strange} show the capability of the ratio $R(\mathrm{p}/\mathrm{n})$ as an useful observable to obtain  precise information on the axial-vector strangeness content in the nucleon. They also indicate, however, that, owing to the actual precision in the axial strangeness given by the COMPASS experiment, a deep and careful analysis of nuclear effects and of the uncertainties linked to nuclear effects and to their description in different nuclear models is mandatory to obtain reliable and precise information on the strange axial form factor of the nucleon.

\section{Comparison with experimental data \label{sec:results}}

As already mentioned in section~\ref{sec:models}, the models adopted to describe $\nu(\nubar)$-nucleus scattering are usually an extension of the models developed to describe electron-nucleus scattering. The comparison with the large amount of available accurate electron-scattering data is a first necessary test of the validity and of the predictive power of a model. In spite of many similar aspects, however, there are also several key differences between electron and neutrino-nucleus scattering. In particular, the kinematic situation is different. In electron-scattering experiments the incident flux and energy are known, the scattered electron is detected, and the energy and momentum transfer, $\omega$ and $\nq$, are clearly determined. In contrast, in neutrino experiments the flux is uncertain to 10--20\% or more, the incident neutrino energy is not known, and there are large uncertainties on the kinematic variables that are critical for the description of QE neutrino-nucleus scattering. In particular, $\omega$ and $\nq$, whose values establish the kinematic region where a specific reaction mechanism is expected to be dominant (for instance, direct one-nucleon knockout in the QE region), are not fixed. The incident neutrino energy can only be  inferred from the particles observed in the final state. Such an energy reconstruction is, however, not trivial, since it relies not only on the available experimental information on the detectors, but on the assumption of a particular reaction mechanism and on reliable nuclear models for its description. In neutrino experiments neutrino energy reconstruction and hence flux unfolding is possible only in model-dependent ways. Therefore, present neutrino-interaction measurements produce flux-integrated differential cross sections as a function of direct observables.

\subsection{The flux-averaged neutrino-nucleus cross section \label{sec:flux}}

The usual quantity measured in neutrino experiments is the differential cross section integrated and then averaged over of the $\nu(\nubar)$ flux of a specific experiment, {\it i.e.},
\begin{equation}
\frac{\diff \sigma } {\diff X } = \frac{1}{\Phi_{\mathrm{tot}}}
\int \frac{\diff \sigma} {{\diff X} {\diff E} }
\Phi(E) {\diff E},
\label{eq.fluxcs}
\end{equation}
where X is a direct observable, usually the kinetic energy of the outgoing nucleon $T_{\mathrm N}$ or $Q^2$, $E$ is the incident $\nu(\nubar)$  energy, $\Phi(E)$ is the $\nu(\nubar)$ flux, and $\Phi_{{\mathrm{tot}}}$ is the total integrated flux factor:
\begin{equation}
\Phi_{\mathrm{tot}}=\int\Phi(E){\diff E}.\label{eq.fluxtot}
\end{equation}

The experimental flux-integrated cross section contains events from a wide range of incoming $\nu(\nubar)$ energies and, as such, it can include contributions of several reaction channels. Therefore, the comparison of the theoretical results with data cannot be made in the same way as in electron scattering, where the incident electron energy is known. The comparison requires the calculation of the cross section for all the values of the incident $\nu(\nubar)$ energy where the flux has significant strength, over many different kinematic situations, for different values of $\omega$ and $\nq$ corresponding not only to the QE region,  but also to other regions, where different reaction mechanisms can be important. As a consequence, models for QE scattering could be unable to describe NCE data unless all other processes contributing to the experimental flux-averaged cross sections (such as, for instance, np-nh contributions, MEC, pion production and reabsorption...) are taken into account. A further complication is given by the fact that the differential cross sections are not function of inferred variables, but of direct observables, which need to be determined by  integrating over the neutrino flux and therefore including all relevant processes.

We point out that, because of the flux dependence of the measured cross sections, experiments with different $\nu(\nubar)$ beams measure different differential cross sections. Therefore, data sets from different beams are not directly comparable and they can only be related through the theoretical interaction models.

\subsection{Comparison of theoretical and experimental results \label{sec:moddata}}

\begin{figure}[tb]
\centering
\includegraphics[width=0.49\textwidth]{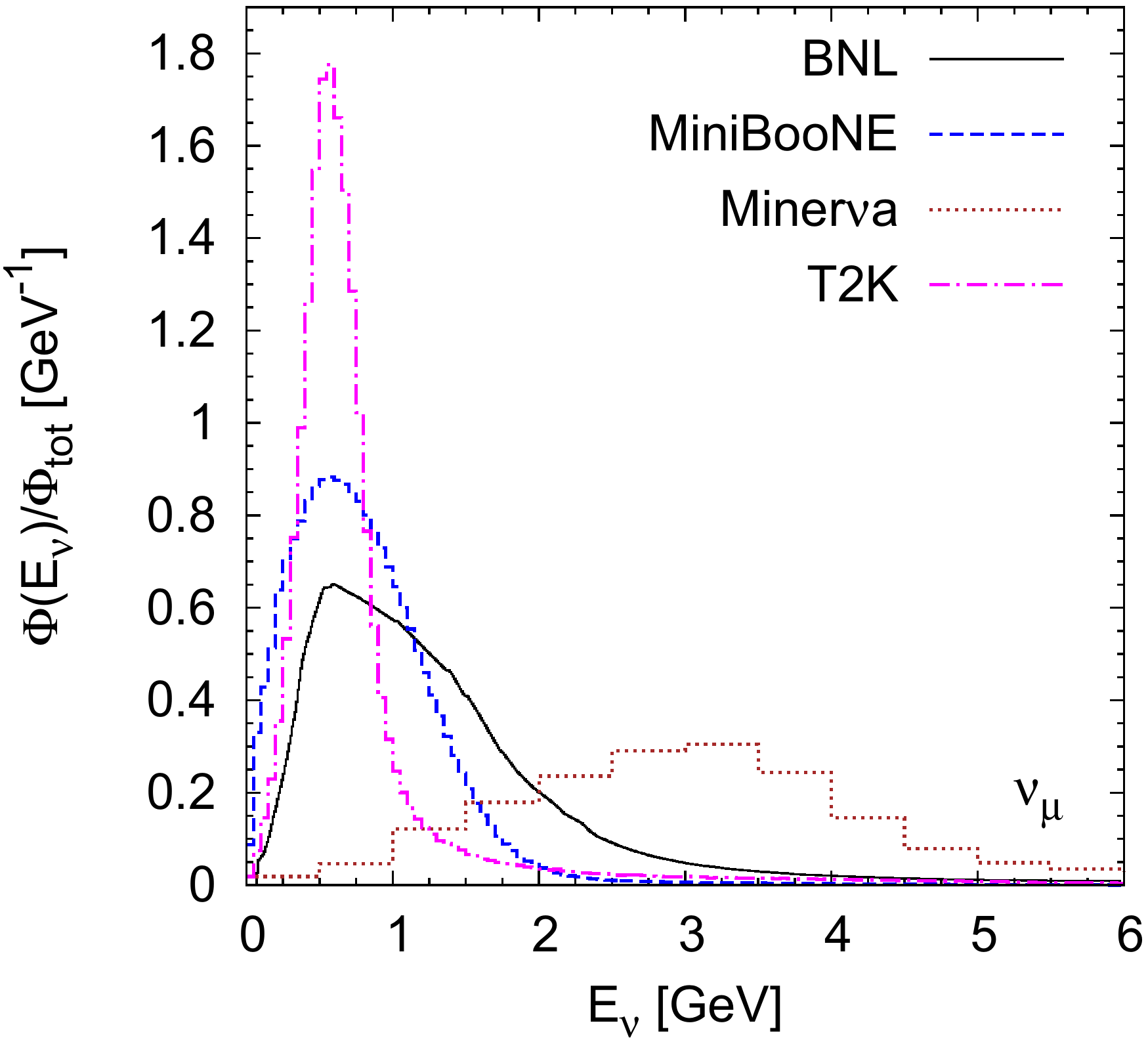}\hfill\includegraphics[width=0.49\textwidth]{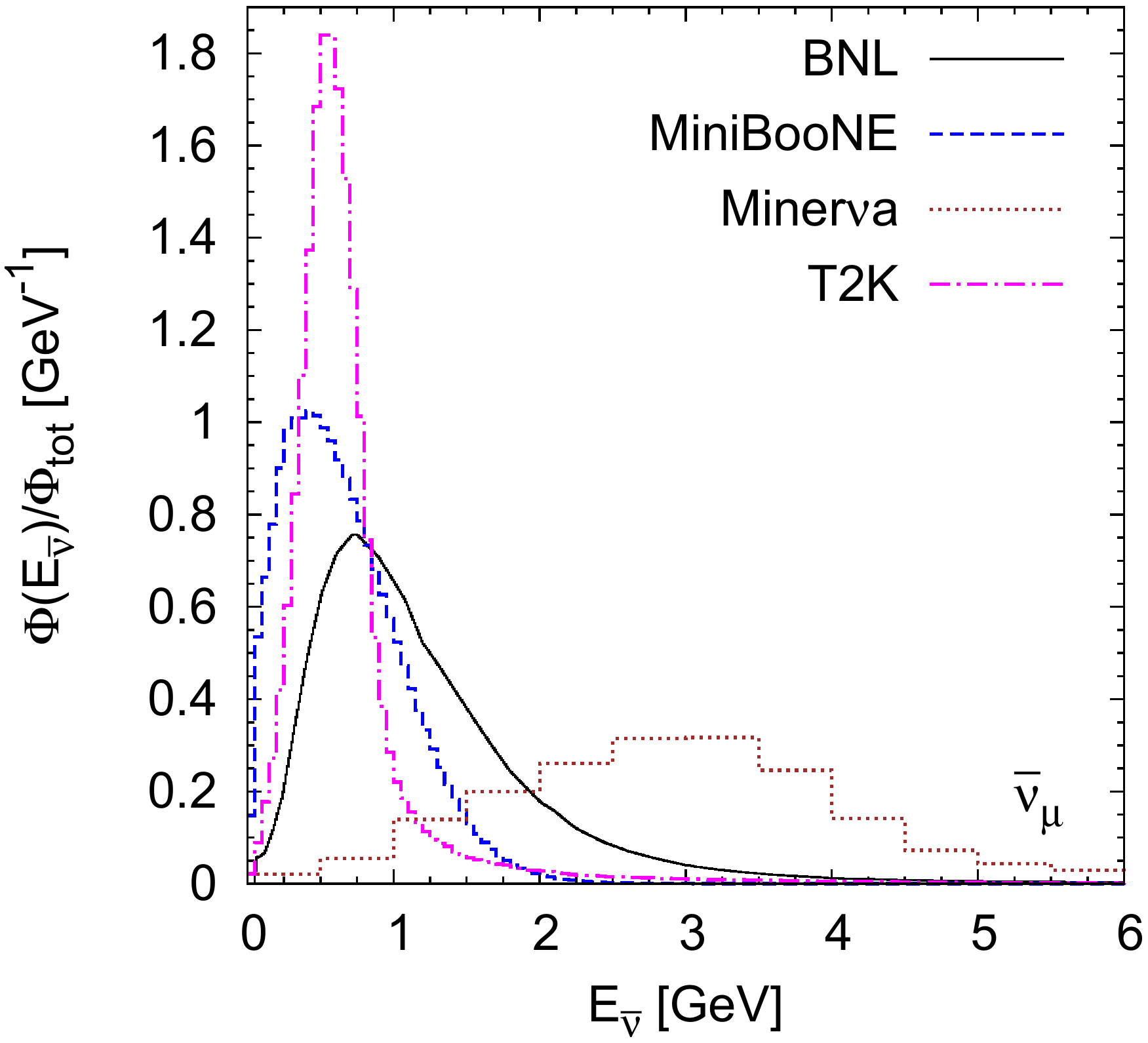}
\vspace{-2mm}
\caption{The predicted $\nu_\mu$ ($\overline{\nu}_\mu$) fluxes at the BNL~\cite{PhysRevD.34.75}, MiniBooNE~\cite{PhysRevD.79.072002}, T2K\cite{PhysRevD.87.012001}, and  MINER$\nu$A~\cite{PhysRevD.94.092005} detectors and corresponding mean energies.\label{fig:fluxes}
}
\end{figure}

The predicted $\nu_\mu$ ($\bar{\nu}_\mu$) fluxes at the BNL~\cite{PhysRevD.34.75}, MiniBooNE~\cite{PhysRevD.79.072002}, T2K~\cite{PhysRevD.87.012001}, and MINER$\nu$A~\cite{PhysRevD.94.092005} detectors are compared in figure~\ref{fig:fluxes}. The neutrino and antineutrino mean energies corresponding to the MiniBooNE and T2K experiments are rather similar ($\sim0.7$--$0.8$~GeV), although the T2K energy flux shows a much narrower distribution. This explains the different role played by 2p-2h MEC effects in the two experiments, these being larger for MiniBooNE (see, \emph{e.g.}, Refs.~\cite{PhysRevD.94.093004, Ivanov:2018nlm}) in the case of CCQE neutrino scattering. On the contrary, the BNL ($\sim 1.2-1.3$~GeV) and MINER$\nu$A ($\sim 3.5-4.0$~GeV) energy fluxes are much more extended to higher energies.

From the neutrino experiments which obtained results on NCE scattering off nuclei, the highest event statistics to date have been collected by the Brookhaven National Laboratory Experiment 734 (BNL E734)~\cite{PhysRevLett.56.1107, PhysRevD.35.785} and MiniBooNE~\cite{miniboonenc, miniboone-ncqe-nubar} Collaborations, allowing for extraction of the differential cross sections. In this section we will present recent results of the comparison between the data of these experiments and the predictions of different theoretical models.

The BNL E734 experiment studying $\nu(\bar{\nu}) p\rightarrow\nu(\bar{\nu}) p$ NCE interactions, where the target was composed in 79\% of protons bound in carbon and aluminium and in 21\% of free protons, was performed using a 170-metric-ton high-resolution target detector in a horn-focused (anti)neutrino at the BNL. The experimental cross-sections show a behavior similar to that of the MiniBooNE experiment, which, using the Cherenkov detector filled with mineral oil (CH$_2$), is sensitive to both $\nu(\bar{\nu}) p\rightarrow\nu(\bar{\nu}) p$ and  $\nu(\bar{\nu}) n\rightarrow\nu(\bar{\nu}) n$ NCE scattering~\cite{miniboonenc, miniboone-ncqe-nubar}.

The NC cross sections depend on the strangeness content of the nucleon, particularly through the axial form factor. However, it has been known for some time (see, \emph{e.g.},~\cite{Alberico:1997vh, ALBERICO1999277, PhysRevC.54.1954}) that the $\Delta s$ dependence of the NCE neutrino-nucleon cross section is, in general, very mild. This results from a cancellation between the effect of $\Delta s$ on the proton and on the neutron contributions, which are affected differently by the axial strangeness: by changing $\Delta s$ from $0$ to a negative value the proton cross section gets enhanced while the neutron one is reduced, so that the net effect on the total cross section is very small. The NCE $\nu(\bar{\nu}) p\rightarrow\nu(\bar{\nu}) p$ differential cross sections measured by the BNL E734 experiment are sensitive to the value of $\Delta s$~\cite{PhysRevC.48.1919} (in this case there is not a cancellation effect) and reported a value of $\Delta s = -0.15 \pm  0.09$~\cite{PhysRevD.35.785, ALBERICO1999277, PhysRevC.48.761}. However, as already mentioned in section~\ref{sec:strange}, measurements of ratios of cross sections offer the possibility for extracting $\Delta s$ at low $Q^2$ with reduced systematic errors. For example, a measurement of $(\nu p\to\nu p)/(\nu_\mu n\to\mu p)$, has been proposed by the FINeSSE experiment~\cite{Bugel:2004yk}. For energies above Cherenkov threshold, a sample of NCE proton-enriched events was obtained in the MiniBooNE experiment~\cite{miniboonenc}, which was used for the measurement of the $(\nu p\to\nu p)/(\nu N\to\nu N)$ ratio, which in turn is sensitive to $\Delta s$. A value of $\Delta s = 0.08 \pm 0.26$ was extracted (based on a Nuance prediction~\cite{CASPER2002161}), in agreement with the results from the BNL E734 experiment~\cite{PhysRevC.48.1919}.

The generally accepted value for the axial mass, extracted from CCQE experiments on deuterium bubble chambers, is  $M_A$ is $1.026 \pm 0.021$ GeV~\cite{Bernard:2001rs}. Recent MiniBooNE data~\cite{PhysRevD.81.092005} for CCQE scattering from nuclear targets have found it useful to employ $M_A = 1.35$~GeV, fitted by a RFG model to the $Q^2$ dependence of their observed yields. The allowed region for the axial vector mass using MiniBooNE NCE data was obtained in Ref.~\cite{miniboonenc}: $M_A = 1.39 \pm 0.11$~GeV. It is in agreement with the shape normalized fits of $\nu_\mu$ CCQE scattering on neutrons bound in carbon obtained by MiniBooNE experiment~\cite{PhysRevD.81.092005}. The value of the axial mass raised a strong debate in connection with the description of the MiniBooNE data. It has been suggested that the increased value of $M_A$ at MiniBooNE can be understood not as the axial mass obtained for free nucleons, but rather as a parameterization of nuclear effects neglected in the RFG model~\cite{Martini:2009uj, Benhar:2010nx}. On the other hand, the BNL E734 experiment reported the value $M_A = 1.06 \pm  0.05$~GeV~\cite{PhysRevD.35.785}, fitted to the $Q^2$ dependence of the measured $\nu(\bar{\nu}) p\rightarrow\nu(\bar{\nu}) p$ differential cross sections. A later reanalysis of that sample data obtained similarly small values of  $M_A$~\cite{PhysRevC.48.761}.

For MiniBooNE NCE interactions, the total charge on all photomultiplier tubes is proportional to the sum of the kinetic energies of all final-state nucleons that are produced in the interaction, which is referred to throughout this paper as $T$. It is important to understand that the nucleon kinetic energy measured in this way is different from the one determined from the track-based reconstruction used in the SciBooNE~\cite{Takei_phdthesis} and BNL E734~\cite{PhysRevD.35.785} experiments. In that case, the reconstructed proton track length is proportional to the kinetic energy of the most energetic proton produced in the event. Also the particle identification at MiniBooNE is based almost entirely on the properties of the measured Cherenkov ring (such as ring sharpness, charge, and time likelihoods), whereas the track-based experiments mostly use the particle's energy loss along the track.

\begin{figure}[tb]
\centering
\includegraphics[width=0.99\textwidth]{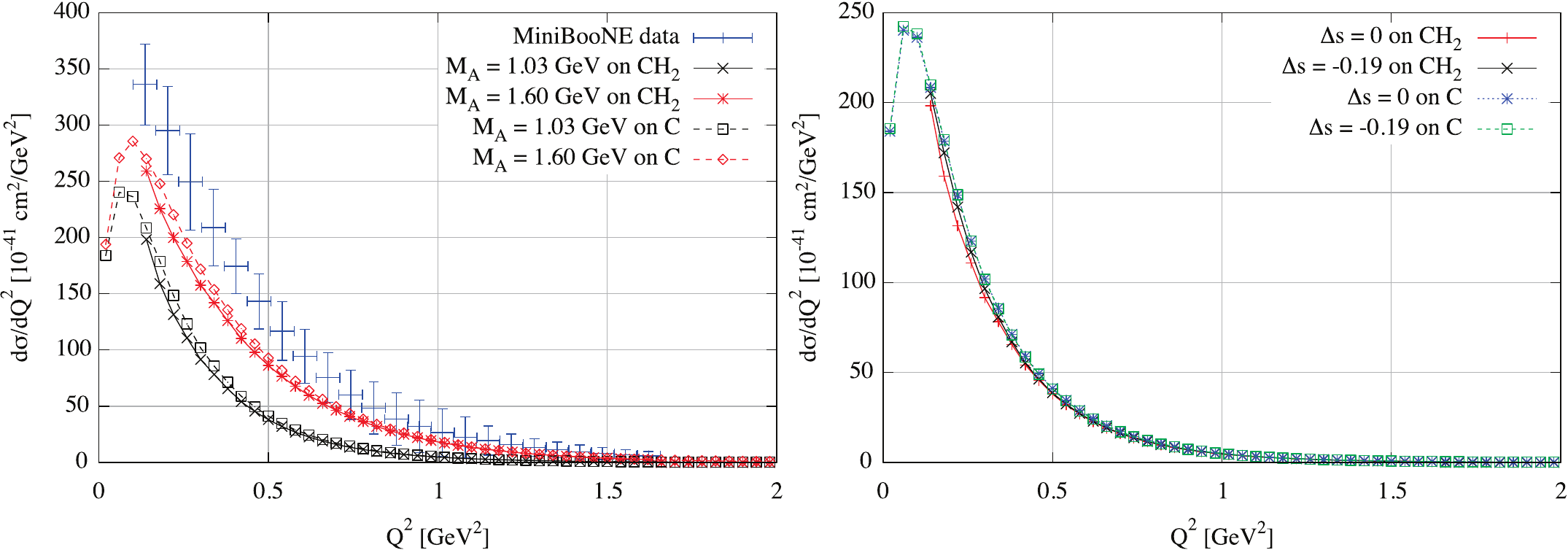}
\vspace{-2mm}
\caption{NCE flux averaged $Q^2$-distribution for different values of the axial mass (left panel) and for different values of $\Delta s$ (right panel). The solid lines correspond to the distribution defined by equation~(\ref{eq.miniboonenc}), while the dashed lines show the results obtained for a pure carbon target. The calculations are performed with the spectral function model~\cite{Benhar:2011wy}. The data points are taken from Ref.~\cite{miniboonenc}. The figures are taken from Ref.~\cite{Benhar:2011wy}. \label{fig:Benhar:2011wy}
}
\end{figure}

The $Q^2$-distribution, averaged over the MiniBooNE flux, are computed in Ref.~\cite{Benhar:2011wy} using the carbon spectral function of Ref.~\cite{BENHAR1994493}. In the left panel of figure~\ref{fig:Benhar:2011wy} the results obtained with different values of the axial mass and $\Delta s = 0$ are compared with the experimental MiniBooNE NCE data of Ref.~\cite{miniboonenc}. In order to compare with MiniBooNE data on CH$_2$, the following differential cross section per nucleon are evaluated:
\beq
\frac{d\sigma}{dQ^2} =
 \frac{1}{7}C_{\nu p,H}\frac{d\sigma_{\nu p,H}}{dQ^2}
+\frac{3}{7}C_{\nu n,C}\frac{d\sigma_{\nu n,C}}{dQ^2}
+\frac{3}{7}C_{\nu n,C}\frac{d\sigma_{\nu n,C}}{dQ^2}.\label{eq.miniboonenc}
\eeq
The $Q^2$-distribution  is the sum of three contributions: scattering on free protons, bound protons in carbon, and bound neutrons in carbon, each of them weighted by an efficiency correction function $C_i$ and averaged over the experimental neutrino flux~\cite{miniboonenc}. It clearly appears in figure~\ref{fig:Benhar:2011wy} that the value of the axial mass yielding a good fit of the MiniBooNE CCQE distribution, $M_A = 1.6$~GeV~\cite{Benhar:2010nx}, does not reproduce NCE data. The results obtained for different values of $\Delta s$ and $M_A = 1.03$ GeV,  in the right panel of figure~\ref{fig:Benhar:2011wy}, show that, due to a cancellation between the proton and neutron contributions, the strange quark content of the nucleon to the cross section of nuclei with equal number of protons and neutrons is vanishingly small. The results discussed in  Refs.~\cite{Benhar:2010nx,Benhar:2011wy}, showing that it is impossible to describe both the CCQE and NCE data sets using the same value of the axial mass, confirm that nuclear effects not included in the oversimplified RFG model cannot be taken into account through a modification of $M_A$.

\begin{figure}[tb]
\centering
\includegraphics[width=0.49\textwidth]{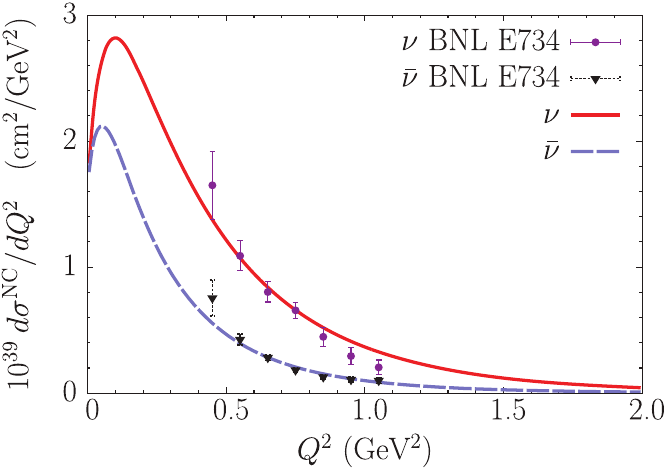}\hfill\includegraphics[width=0.49\textwidth]{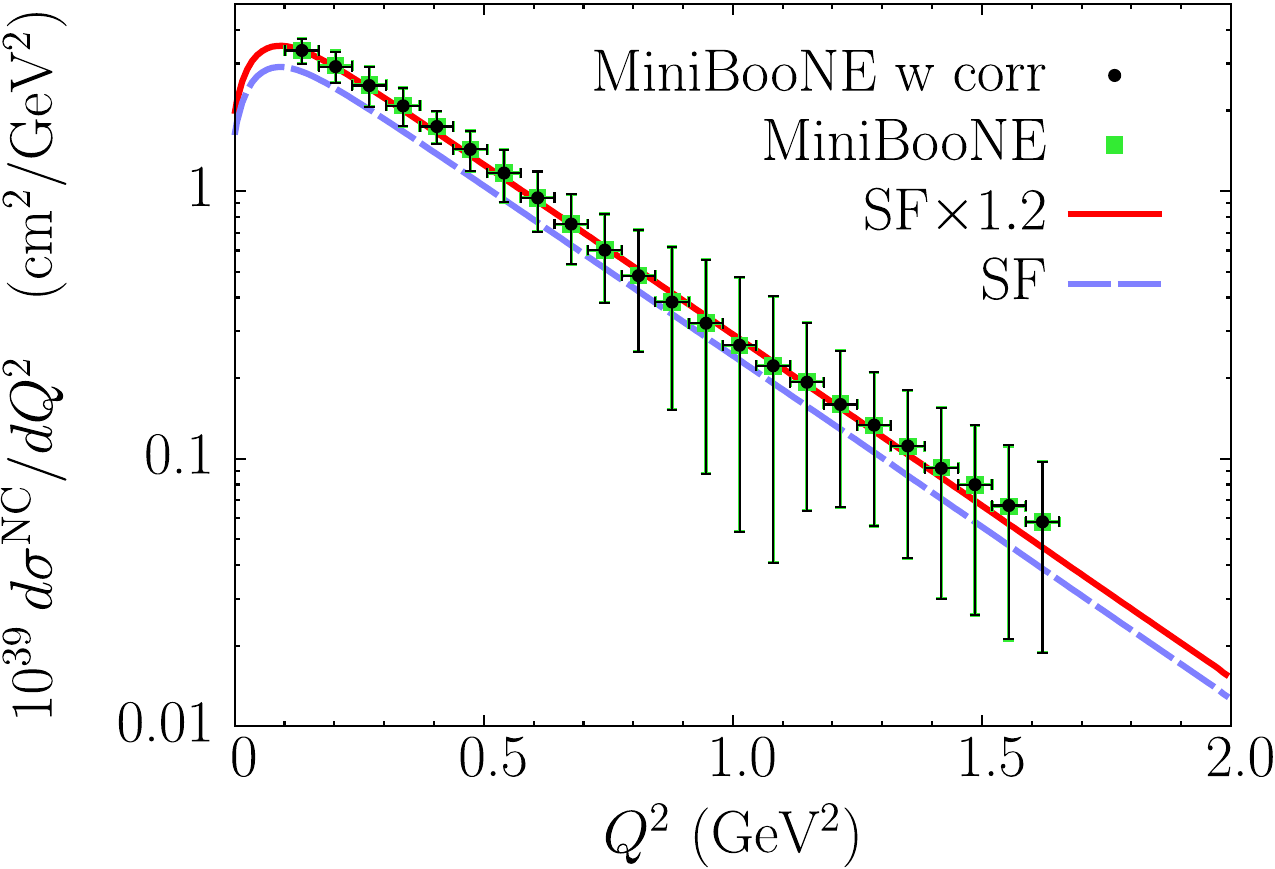}
\vspace{-2mm}
\caption{Left panel: differential cross section  $d\sigma^{NC}/dQ^2$ for NCE $\nu$ and $\bar{\nu}$ scattering in the BNL E734 experiment~\cite{PhysRevLett.56.1107, PhysRevD.35.785}. The error bars do not include the normalization uncertainty of 11.2\% (10.4\%) in the $\nu$ ($\bar{\nu}$) case. Right panel: differential cross section $d\sigma^{NC}/dQ^2$ for NCE neutrino scattering off CH$_2$ averaged over the MiniBooNE flux. The spectral function calculations (dashed line) are compared to the data~\cite{miniboonenc} with (circles) and without (squares) the efficiency correction to the data (see~\cite{Ankowski:2012ei}). The error bars do not account for the normalization uncertainty of 18.1\%. Taken from Ref.~\cite{Ankowski:2012ei}. \label{fig:Ank2012}}
\end{figure}

In Ref.~\cite{Ankowski:2012ei} the differential and total cross sections for energies ranging from a few hundreds of MeV to $100$~GeV have been obtained and compared with the data from the BNL E734, MiniBooNE, and NOMAD experiments (see Ref.~\cite{Lyubushkin:2009}). The spectral function approach is applied to describe nuclear effects in $\nu(\nubar)$ scattering off carbon nucleus, treating in a consistent manner NCE and CCQE interactions. In the calculations an effective value of $M_A=1.23$~GeV has been used, as determined from the shape of the $Q^2$ distribution of CCQE $\nu_{\mu}$ events by the MiniBooNE Collaboration~\cite{PhysRevLett.100.032301}. The SF approach provides a fairly good description of the NCE neutrino and antineutrino differential cross sections $d\sigma^{NC}/dQ^2$ measured with the BNL E734 experiment (left panel of figure~\ref{fig:Ank2012}) and a very accurate description of the shape of the NCE neutrino differential cross section $d\sigma^{NC}/dQ^2$ obtained from the MiniBooNE experiment (right panel of figure~\ref{fig:Ank2012}). However, the absolute value of the calculated cross section $d\sigma^{NC}/dQ^2$ underestimates the MiniBooNE data by 20\%. The same discrepancy is observed for the CCQE $\nu_\mu$ interaction in comparison with the differential cross section $d\sigma^{CC}/dQ^2$ and the flux-unfolded total cross section reported by the MiniBooNE Collaboration. The conclusion of this work~\cite{Ankowski:2012ei} is that nuclear effects in NCE and CCQE scattering seem to be very similar, though, according to  the author, the discrepancy between the results from the MiniBooNE and NOMAD experiments seems more likely to be ascribable to underestimated flux uncertainty in the MiniBooNE data analysis.

In theoretical neutrino physics, the differential cross sections are often considered with respect to the four-momentum squared $Q^2$,
\begin{equation}
Q^2=-q^2 = |\nq|^2 - \omega^2=(\nk-\nk^{\prime})^2-
(E-E^{\prime})^2 .\label{eq:Q2lept}
\end{equation}
However, invariant $Q^2$ cannot be calculated from measurable quantities and therefore in data analysis the reconstructed $Q^2_{QE}$ (assuming that the target nucleon is at rest) is defined as
\begin{equation}
Q^2_{QE} = 2 m T, \label{eq:recQ2NC}
\end{equation}
where $T$ is the sum of the kinetic energies of all final state nucleons that are produced in the interaction $ T = \sum\limits_i T_i$.

\begin{figure}[tb]
\centering
\includegraphics[width=0.99\textwidth]{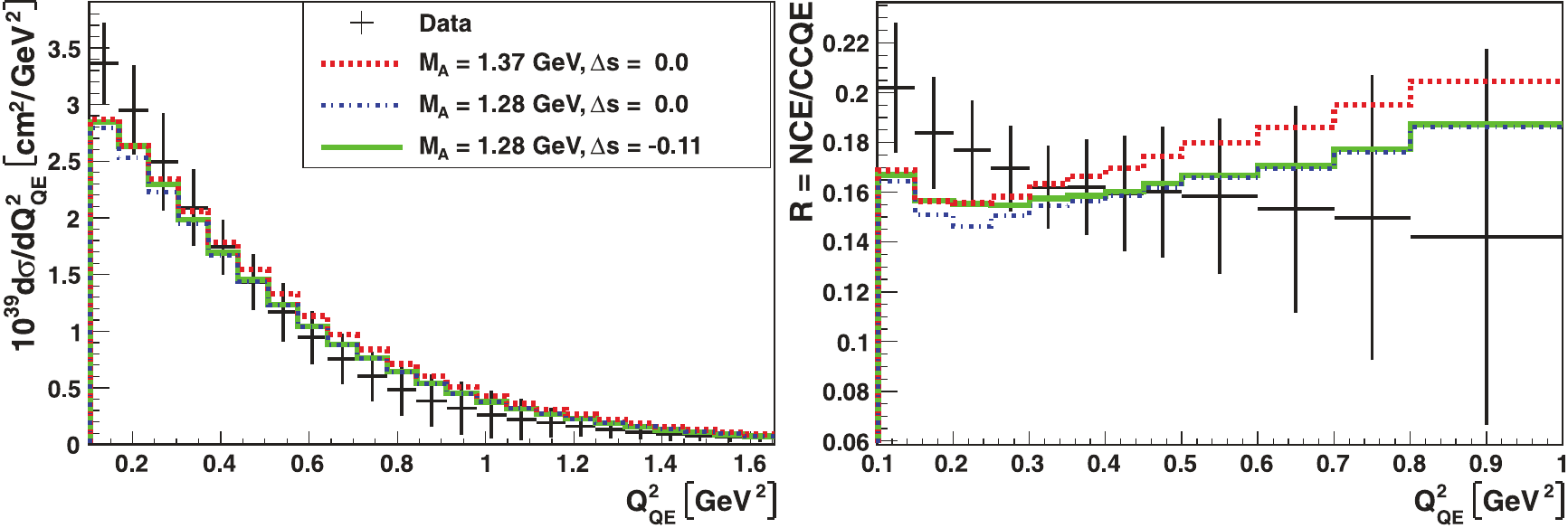}
\vspace{-2mm}
\caption{Left panel: Flux-averaged $d\sigma/dQ^2_{QE}$ cross section  for neutrino scattering on CH$_2$ as a function of $Q^2_{QE}$. Right panel: NCE/CCQE cross section ratio as a function of $Q^2_{QE}$. Calculations are performed with the RDWIA~\cite{Butkevich:2011fu}. The MiniBooNE data are from~\cite{miniboonenc}. Taken from Ref.~\cite{Butkevich:2011fu}.
\label{fig:Butkevich:2011fua}}
\end{figure}

The NCE $\nu$ and $\nubar$ scattering on carbon and CH$_2$ targets are studied in Ref.~\cite{Butkevich:2011fu} within the framework of the RDWIA using a relativistic optical potential. Using the RDWIA with the Booster Neutrino Beamline flux~\cite{PhysRevD.79.072002}, the axial mass is extracted from a ``shape-only'' fit of the measured flux-averaged $d\sigma/dQ^2_{QE}$ differential cross section. The extracted value of $M_A = 1.28 \pm 0.05$ GeV is in agreement within errors with the MiniBooNE result of $M_A = 1.39 \pm 0.11$~GeV. It can be seen in figure~\ref{fig:Butkevich:2011fua} that there is a good overall agreement within errors between the RDWIA predictions and the MiniBooNE data of the NCE flux-averaged differential cross section $d\sigma /dQ_{QE}^2$ on CH$_2$ and of the NCE/CCQE cross section ratio in the range of $0.25 < Q_{QE}^2 < 1.65$~GeV$^2$. For low values of $Q_{QE}^2$, $Q_{QE}^2 < 0.25$~GeV$^2$, the calculations underestimate the data by at most 17\%.

\begin{figure}[bt]
\centering
\includegraphics[width=0.7\textwidth]{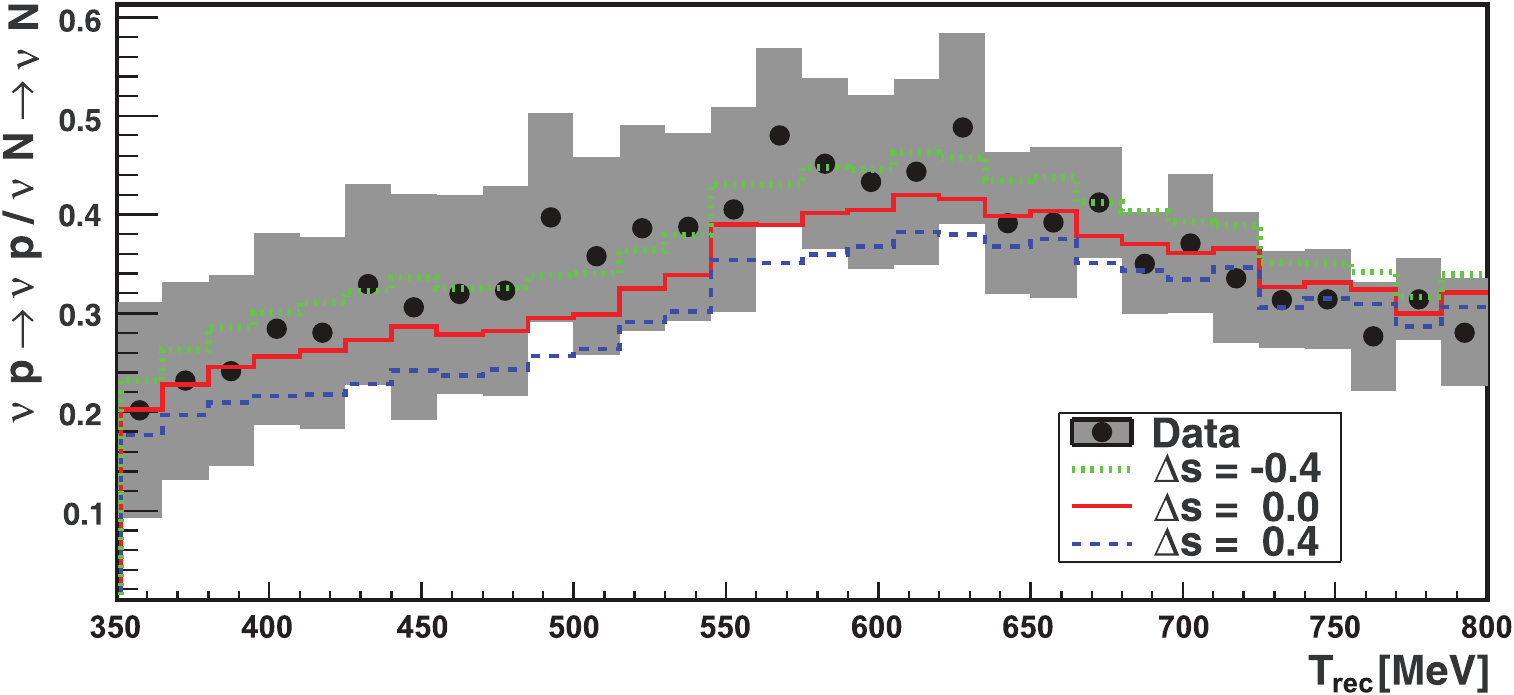}
\vspace{-2mm}
\caption{(MiniBooNE $(\nu p \rightarrow \nu p)/(\nu N \rightarrow \nu N)$ ratio as a function of $T_{rec}$. The predictions for $M_A = 1.28$~GeV and $\Delta s = 0.0$, $0.4$, and $-0.4$ are shown. The MiniBooNE data are from~\cite{miniboonenc}. Taken from Ref.~\cite{Butkevich:2011fu}. \label{fig:Butkevich:2011fub}}
\end{figure}

In addition to the $\nu N \rightarrow \nu N$ differential cross section, the MiniBooNE Collaboration has published the $(\nu p \rightarrow \nu p) / (\nu N \rightarrow \nu N)$ ratio at $Q^2 > 0.7$~GeV$^2$ (above the Cherenkov threshold for protons in mineral oil)~\cite{PhysRevD.81.092005}. The ratio was reported as a function of the MiniBooNE reconstructed nucleon kinetic energy $T_{rec}$. Also the migration matrices were published in Ref.~\cite{NCEMiniBooNE}, and these carry the detector resolution and efficiency information. Using them, one can smear the predicted cross sections and obtain the predicted event rates in the MiniBooNE detector as a function of $T_{rec}$. The procedure for carrying out calculations of event rates in terms of the MiniBooNE reconstructed energy is described in Appendix B of Ref.~\cite{Perevalov_phdthesis}. The comparison of the RDWIA calculations with the MiniBooNE data is presented in Ref.~\cite{Butkevich:2011fu}. The predictions of the event rates for different values of $\Delta s$ covering the range from $-0.4$ to $0.4$ are shown in figure~\ref{fig:Butkevich:2011fub}. The $\chi^2$ distribution between the data and the RDWIA results is obtained using the full error matrix for the ratio published in Ref.~\cite{NCEMiniBooNE}. By using the MiniBooNE data for the high-energy $\nu p \rightarrow \nu p$ to $\nu N \rightarrow \nu N$ ratio, the value of $\Delta s = -0.11 \pm 0.36 $ (with $\chi^2_{min} = 33.4$ for $29$ degrees of freedom) has been extracted; this value is consistent with other measurements of $\Delta s$.

\begin{figure}[ht]
\centering
\includegraphics[width=0.49\textwidth]{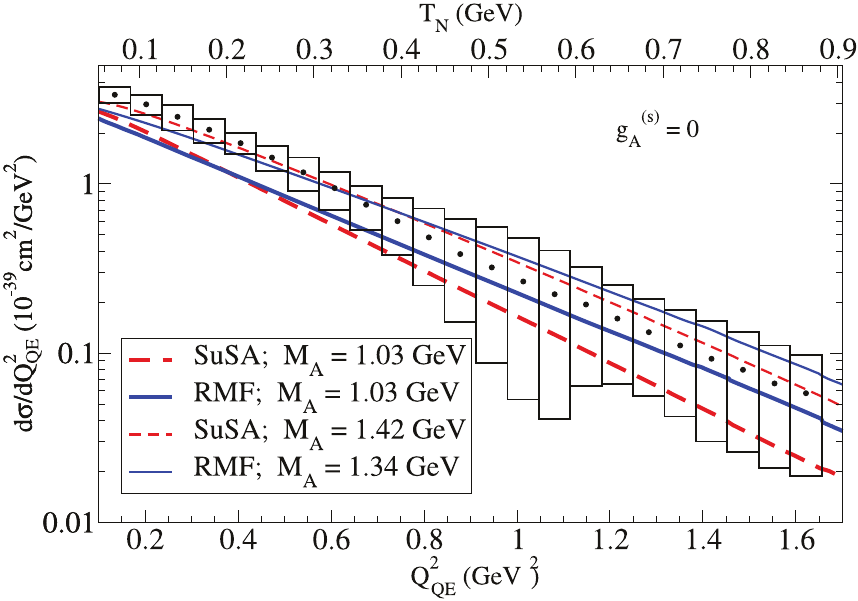}\hfill
\includegraphics[width=0.49\textwidth]{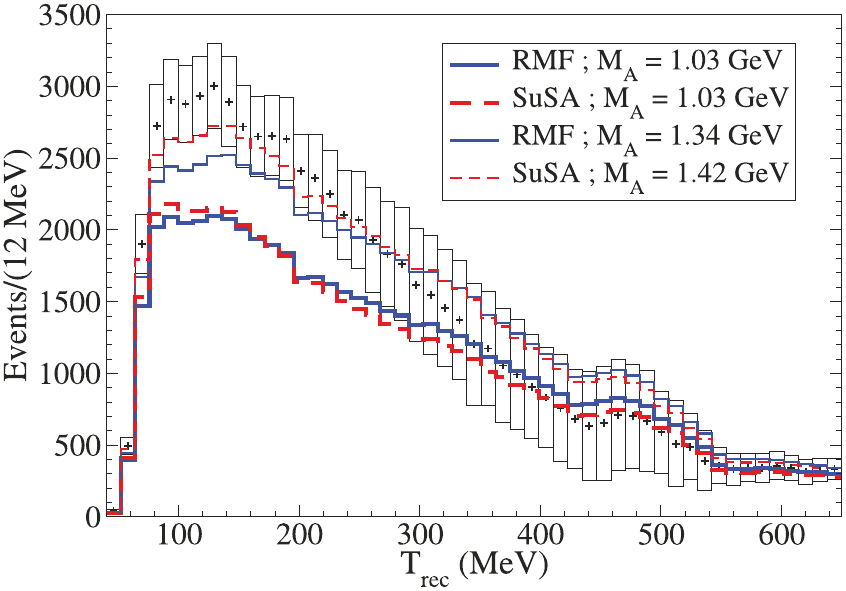}
\vspace{-2mm}
\caption{NCE flux-averaged cross section computed within the RMF (solid blue lines) and SuSA (dashed red lines) models, compared with MiniBooNE data~\cite{miniboonenc} as a function of true energy (left panel) and of the reconstructed energy (right panel), for different values of $M_A$. Taken from Ref.~\cite{GONZALEZJIMENEZ20131471}.}
\label{fig:Rauletal_2013_RMF_SuSA}
\end{figure}

The predictions of the two relativistic SuSA and RMF models are compared in Ref.~\cite{GONZALEZJIMENEZ20131471} with the MiniBooNE NCE neutrino cross sections. If the calculations are performed with the standard value of the axial mass, $M_A = 1.03$ GeV, both models underpredict the experimental cross section for both CCQE an NCE scattering. In Ref.~\cite{GONZALEZJIMENEZ20131471} the axial mass is used as an effective parameter to incorporate nuclear effects not included in the models (such as, for instance,  multi-nucleon knockout). The dependence of the cross section upon $M_A$ at strangeness $\Delta s = g_A^{(s)}=0$ is illustrated in figure~\ref{fig:Rauletal_2013_RMF_SuSA}, where the results obtained with the standard value are compared with the ones obtained with the value that provides the best fit to the cross section within either SuSA or RMF models. The $g_A^{(s)}$-dependence of the NCE neutrino-nucleon cross section is very mild, this is the reason to fix the axial strangeness to zero in the previous analysis. The 1-$\sigma$ allowed regions of the axial mass for the two models are
\begin{eqnarray}
\label{eq:MA_RMF}
M_A&=& 1.34\pm 0.06~\textrm{GeV \ \ for RMF}
\\
\label{eq:MA_SuSA}
M_A&=& 1.42\pm 0.06~\textrm{GeV \ \ for SuSA},
\end{eqnarray}
corresponding to $\chi^2/\mathrm{DOF}=16.5/22$ and $\chi^2/\mathrm{DOF}=4.7/22$, respectively. These have to be compared with $\chi^2/\mathrm{DOF}=46.2/22$ (RMF) and $\chi^2/\mathrm{DOF}=45.3/22$ (SuSA) for $M_A=1.03$~GeV.

In figure~\ref{fig:Rauletal_2013_RMF_SuSA} the RMF and SuSA results are compared with the MiniBooNE data as functions of the true (left panel) and reconstructed (right panel) energies. Whenever a physical quantity is measured, there are distortions to the original distribution in the observed quantity. Experimentalists correct the data distribution using unfolding techniques. There is an alternative method, which is to report them in the reconstructed nucleon energy without applying the unfolding procedure (and corresponding errors). To produce the reconstructed energy results in Ref.~\cite{GONZALEZJIMENEZ20131471} the folding procedure detailed in Appendix~B of Ref.~\cite{Perevalov_phdthesis} is used. Both models give a reasonably good representation of the data when the non-standard value of the axial mass is used. Moreover, it is important to note that the SuSA cross section reproduces quite well the slope of the experimental data, better than the RMF one, which has a smaller $Q^2$ slope and falls slightly below the error bars for lowest $Q^2$ data. It is, however, important to observe that neither of the two models is expected to describe correctly the low-$Q^2$ region, where collective effects play a dominant role. The values of the axial mass obtained with both models are compatible, within 1-$\sigma$, with the value $M_A=1.35$~GeV employed by the MiniBooNE collaboration to fit their RFG model to the CCQE data.

Following the procedure to convert the NCE `true energy' cross section into the NCE reconstructed energy distributions~\cite{Perevalov_phdthesis}, the corresponding theoretical ratio is calculated within the SuSA and RMF models. As expected, the ratio, unlike the cross section, is sensitive to axial strangeness. $\chi^2$-fit to the axial strangeness parameter is performed (the axial mass is fixed to the value given in equations~(\ref{eq:MA_RMF}) and~(\ref{eq:MA_SuSA})). The 1-$\sigma$ allowed regions turn out to be
\begin{eqnarray}
\label{eq:gas_RMF}
g_A^{(s)} &=& +0.04\pm0.28 \textrm{ \ \ for RMF}
\\
\label{eq:gas_SuSA}
g_A^{(s)} &=& -0.06\pm0.31 \textrm{ \ \ for SuSA},
\end{eqnarray}
corresponding to $\chi^2/\mathrm{DOF} = 33.6/29$ and $\chi^2/\mathrm{DOF} = 31.3/29$, respectively.

\begin{figure}[th]\centering
\includegraphics[width=0.49\textwidth]{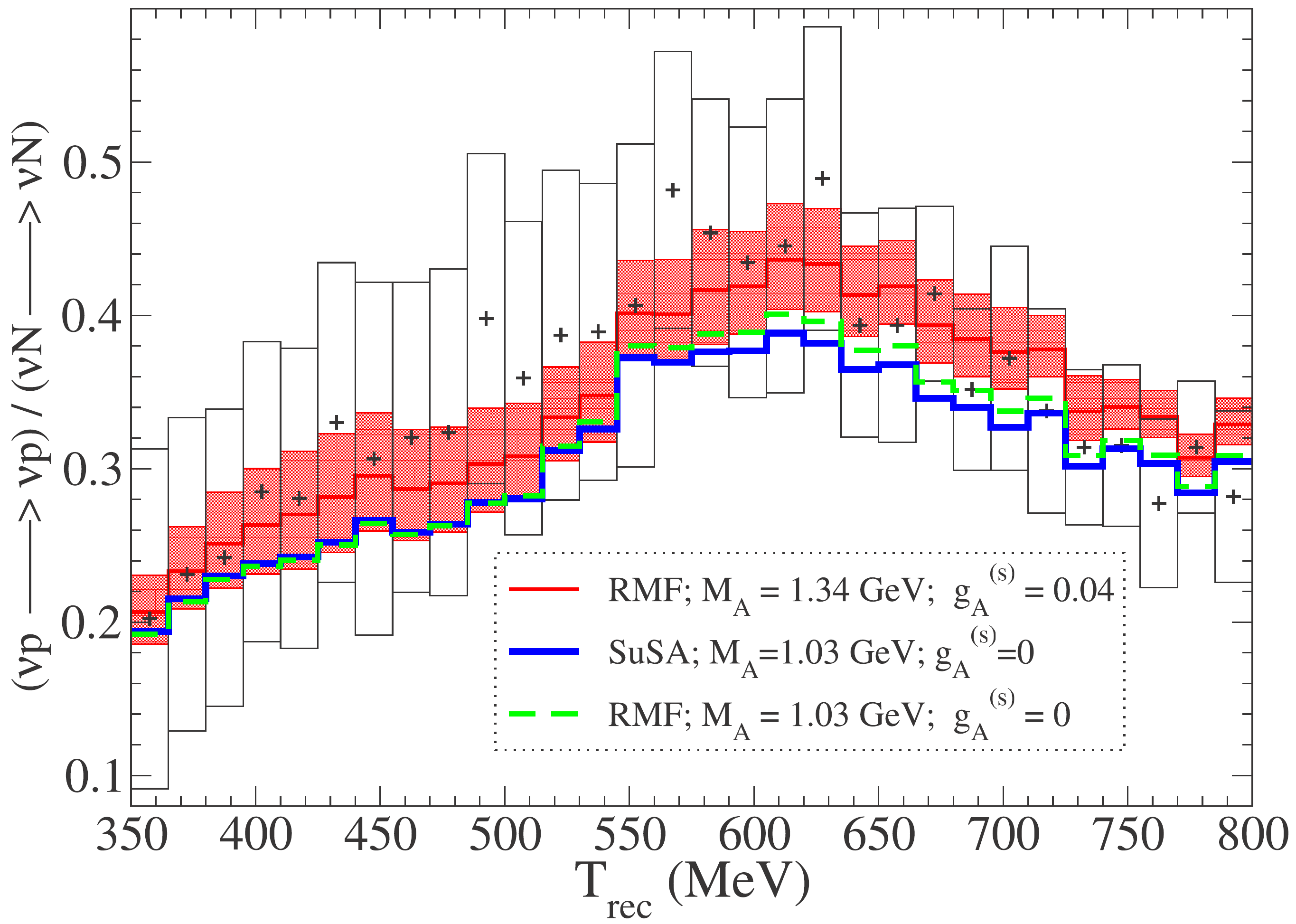}\hfill
\includegraphics[width=0.49\textwidth]{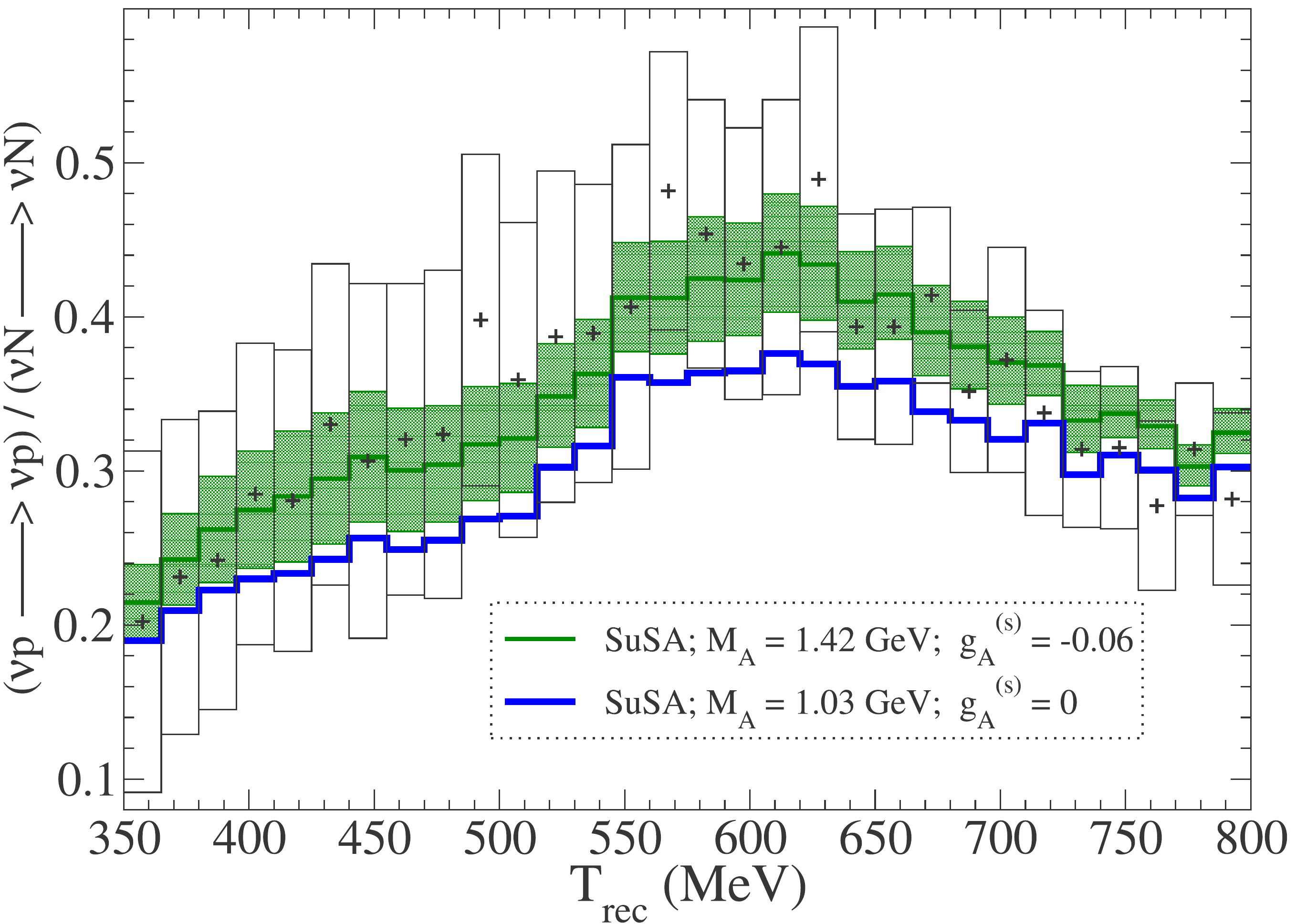}
\vspace{-2mm}
\caption{Ratio $(\nu p \to \nu p)/(\nu N \to \nu N)$ computed within the RMF and SuSA models. Shadowed areas represent the 1-$\sigma$ region allowed for $g_A^{(s)}$ (see text). The ratio computed with the best value of $g_A^{(s)}$ is presented, as well as those obtained with the standard axial mass and no strangeness. Data from Ref.~\cite{miniboonenc}. Taken from Ref.~\cite{GONZALEZJIMENEZ20131471}.
}
\label{fig:Rauletal_2013_ratio_RMF_SuSA}
\end{figure}

The ratio computed by using the above sets of parameters, as well as with the standard axial mass and no strangeness, are presented in figure~\ref{fig:Rauletal_2013_ratio_RMF_SuSA}. From the comparison between the two curves in the left panel, having standard parameters $M_A=1.03$~GeV and $g_A^{(s)}=0$ (green and blue lines), it appears that the dependence upon the nuclear model is essentially canceled in the ratio, confirming that this is a good observable for determining the axial strangeness content of the nucleon. Within the error bars, the obtained values of $g_A^{(s)}$ are compatible with the ones of the previous analysis. Of course, before drawing definitive conclusions on the allowed value of $g_A^{(s)}$, an extended analysis of the nuclear effects that are being effectively incorporated in the increased value of $M_A$ should be performed. However, it is worth mentioning that the ratio shown in figure~\ref{fig:Rauletal_2013_ratio_RMF_SuSA} shows little sensitivity to a possible $\mathrm{np}$ charge-exchange due to FSI. For instance, a 20\% of charge-exchange would not affect the results displayed in figure~\ref{fig:Rauletal_2013_ratio_RMF_SuSA} by more than a few percent, for any reasonable value of $g_A^{(s)}$.

\begin{figure}[tb]
\centering
\includegraphics[width=0.49\textwidth]{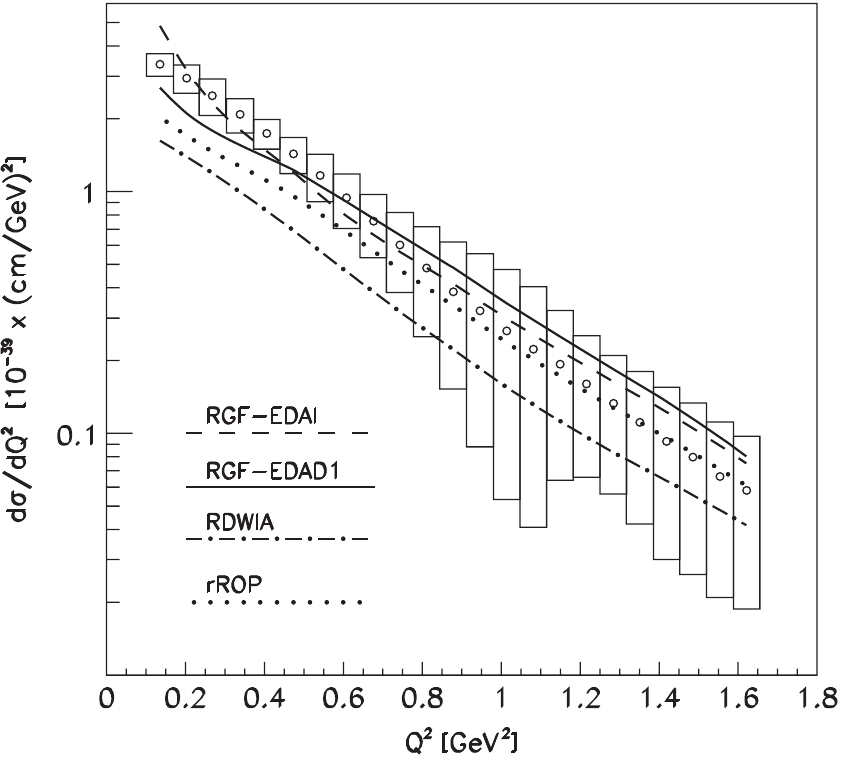}
\hfill
\includegraphics[width=0.49\textwidth]{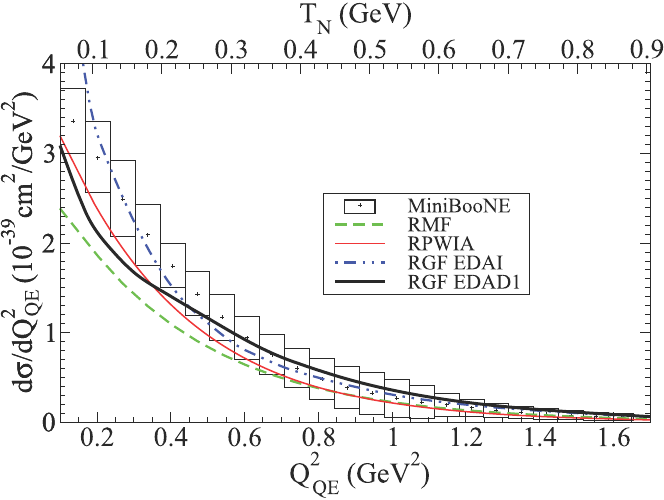}
\vspace{-2mm}
\caption{Left panel: NCE flux-averaged ($\nu N \rightarrow \nu N$) cross section as a function of $Q^2$, calculated with the RGF-EDAD1 (solid line) and RGF-EDAI (dashed line). The dotted and dot-dashed lines are rROP and RDWIA results, calculated with the EDAI potential, respectively. Taken from Ref.~\cite{Meucci:2011nc}. Right panel: NCE flux-averaged ($\nu N \rightarrow \nu N$) cross section as a function of $Q^2$ calculated in the RPWIA (thin solid lines), RMF (dashed lines), RGF-EDAD1 (thick solid lines), and RGF-EDAI (dash-dotted lines). The vector and axial-vector strange form factors have been fixed to zero. The data are from MiniBooNE~\cite{PhysRevD.81.092005}. Taken from Ref.~\cite{PhysRevC.88.025502}.
\label{fig:Meucci2011}}
\end{figure}

\begin{figure}[tb]
\centering
\includegraphics[width=0.49\textwidth]{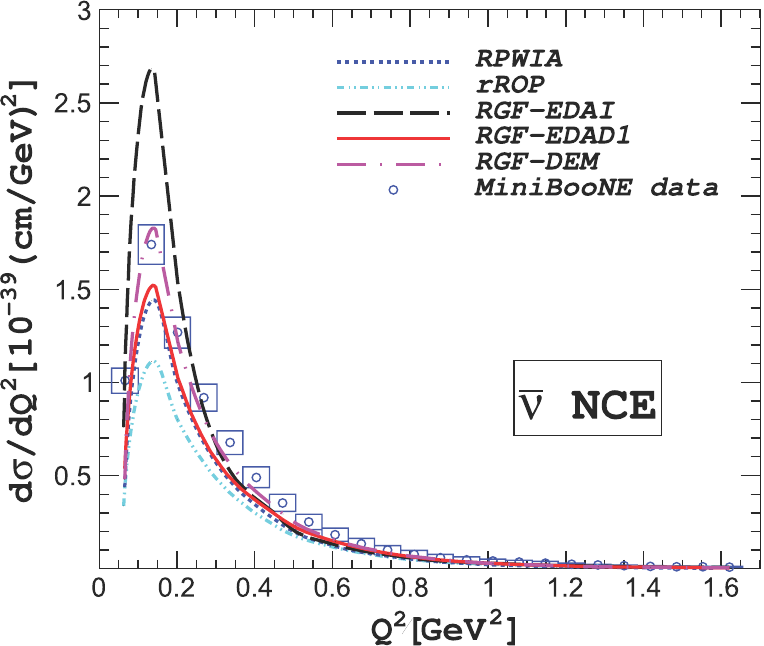}
\hfill
\includegraphics[width=0.49\textwidth]{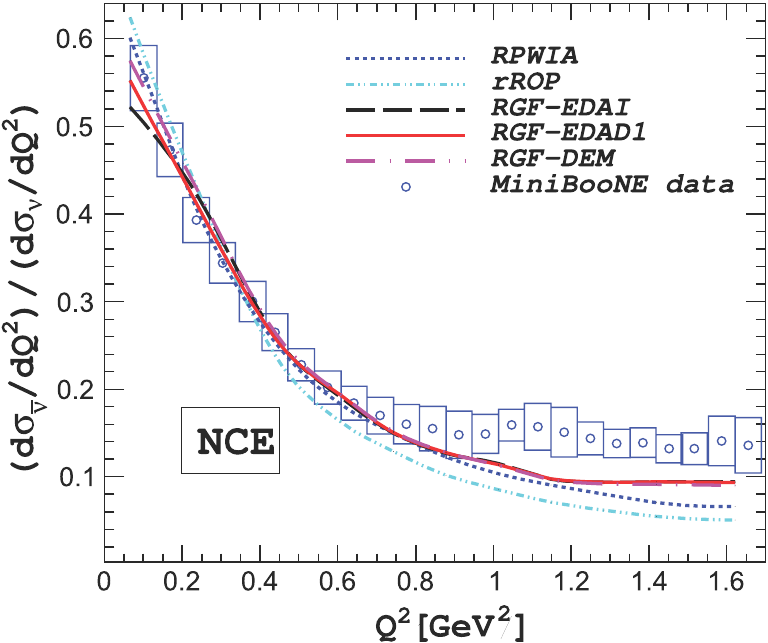}
\vspace{-2mm}
\caption{Left panel: NCE flux-averaged ($\bar{\nu} N \rightarrow \bar{\nu} N$) cross section as a function of $Q^2$. Right panel: Ratio of the $\bar{\nu}$ to $\nu$ NCE scattering cross section with total error. The data are from~\cite{miniboone-ncqe-nubar}. Taken from Ref.~\cite{Meucci:2014pka}.
\label{fig:Meucci2014}}
\end{figure}

The sensitivity to FSI of NCE $\nu(\nubar)$ scattering cross sections are investigated in~\cite{Meucci:2011nc, PhysRevC.88.025502, Meucci:2014pka}, where the RGF cross sections are compared to the MiniBooNE data. As already mentioned in section~\ref{sec:fsi}, we note that the RGF is appropriate for an inclusive process like the CCQE scattering where only the final lepton is detected, whereas NCE scattering is a semi-inclusive process where only the final nucleon can be detected. As a consequence, the RGF may include channels that are not present in the experimental NCE measurements, but it can also recover important contributions which are not taken into account by other models based on the IA. From this point of view, the RDWIA can represent a lower limit and the RGF an upper  limit to the semi-inclusive NCE cross sections. In figure~\ref{fig:Meucci2011} the RDWIA (left panel) and RMF (right panel) results generally underpredict the experimental MiniBooNE cross section for neutrino scattering, while the RGF results obtained with different parameterizations of the phenomenological relativistic optical potential are in reasonable agreement with the data using the standard value of the axial mass, \emph{i.e.}, $M_A = 1.03$ GeV.

The comparison of the theoretical predictions of the RGF model with the MiniBooNE antineutrino data has been presented in Ref.~\cite{Meucci:2014pka}. The results obtained with different parameterizations of the relativistic optical potential are shown in figure~\ref{fig:Meucci2014} together with the corresponding RPWIA and rROP ones. The RGF results in figures~\ref{fig:Meucci2011} and~\ref{fig:Meucci2014} describe well the NCE flux-averaged MiniBooNE cross sections, for both $\nu$ and $\nubar$ scattering, without the need to increase the standard value of the axial mass. Visible differences are produced, at low values of $Q^2$, by the use of different parameterizations of the optical potential. The sensitivity to the parameterization of the optical potential and the effects of FSI are strongly reduced in the ratio of the $\nubar$ to $\nu$ NCE scattering cross sections. The results are shown in the left panel of figure~\ref{fig:Meucci2014} in comparison with the experimental ratio measured by the MiniBooNE Collaboration. All the RGF calculations  give very close and practically equivalent results, small differences can be seen only at very low $Q^2$. The experimental ratio is reasonably described when $Q^2 \leq 1$~(GeV$/c)^2$ and it is slightly understimanted when $Q^2 \geq 1$~(GeV$/c)^2$. This is because the antineutrino cross section is underestimated for  large $Q^2$, whereas the neutrino cross section is within the error bars in the entire range of $Q^2$.

It is not easy to disentangle the role of specific contributions which may be neglected in the RDWIA or spuriously added in the RGF, in particular if we consider that both RDWIA and RGF calculations make use of phenomenological optical potentials obtained through a fit to elastic proton-nucleus scattering data. In order to clarify the content of the enhancement of the RGF cross sections compared to those of the IA models, a careful evaluation of all nuclear effects and of the relevance of multinucleon emission and of some non-nucleonic contributions~\cite{PhysRevC.81.064614} is required. At present, lacking a phenomenological OP that exactly fulfills the dispersion relations in the whole energy region of interest, the RGF predictions are not univocally determined from the elastic phenomenology. Different phenomenological potentials, with different imaginary parts, are able to give equivalently good descriptions of elastic proton-nucleus scattering data, but may give differences and theoretical uncertainties in the numerical predictions of the RGF model. A better determination of a relativistic OP, which closely fulfills the dispersion relations, deserves further investigation. A step forward in this direction has been done in Ref.~\cite{Ivanov:2016pon}, where a new microscopic global relativistic folding optical potential (GRFOP) has been obtained for $^{12}$C and tested within the RGF model for QE electron and $\nu(\nubar)$ scattering. The use of the microscopic  GRFOP reduces the theoretical uncertainties in the predictions of the RGF model and confirms the reasonably good agreement in comparison with electron and (anti)neutrino scattering data. The GRFOP gives results generally between the EDAI and EDAD1 ones and in many cases in better agreement with the experimental data.

\begin{figure}[tb]
\centering
\includegraphics[width=0.49\textwidth]{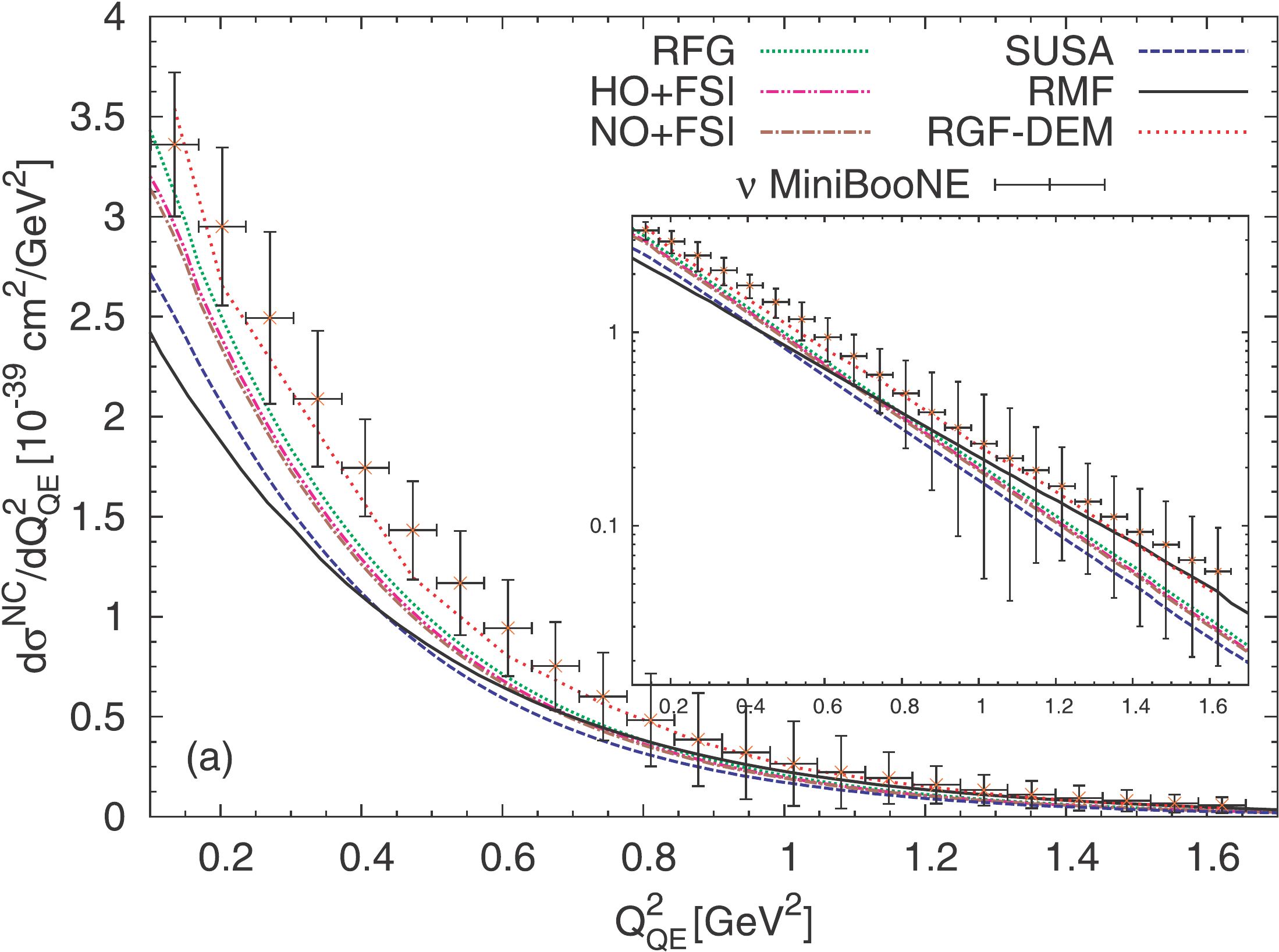}
\hfill
\includegraphics[width=0.49\textwidth]{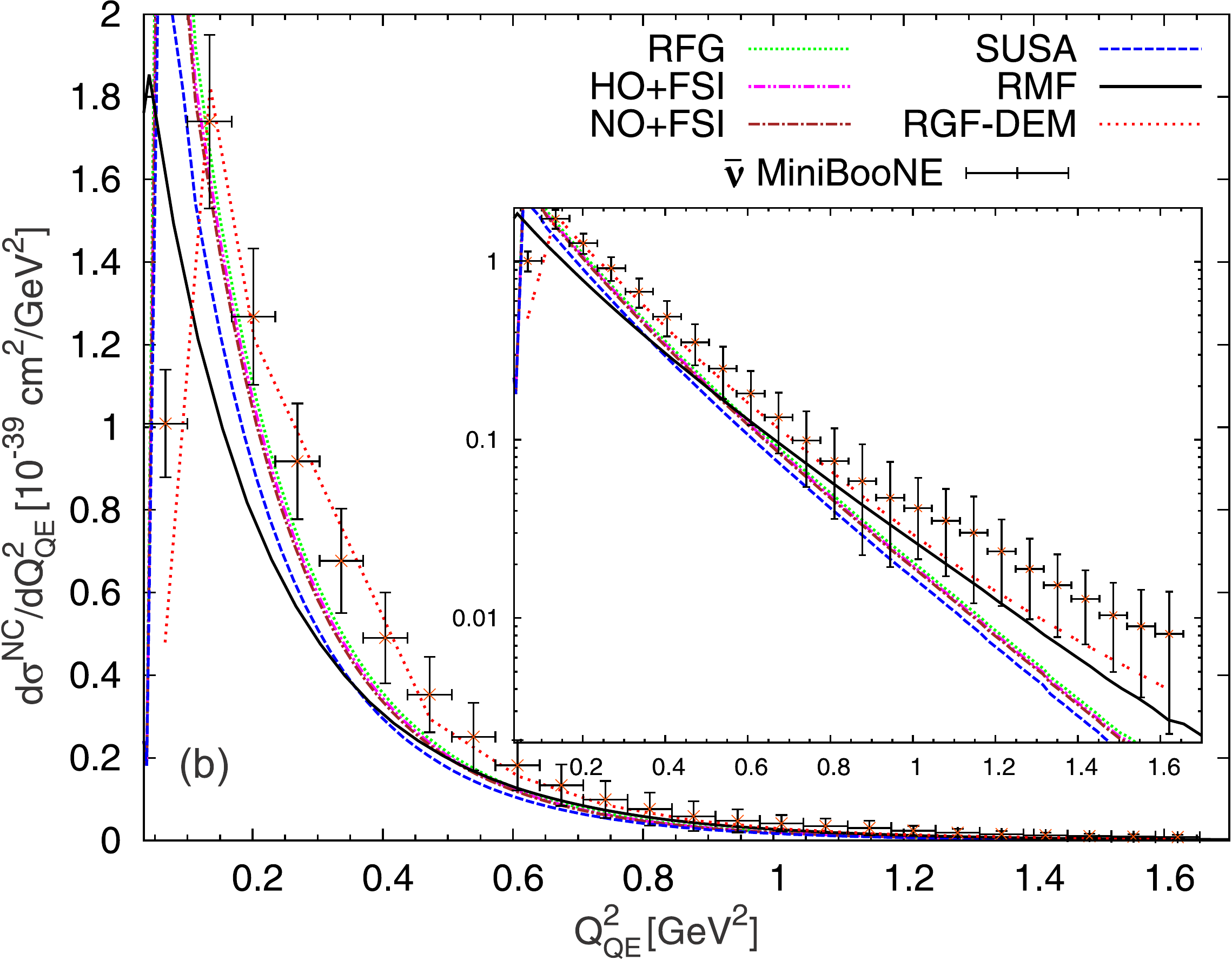}
\vspace{-2mm}
\caption{NCE neutrino [panel (a), $\nu N\to\nu N$] and antineutrino [panel (b),  $\overline{\nu}N\to\overline{\nu}N$] flux-averaged differential cross section computed using the RFG, HO+FSI, NO+FSI, SuSA scaling functions, RGF and RMF models and compared with the MiniBooNE data~\cite{miniboonenc, miniboone-ncqe-nubar}. The results have been obtained using  the world-average axial mass $M_A = 1.032$~GeV and strangeness $\Delta s = 0$. The error bars do not account for the normalization uncertainty of 18.1\% (19.5\%) in the $\nu$($\overline{\nu}$) case. Taken from Ref.~\cite{Ivanov:2015wpa}.}
\label{fig:Martin2015etal_dsdq2}
\end{figure}

The main objective of work of Ref.~\cite{Ivanov:2015wpa} centers on the use of a realistic spectral function, which accounts for short-range \emph{NN} correlations and also has a realistic energy dependence (see section~\ref{sec:models} for more details).  This function gives a scaling function in accordance with electron scattering data and it can be used for a wide range of neutrino energies. Therefore, the use of this spectral function to describe the general reaction mechanism involved in NC neutrino-nucleus scattering processes can provide very valuable information that can be compared with the results obtained with other theoretical approaches. The comparison between the results of different  models (RFG, HO + FSI, NO + FSI, and SuSA scaling functions, RMF, and RGF) and the experimental NCE flux-averaged MiniBooNE $\nu(\nubar)$ cross sections is presented in figure~\ref{fig:Martin2015etal_dsdq2}. All models except RGF underestimate the neutrino data in the region $0.1 < Q^2 < 0.7$~GeV$^2$ and are within the error bars for higher $Q^2$. On the other hand, all models underestimate the antineutrino data at high $Q^2$. This is clearly shown in the insets in figure~\ref{fig:Martin2015etal_dsdq2}, where the cross sections are represented on a logarithmic scale. The RGF-DEM results are larger than the results of the other models and in generally good agreement with the data over the entire $Q^2$ region considered in the figure. The enhancement of the RGF cross sections is due to the contribution of final-state channels that are recovered by the imaginary part of the optical potential and that are not included in the other models.

\begin{figure}[tb]
\centering
\includegraphics[width=0.99\textwidth]{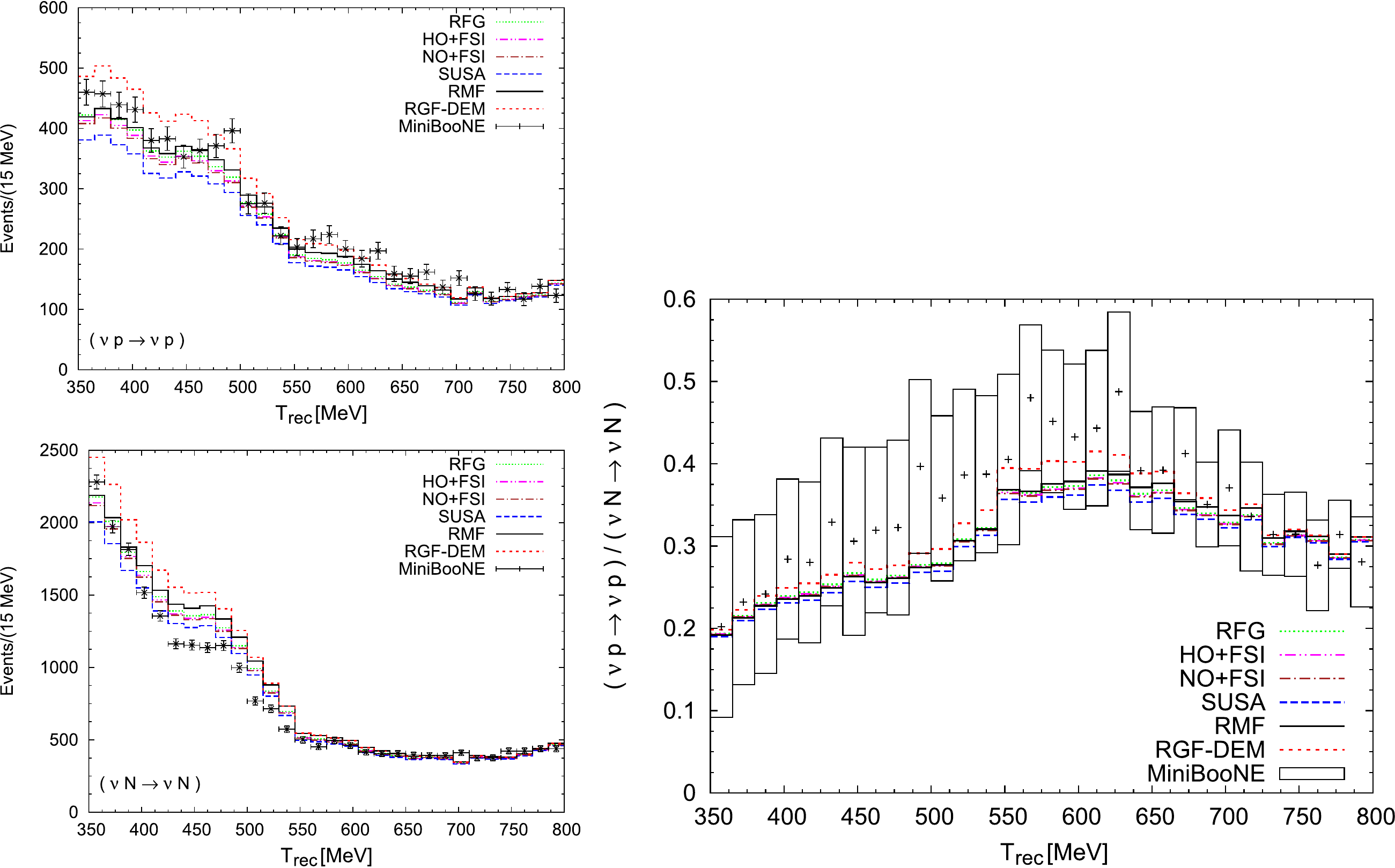}
\vspace{-2mm}
\caption{RFG, HO+FSI, NO+FSI, SUSA, RGF, and RMF predictions, after the folding procedure, compared with the histograms of the numerator (top-left panel) and denominator (bottom-left panel) entering the ratio between $\nu$ scattering from proton and nucleon (proton plus neutron). The error bars in the left panel represent only the statistical uncertainty, computed as the square root of the event number. The corresponding ratio is shown in the right panel of the figure. The standard values of the axial mass and zero strangeness have been assumed in all the calculations. Data are taken from~\cite{miniboonenc, Perevalov_phdthesis, NCEMiniBooNE}. Taken from Ref.~\cite{Ivanov:2015wpa}.} \label{fig:Martin2015etal_ratio}
\end{figure}

In figure~\ref{fig:Martin2015etal_ratio} the spectra corresponding to the numerator and denominator entering the ratio between $\nu$ scattering from proton and nucleon (proton plus neutron) (left panels) and the ratio, computed by dividing the two samples (right panels) within the various models, are presented in  comparison with the MiniBooNE data~\cite{miniboonenc}. The experimental numerator and denominator in the left panels are taken from~\cite{NCEMiniBooNE}, where the data are reported without the corresponding errors (only statistical errors are included in the figure). The standard value of the axial mass and zero strangeness have been assumed in the calculations. Here it is important to note that the dispersion between the results of different models tends to cancel when this ratio is considered.  Actually, there are some differences in the figure between the results of the different models for the numerator and the denominator, which are well described by  all the models for kinetic energies of the outgoing nucleon $T_{\mathrm N} > 350$~MeV, while all the  results for the ratio are within the error bars for all values  of $T_{\mathrm N}$.  This result clearly shows that the proton/nucleon ratio is insensitive to nuclear model effects and to FSI, and hence it may provide information that improves our present knowledge of the electroweak nucleon structure, in particular, of the nucleon strangeness.

The theoretical results presented in Ref.~\cite{Ivanov:2015wpa} show that the inclusion of FSI effects in the spectral-function-based calculations leads to a slight depletion of the cross section being in close agreement with the RFG prediction. The inclusion of FSI effects in the RGF model leads to larger cross sections, in good agreement with the data. On the contrary, the SuSA and, in particular, the RMF approaches lead to significantly smaller differential cross sections at low values of $Q^2$ ($\leq0.6 - 0.8$GeV$^2$), also departing from the data.  Another point of relevance when comparing the different models is the softer $Q^2$ dependence (with a smaller slope) shown by the RMF cross section. Whereas it is clearly below the other curves at low $Q^2$ (up to $0.5 - 0.6$~GeV$^2$), it crosses them, providing the largest contribution, at higher $Q^2$. This result can be taken as an indication of the particular sensitivity of NC processes to the specific description of FSI effects. All calculations are based on the IA, \emph{i.e.}, they do not include effects beyond the one-body approach, such as, for example, 2p-2h contributions induced by MEC. These ingredients have been shown to be very important in the analysis of neutrino-nucleus scattering processes. In particular, they produce a significant enhancement of the cross section at low to moderate values of the transferred four-momentum. This is consistent with theoretical predictions that clearly underestimate data (with exclusion on the RGF model) for such kinematic regions.

\begin{figure}[tb]
\centering
\includegraphics[width=0.6\textwidth]{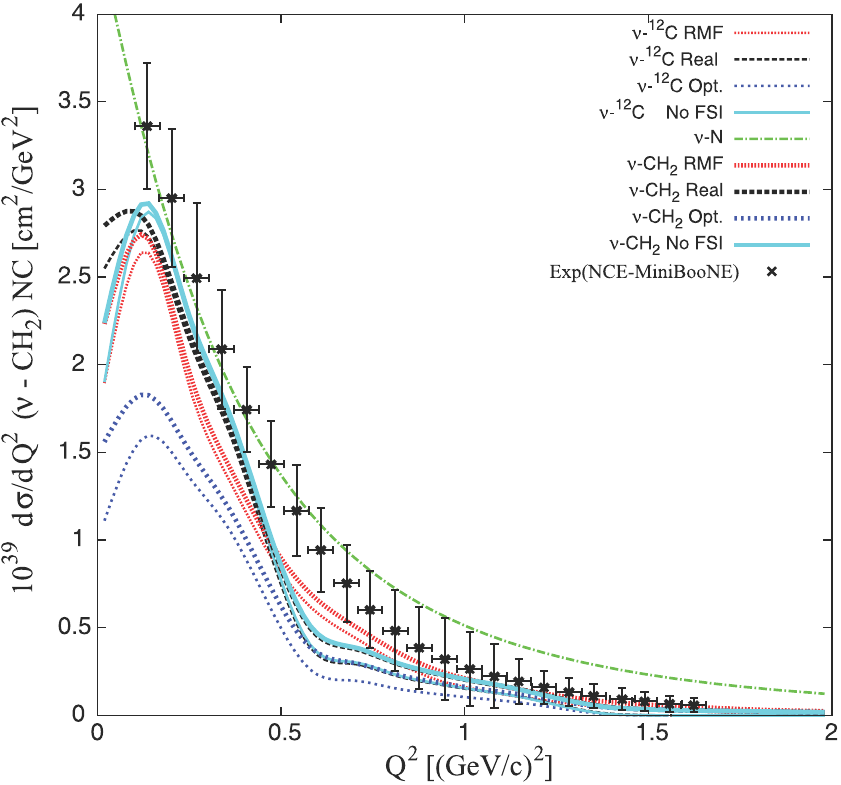}
\vspace{-2mm}
\caption{Flux-averaged differential cross sections $d\sigma/dQ^2$ per nucleon. The solid curve is obtained with  the RPWIA (No FSI), the dotted curve with the RDWIA and the EDAD1 ROP (Opt), the dashed curve with only the real part of the ROP (Real), and the short-dotted curve with the RMF. Thin curves are the results only for the $^{12}$C target and thick curves are for the CH$_2$ target, including hydrogen. Data are taken from~\cite{miniboonenc}. For a reference, results for $\nu$-N are also presented as a dash-dotted (green) curve. Taken from Ref.~\cite{PhysRevC.88.044615}.} \label{fig:Kim2013}
\end{figure}

Within the framework of a relativistic single-particle model~\cite{Kim:2007ez} the effect stemming from the FSI between the knocked-out nucleon and the residual nucleus on the inclusive $\nu(\overline{\nu})$--$^{12}$C scattering via the NC reaction has been investigated in Ref.~\cite{PhysRevC.88.044615}. An example of the results is presented in figure~\ref{fig:Kim2013} in comparison with the MiniBooNE data~\cite{miniboonenc}. The RPWIA results (labeled as No FSI in the figure) are compared with the corresponding results where the wave functions of the knocked-out nucleons are obtained by solving the Dirac equation with three different potentials: the phenomenological EDAD1 ROP (labeled as Opt.), only the real part of the EDAD1 potential (labeled as Real), and the same RMF potential as for the bound nucleons, generated by TIMORA~\cite{HOROWITZ1981503}) (labeled as RMF).  The differential cross section per nucleon on CH$_2$ target are evaluated using equation~(\ref{eq.miniboonenc}), where the efficiency correction functions are taken as unity. The standard value of the axial mass has been used in the calculations. The calculated cross sections underestimate the MiniBooNE data by $\sim 20-30\%$, which is consistent with other similar calculations.

\begin{figure}[tb]
\centering
\includegraphics[width=0.6\textwidth]{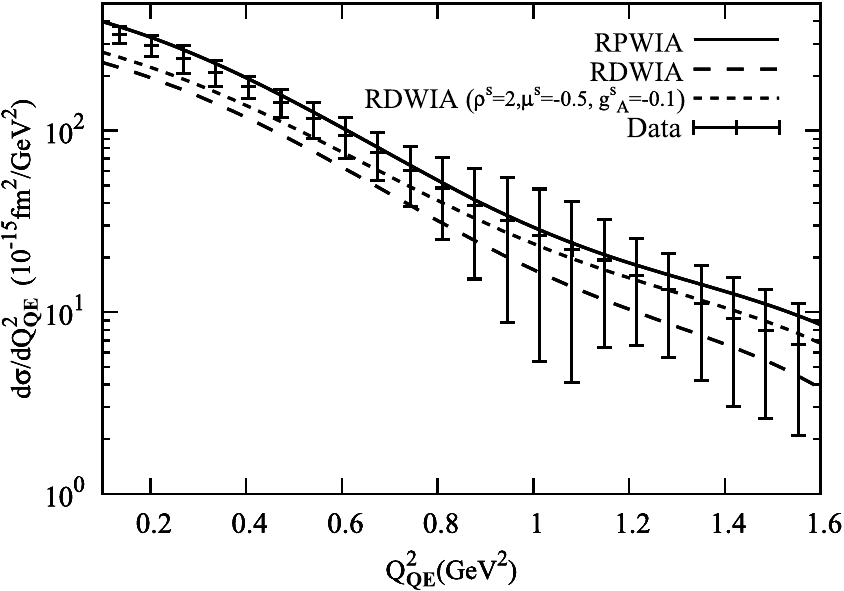}
\vspace{-2mm}
\caption{Neutral-current quasi-elastic flux-averaged differential cross section for neutrino scattering on CH$_2$ calculated in the RPWIA and RDWIA of Ref.~\cite{Hedayatipoor_2017} and compared with the MiniBooNE data~\cite{miniboonenc}. Taken from Ref.~\cite{Hedayatipoor_2017}.\label{fig:Hedayati_2018}}
\end{figure}

For completeness, recent RPWIA and RDWIA results~\cite{Hedayatipoor_2017} for NCE neutrino scattering on carbon in comparison with the MiniBooNE data are presented in figure~\ref{fig:Hedayati_2018}. The standard value of the axial mass has been used in the calculations. The RPWIA cross section obtained without strangeness reproduces the data quite well, in contrast with the RPWIA results~\cite{PhysRevC.88.025502} shown in the right panel of figure~\ref{fig:Meucci2011}; the RDWIA cross section underestimates the data in the low-$Q^2$ region up to about $0.7$~GeV$^2$. The RDWIA cross section with no strangeness is similar to the RMF calculation~\cite{PhysRevC.88.025502} presented in figure~\ref{fig:Martin2015etal_dsdq2}.

\begin{figure}[tb]
\centering
\includegraphics[width=0.99\textwidth]{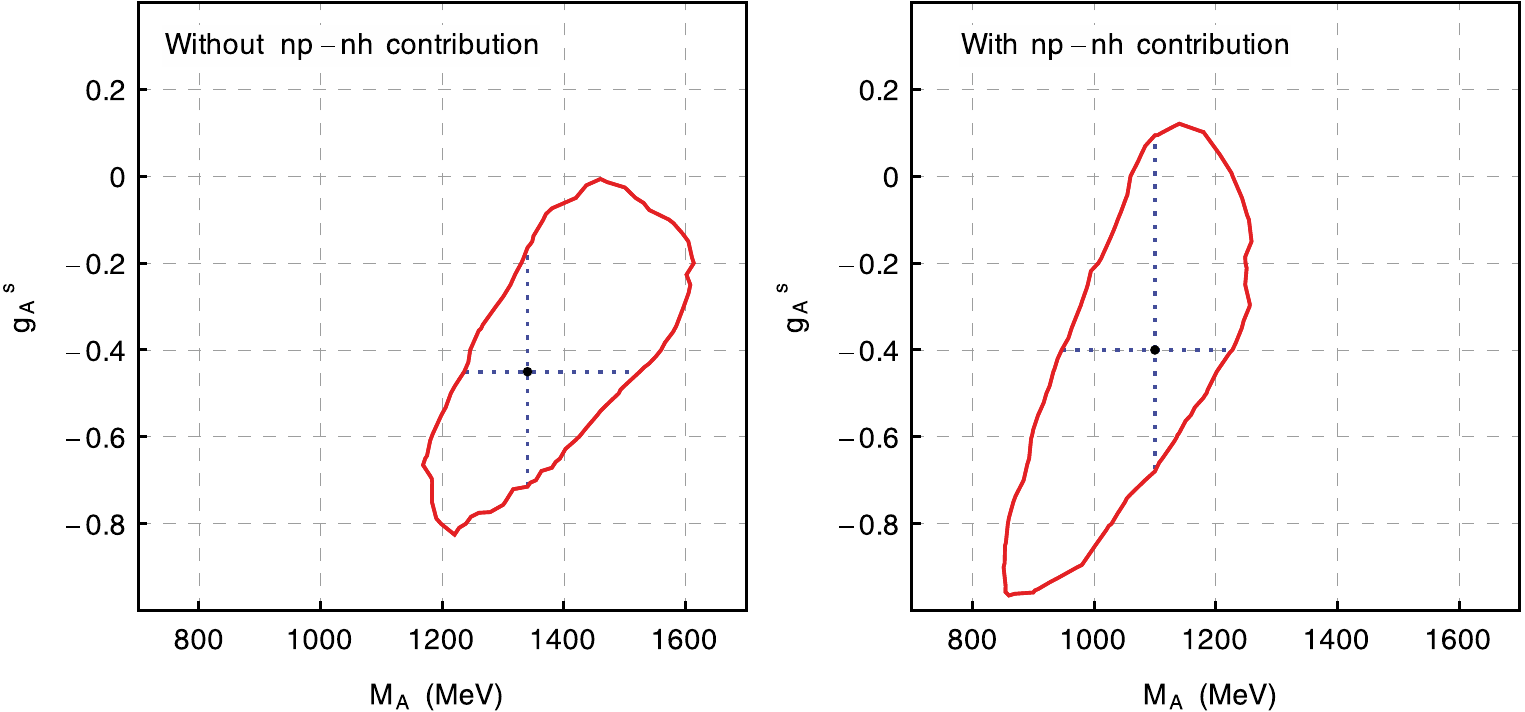}
\vspace{-2mm}
\caption{$1\sigma$ error contour for ($M_A$, $g^s_A$) parameters obtained from $\chi^2$ [see equation~(17), Ref.~\cite{PhysRevC.88.024612}], but only for the total reconstructed kinetic energy of the final state nucleons. Dots denote $\chi^2$ minima. Taken from Ref.~\cite{PhysRevC.88.024612}. \label{fig:PhysRevC.88.024612_Fig.2}}
\end{figure}

\begin{figure}[tb]
\centering
\includegraphics[width=0.99\textwidth]{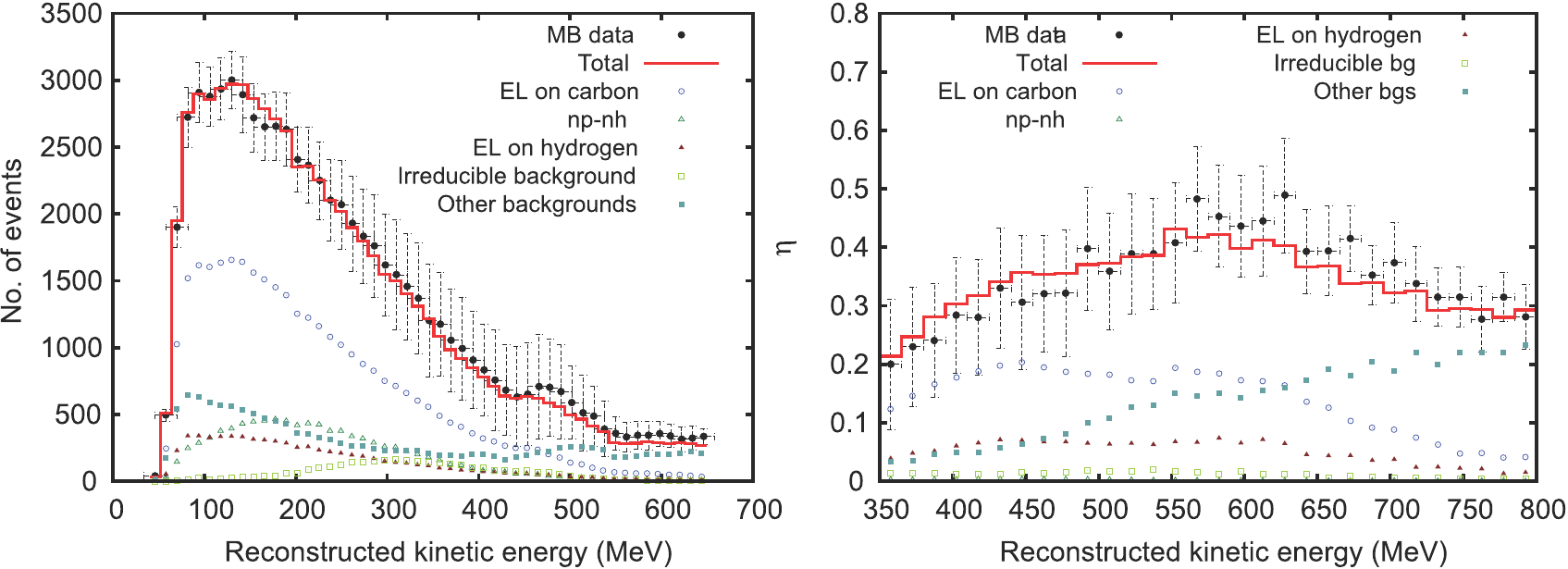}
\vspace{-2mm}
\caption{Left panel: The distribution of the total reconstructed kinetic energy of the final state nucleons, broken down to individual contributions from elastic scattering on carbon, np-nh, elastic scattering on hydrogen, irreducible  background, and other backgrounds. The NuWro results are obtained with the $M_A = 1.10$~GeV and $g^s_A = -0.4$ values. Right panel: The same as in the left panel but for $(\nu p \rightarrow \nu p)/(\nu N \rightarrow \nu N)$ ratio. Taken from Ref.~\cite{PhysRevC.88.024612}. \label{fig:PhysRevC.88.024612_Fig.34}}
\end{figure}

The results of the analysis of the MiniBooNE data for NCE neutrino scattering off CH$_2$ with the NuWro Monte Carlo generator are presented in Ref.~\cite{PhysRevC.88.024612}. A simultaneous fit to two theoretical model parameters obtained the values $M_A = 1.10^{+0.13}_{-0.15}$~GeV and $g^s_A = -0.4^{+0.5}_{-0.3}$, consistent with the older evaluations based on the neutrino-deuteron scattering data. In the analysis of the NCE data the effective transverse enhancement (TE) model~\cite{Bodek2011} is used, because to date this is the only NC np-nh model available in NuWro.  The results for the simultaneous two dimensional fits without and with the np-nh contribution included in the NuWro simulations together with 68\% confidence regions are shown in figure~\ref{fig:PhysRevC.88.024612_Fig.2}. The inclusion of np-nh contributions makes the best fit result for $M_A$ consistent with the world average and confirms that the difference between recent and older axial mass measurements can be explained by taking into account two-body current contributions. The value of the strange quark contribution is consistent with zero. It would be interesting to repeat this analysis using one of the microscopic models of the NC two-body current contribution. The best fit for the distribution of the total reconstructed kinetic energy of the final-state nucleons and the overall NuWro prediction (broken down to individual contributions from elastic scattering on carbon, elastic scattering on hydrogen, two-body current, irreducible background, and other backgrounds) is demonstrated in figure~\ref{fig:PhysRevC.88.024612_Fig.34}.

\begin{figure}[tb]
\centering
\includegraphics[width=0.6\textwidth]{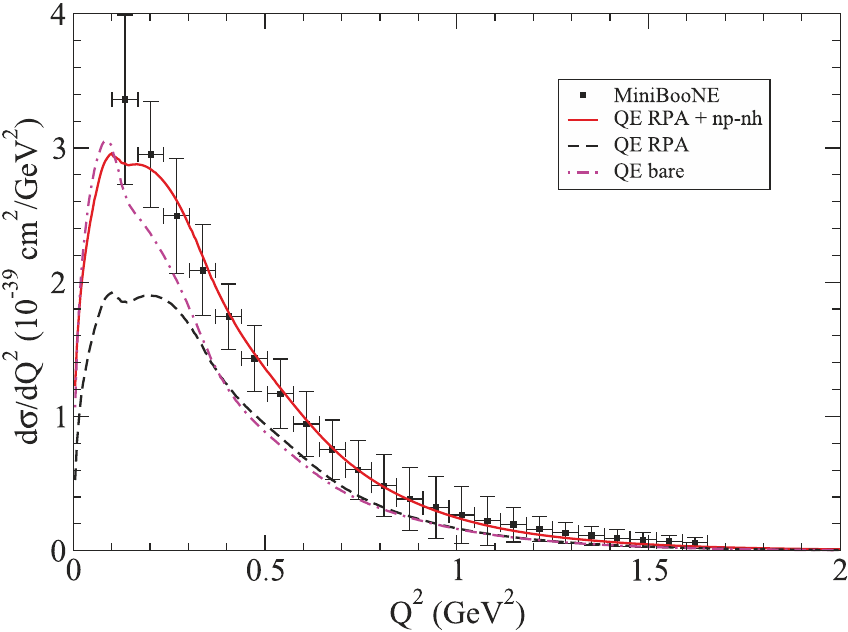}
\vspace{-2mm}
\caption{MiniBooNE flux-averaged NCE $Q^2$ distribution per nucleon. (Dashed curve) Pure quasielastic (1p-1h); (solid curve) with the inclusion of np-nh component; (dot-dashed line) bare distribution. The experimental MiniBooNE points are taken from Ref.~\cite{miniboonenc}. Taken from Ref.~\cite{Martini:2011wp}.\label{fig:Martini_NUQE_2011_fig11}}
\end{figure}

The double differential neutrino-carbon quasielastic cross sections, as well as the $Q^2$ distribution for charged and neutral current, are investigated in Ref.~\cite{Martini:2011wp}  within the RPA approach, incorporating relativistic corrections in the nuclear response functions and including the multinucleon component. In the case of NCE scattering, the final lepton, a neutrino, is not observed and the variable $Q^2$ is obtained indirectly from the kinetic energy of the ejected nucleons. In this case it is not quite clear how the multinucleon component shows up in the experimental data. However, the same problem of the axial mass also seems to emerge from these data~\cite{Benhar:2011wy, Butkevich:2011fu}. The data are for CH$_2$ instead of pure carbon as in the theory, but the difference between the two cases has been shown to be small~\cite{Benhar:2011wy}. The comparison of the theoretical predictions with data is shown in figure~\ref{fig:Martini_NUQE_2011_fig11}. It turns out that the combination of the RPA quenching and the np-nh piece leads to a good agreement with data without the need to increase the standard value of the axial mass $M_A = 1.03$~GeV.

\section{Concluding remarks \label{sec:conclusions}}

After the discovery of neutrino oscillations at the end of the twentieth century~\cite{PhysRevLett.81.1562, PhysRevLett.87.071301, PhysRevLett.90.021802, nobelprize2015} neutrino physics is entering the age of precision measurements. Qualitative improvements on the determination of neutrino properties require percent-level control of systematic uncertainties. The treatment of nuclear effects in the interaction of neutrinos with the nuclei in the neutrino detectors is one of the main sources of systematic uncertainty. Neutrino-nucleus scattering has therefore gained in recent years a wide interest. Both CC and NC scattering must be studied to obtain a complete information on neutrino-nucleus interactions.

Weak neutral currents were predicted and experimentally confirmed shortly thereafter in 1973~\cite{nobelprize1979, HASERT1973138}. Almost fifty years later, several experiments are now able to produce high-statistics cross section measurements that can be used to test the reliability of the models in the description of $\nu(\nubar)$-nucleus interactions.

In this paper we have reviewed the recent developments in the theoretical description of neutral current quasielastic (NCQE) $\nu(\nubar)$-nucleus scattering. Quasielastic scattering is the main interaction mechanism in the energy region at the core of the energy distribution for many neutrino experiments. We have given an overview of the available theoretical models and we have discussed the current status of the comparison between the numerical predictions of the models and the experimental data.

In the following we draw our concluding remarks and summarize the main challenges and open questions in the theoretical description of NCQE $\nu(\nubar)$-nucleus scattering.

\begin{itemize}

\item Most of the models applied to QE $\nu(\nubar)$-nucleus scattering were originally developed for QE electron-nucleus  scattering. In spite of many similar aspects, the two situations are different, for instance, in the nuclear current and in the kinematic situation, and it is not guaranteed that a model able to describe electron-scattering data will be able to describe neutrino-scattering data with the same accuracy. The analysis of the electron scattering data can serve, however,  not only to determine the vector current part of the neutrino cross sections (related to the electromagnetic response through CVC), but also to constrain nuclear effects more stringently than from neutrino scattering data. Therefore, only models that have been validated from the comparison with electron scattering data should be extended to $\nu(\nubar)$-nucleus scattering.

\item It would in principle be desirable to have consistent models for QE NC and CC $\nu(\nubar)$-nucleus scattering, but the two cases present  some differences. In CC scattering the inclusive process, where only the charged final lepton is detected (similar to the inclusive \ee scattering), can be considered. In contrast, in NC neutrino-scattering experiments only the emitted nucleon can be detected and the cross section is integrated over the energy and angle of the final lepton. Also the state of the residual $(A-1)$-nucleon system is not determined and the cross section is summed over all the available final nuclear states. The cross sections are therefore inclusive in the leptonic sector and semi-inclusive in the hadronic sector. This means that while the extension of the models developed for the inclusive \ee scattering to CCQE scattering is appropriate, the extension of the same models to the semi-inclusive NCQE scattering can be less appropriate. The models can include contributions  that are present in the inclusive but not in the semi-inclusive cross sections.

\item The word `quasielastic' has not the same meaning in electron and neutrino-scattering experiments. In electron-scattering experiments the incident flux and energy are known, the energy and momentum transfer, $\omega$ and $\nq$, are clearly determined and establish the QE kinematic region where one-nucleon knockout is expected to be the dominant reaction mechanism. In neutrino experiments the flux is uncertain, the beam energy is not known, and $\omega$ and $\nq$ are not fixed. The beam energy reconstruction, and hence flux unfolding, is possible only in model-dependent ways. Therefore measurements produce flux-integrated and then averaged differential cross sections, which contain events for a wide range of incoming energies and for different values of $\omega$ and $\nq$, corresponding not only to the QE region,  but also to other kinematic regions where different reaction mechanisms can be important. In $\nu(\nubar)$-nucleus  scattering QE usually refers to events where there are no pions in the observed final state and the definition QE-like or $0\pi$ events would be more appropriate. As a consequence, models for QE scattering could be unable to describe CCQE and NCE data unless all other processes contributing to the experimental cross sections are taken into account.

\item Models based on the IA and including only one-nucleon knockout contributions usually underpredict the experimental CCQE and NCE cross sections, unless a value of the axial mass $M_A$ significantly larger than that obtained from deuteron data is adopted in the calculations. It is generally impossible to describe, within the same model, both CCQE and NCE data using the same value of $M_A$ (although there are exceptions, for example, the oversimplified  RFG model can describe the MiniBooNE CCQE and NCE data with $M_A\approx1.35$~GeV). It has been suggested that the increased value of $M_A$ required to reproduce the experimental cross sections can be understood as a parameterization of nuclear effects neglected in models based on the IA. These effects, which cannot be parameterized by a simple enhancement of $M_A$, can be different in NC and CC scattering and for different data sets and should be included explicitly and carefully in the theoretical models.

\item Different models including contributions beyond IA can describe the experimental cross sections. We note that some of these models have been used for CCQE and have not yet been extended to NCQE scattering, or have been developed for the inclusive but not for the semi-inclusive scattering. Nuclear effects beyond IA, such as FSI, rescattering, multi-nucleon processes, non nucleonic contributions, np-nh excitations, two-body MEC, are accounted for in the models within different frameworks and approximations, and it is not easy to evaluate the role of the various approximations in the different models. The relevance of a specific contribution obviously depends on the assumptions and approximations that have been adopted in the models. Although it would  in principle be desirable that all the necessary contributions are included in a model consistently, we are aware that in practice a model able to account for all nuclear effects in a complete and consistent manner represents a tremendous and challenging task, perhaps a too ambitious goal.

\item  Even if a complete and fully consistent model appears a tremendous task, several issues should be addressed. It is generally accepted that multinucleon processes and two-body currents give a significant contribution to the description of data, both for CC and NC scattering. These contributions should anyhow be added to the `pure' QE contribution given by the IA as consistently as possible within the models. Moreover, in the case of NCE scattering the variable $Q^2_{QE}$ is obtained indirectly from the kinetic energy of the ejected nucleons [see equation~(\ref{eq:recQ2NC})], and it is not quite clear how the multinucleon component shows up in the experimental data.

\item There are uncertainties also in the description of the `pure' QE $\nu(\nubar)$-nucleus interaction. For instance, long- and short-range correlations, Fermi momentum description, binding energy corrections, FSI, the modeling of the axial-vector and strange form factors need the identification of parameter uncertainties and of the errors they bring along in the theoretical results. Moreover, a deep understanding of the theoretical results  and of  the discrepancies among them would require a more careful treatment of interferences between different nuclear effects and a meticulous study of possible double counting.

\item Flux-integrated cross sections, in terms of leptonic variables (CCQE processes) and outgoing nucleons (NCQE processes), provide only partial information on the kinematics. Precise measurements and predictions of hadronic final states are clear next steps. New higher-resolution neutrino detectors, such as liquid argon Time Projection Chamber (TPC), will be able to give more detailed information on final states. Moreover, the experimental cross sections require a good knowledge of the incident neutrino flux. Improved hadro-production measurements are an essential ingredient to ensure reliable flux predictions. This allows to fix the kinematics of given neutrino interactions, to further constrain interaction models, and to reduce as much as possible the model-dependence of their results, which eventually allows higher-precision neutrino-oscillation experiments.

\item Future neutrino experiments will provide large amounts of data. For a correct interpretation of the data it is important to develop models able to cover all the experimental needs, from 200~MeV to 10~GeV (or more), and adequate MC implementation of all reaction channels in the entire neutrino energy range. A better quantitative evaluation of the $\nu_{\mu}$ and $\nu_e$ cross sections will be very important for CP violation measurements in the leptonic sector in future experiments.

\end{itemize}

\ack{
  We are deeply grateful to Anton N. Antonov for many helpful discussions, for his valuable advice and careful reading of the draft.

}

\bibliography{nc_rev}

\end{document}